\documentclass[12pt]{article}
\usepackage{graphicx}
\usepackage[margin=1.25in]{geometry}
\usepackage[usenames,dvipsnames]{color}
\usepackage{url}

\newcommand\pubnumber{SLAC--PUB--17142}
\newcommand\pubdate{August, 2017}

\def\SLAC{SLAC,
    Stanford University, Menlo Park, CA 94025, USA}
\def\doeack{\footnote{Work supported by the US Department of Energy,
                     contract DE--AC02--76SF00515.}}

\def\Title#1{\begin{center} {\Large #1 } \end{center}}
\def\Author#1{\begin{center}{ \sc #1} \end{center}}
\def\Address#1{\begin{center}{ \it #1} \end{center}}

\newcommand\pubblock{\rightline{\begin{tabular}{l} \pubnumber\\
         \pubdate \end{tabular}}}
\newenvironment{Abstract}{\begin{quotation} \begin{center}
                       ABSTRACT
     \end{center}\bigskip  }{\end{quotation}}
\newenvironment{Presented}{
      \begin{center}\begin{large}}{\end{large}\end{center} }

\def\beq{\begin{equation}}
\def\eeq#1{\label{#1}\end{equation}}
\def\eeqn{\end{equation}}

\newenvironment{Eqnarray}%
   {\arraycolsep 0.14em\begin{eqnarray}}{\end{eqnarray}}
\def\beqa{\begin{Eqnarray}}
\def\eeqa#1{\label{#1}\end{Eqnarray}}
\def\eeqan{\end{Eqnarray}}
\def\CR{\nonumber \\ }

\def\leqn#1{(\ref{#1})}

\let\bar=\overbar

\def\etal{{\it et al.}}

\def\VEV#1{\left\langle{ #1} \right\rangle}
\def\bra#1{\left\langle{ #1} \right|}
\def\ket#1{\left| {#1} \right\rangle}

\def\lsim{\mathrel{\raise.3ex\hbox{$<$\kern-.75em\lower1ex\hbox{$\sim$}}}}
\def\gsim{\mathrel{\raise.3ex\hbox{$>$\kern-.75em\lower1ex\hbox{$\sim$}}}}

\def\L{{\cal L}}
\def\M{{\cal M}}

\def\L{{\cal L}}

\def\tr{{\mbox{\rm tr}}}
\def\half{\frac{1}{2}}

\def\third{\frac{1}{3}}
\def\tthird{\frac{2}{3}}

\def\del{\partial}
\def\Dslash{\not{\hbox{\kern-4pt $D$}}}
\def\dslash{\not{\hbox{\kern-2pt $\del$}}}

\def\Dlr{\mathrel{\raise1.5ex\hbox{$\leftrightarrow$\kern-1em\lower1.5ex\hbox{$D$}}}}

\def\ee{e^+e^-}
\def\sstw{\sin^2\theta_w}

\def\msb{{\bar{\scriptsize M \kern -1pt S}}}

\def\drb{{\bar{\scriptsize D \kern -1pt R}}}

\def\eps{\epsilon}

\makeatletter
\def\section{\@startsection{section}{0}{\z@}{5.5ex plus .5ex minus
 1.5ex}{2.3ex plus .2ex}{\large\bf}}
\def\subsection{\@startsection{subsection}{1}{\z@}{3.5ex plus .5ex minus
 1.5ex}{1.3ex plus .2ex}{\normalsize\bf}}
\def\subsubsection{\@startsection{subsubsection}{2}{\z@}{-3.5ex plus
-1ex minus  -.2ex}{2.3ex plus .2ex}{\normalsize\sl}}

\renewcommand{\@makecaption}[2]{%
   \vskip 10pt
   \setbox\@tempboxa\hbox{\small #1: #2}
   \ifdim \wd\@tempboxa >\hsize     
       \small #1: #2\par          
     \else                        
       \hbox to\hsize{\hfil\box\@tempboxa\hfil}
   \fi}

 \def\citenum#1{{\def\@cite##1##2{##1}\cite{#1}}}
 
\newcount\@tempcntc
\def\@citex[#1]#2{\if@filesw\immediate\write\@auxout{\string\citation{#2}}\fi
  \@tempcnta\z@\@tempcntb\m@ne\def\@citea{}\@cite{\@for\@citeb:=#2\do
    {\@ifundefined
       {b@\@citeb}{\@citeo\@tempcntb\m@ne\@citea\def\@citea{,}{\bf ?}\@warning
       {Citation `\@citeb' on page \thepage \space undefined}}%
    {\setbox\z@\hbox{\global\@tempcntc0\csname b@\@citeb\endcsname\relax}%
     \ifnum\@tempcntc=\z@ \@citeo\@tempcntb\m@ne
       \@citea\def\@citea{,}\hbox{\csname b@\@citeb\endcsname}%
     \else
      \advance\@tempcntb\@ne
      \ifnum\@tempcntb=\@tempcntc
      \else\advance\@tempcntb\m@ne\@citeo
      \@tempcnta\@tempcntc\@tempcntb\@tempcntc\fi\fi}}\@citeo}{#1}}
\def\@citeo{\ifnum\@tempcnta>\@tempcntb\else\@citea\def\@citea{,}%
  \ifnum\@tempcnta=\@tempcntb\the\@tempcnta\else
  {\advance\@tempcnta\@ne\ifnum\@tempcnta=\@tempcntb \else\def\@citea{--}\fi
    \advance\@tempcnta\m@ne\the\@tempcnta\@citea\the\@tempcntb}\fi\fi}
\makeatother

\def\VmA{ $V$--$A$}

\begin{document}
\begin{titlepage}
\pubblock

\vfill
\Title{Lectures on the Theory of the Weak Interaction}
\vfill
\Author{Michael E. Peskin\doeack}
\bigskip
\Address{\SLAC}
\bigskip
\vfill
\begin{Abstract}
I review aspects of the theory of the weak interaction in a set of
lectures
originally presented at the 2016 CERN-JINR European  School of
Particle Physics.   The
topics discussed are:  (1) the experimental basis of the $V$--$A$
structure
of the weak interaction; (2) precision electroweak measurements at
the $Z$ resonance; (3) the Goldstone Boson Equivalence Theorem; (4)
the Standard Model theory of the Higgs boson; (5)  the future program
of precision study of the Higgs boson.
\end{Abstract}
\vfill
\begin{Presented}
Lectures presented at the CERN-JINR\CR
European School of Particle
Physics\CR 
Skeikampen,
Norway,  
   June 15-28, 2016 
\end{Presented}
\vfill

\newpage

\hbox to\hsize{ \ }

\newpage

\tableofcontents
\end{titlepage}

\def\thefootnote{\fnsymbol{footnote}}
\setcounter{footnote}{0}

\newpage

\section{Introduction}

Today, all eyes in particle physics are on the Higgs boson.   This
particle has been central to the structure of our theory of weak
interactions ever since Weinberg and Salam first wrote down what we
now call the Standard Model of this interaction in
1967~\cite{Weinberg,Salam}.  As our understanding of particle physics
developed over the following decades. what lagged behind was our
knowledge of this particle and its interactions.  Increasingly, the
remaining mysteries of particle physics became centered on this
particle and the Higgs field of which it is a part.

In 2012, the Higgs boson was finally discovered by the ATLAS and CMS 
experiments at the LHC~\cite{ATLAS,CMS}.  Finally, 
we have the opportunity to study
this particle in detail and to learn some of its secrets by direct
observation.  Many students at this summer school, and many others
around the world, are involved in this endeavor.   So it is worthwhile
to review the theory of the Higgs boson and the broader theory of weak
interactions in which it is embedded.   That is the purpose of these
lectures.

To learn where we are going, it is important to understand thoroughly
where we have been.  For this reason, the first half of this lecture
series is 
devoted to historical topics. In Section 2, I  review the basic
formulae of the Standard Model and set up my notation. An important
property of the Standard Model is that, unexpectedly at first sight, 
charge-changing weak interactions couple only to left-handed-polarized
fermions.  This structure, called the \VmA\ interaction, is the reason
that we need the Higgs field in the first place.   In 
Section~3, I  review the most convincing experimental tests of \VmA.
Section~4 reviews the precision measurements on the weak
interaction made possible by the $\ee$ experiments of the 1990's at
the $Z$ resonance.  These experiments confirmed the basic structure
of the Standard Model and made the Higgs field a necessity.

One aspect of the Higgs field that is subtle and difficult to
understand but very powerful it is application is the influence of 
the Higgs field on  the high-energy dynamics of vector bosons $W$ and
$Z$.   Section~5 is devoted to this topic.   The physics of $W$ and
$Z$ bosons at high energy is full of seemingly mysterious enhancements
and cancellations.  The rule that explains these is the connection to
the Higgs field through a result called the Goldstone Boson
Equivalence Theorem, first enunciated by Cornwall, Levin, and Tiktopoulos and
Vayonakis~\cite{Cornwall,Vayonakis}.    In Section~5, I explain this
theorem and illustrate the way it controls the energy-dependence of a 
number of interesting high-energy processes.

In Sections~6 and 7, I turn to the study of the Higgs boson itself.
Section~6 is devoted to the Standard Model theory of the Higgs boson.
I will review the general properties of the Higgs boson and explain in
some details its expected pattern of decay models.  Section~7 is
devoted to the remaining mysteries of the Higgs boson and the
possibility of their elucidation through a future program of precision
measurements.   

\section{Formalism of the Standard Model}

To begin, I write the formalism of the Standard Model (SM) in a form convenient
for the analysis given these lectures.   The formalism of the SM is 
standard material for students of 
particle physics, so I assume that you have seen this before.  It
is explained more carefully in many textbooks (for 
example, \cite{PeskinSchr,Schw}).

\subsection{Gauge boson interactions}

The SM is a gauge theory based on the symmetry group $SU(2)\times
U(1)$.   A gauge theory includes interactions mediated by vector
bosons, one boson for each generator of the gauge symmetry $G$.   The coupling
of spin 0 and spin $\half$ particles to these vector bosons is highly
restricted by the requirements of gauge symmetry.    The interactions
of these fermions and scalars with one another is much less
restricted, subject only to the constraints of the symmetry $G$ as a
global symmetry.   Thus, the theory of fermions and vector bosons is
extremely tight, while the introduction of a scalar field such as the
Higgs field introduces a large number of new and somewhat uncontrolled
interaction terms.

The SM contains 4 vector bosons corresponding to the 3 generators of
SU(2) and 1 generator of $U(1)$.  I will call these 
\beq
            A^a_\mu \ , \qquad   B_\mu \ , 
\eeqn
with $a = 1,2,3$.   These couple to fermion and scalar fields only through the
replacement of the  derivatives by covariant derivative
\beq
   \del_\mu \to    D_\mu  =   (\del_\mu - i g A^a_\mu t^a) \ , 
\eeqn
where $t^a$ is the generator of $G$ in the representation to which the
fermions or scalars are assigned.    For the SM,  fermion and scalar
fields are assign $SU(2)$, or weak isospin,  quantum numbers 0 or $\half$
and a $U(1)$, or hypercharge, quantum number $Y$.   The covariant
derivative is then written more explicitly as 
\beq
    D_\mu =  \del_\mu - i g A^a_\mu t^a  - i g' B_\mu Y \ , 
\eeq{covderiv}
with 
\beq
      t^a = 0  \ \mbox{for}\ I = 0 \ , \quad  
  t^a = {\sigma^a\over 2}  \ \mbox{for}\ I = \half \ .
\eeqn 

This formalism makes precise predictions for the coupling of the weak
interaction vector bosons to quarks and leptons, and to the Higgs
field.   To obtain the masses of the vector bosons, we need to make
one more postulate:   The Higgs field obtains a nonzero value in the
ground state of nature, the vacuum state, thus spontaneously breaking
the $SU(2)\times U(1)$ symmetry.   This postulate is physically very
nontrivial.   I will discuss its foundation and implications 
 in some detail in Section~7.  However, for now, I will consider this
 a known aspect of the SM.

We assign the Higgs field $\varphi$ the $SU(2)\times U(1)$ quantum
numbers $I = \half$, $Y= \half$.    The Higgs field is thus a spinor
in isospin space, a 2-component complex-valued vector of fields
\beq
     \varphi =  \pmatrix{ \varphi^+\cr \varphi^0\cr}
\eeqn
The action of an $SU(2)\times U(1)$ transformation on this field is
\beq
       \varphi \to \exp[ i \alpha^a {\sigma^a\over 2} + i \beta
       \half]\ \pmatrix{ \varphi^+\cr \varphi^0\cr}\ .
\eeq{SUtwotrans}
If $\varphi$ obtains a nonzero vacuum value, we can rotate this by an
$SU(2)$ symmetry transformation into the form
\beq
\VEV{\varphi} =   {1\over \sqrt{2} } \pmatrix{0\cr v \cr} \ . 
\eeq{Higgsvev}
where $v$ is a nonzero value with the dimensions of GeV.  Once
$\VEV{\varphi}$ is in this form, any $SU(2)\times U(1)$ transformation
will disturb it, except for the particular direction
\beq
               \alpha^3 = \beta \ , 
\eeqn
which corresponds to  a $U(1)$ symmetry generated by $Q = (I^3 + Y)$.
We say that the $SU(2)\times U(1)$ symmetry generated by $(I^a,Y)$ is 
spontaneously broken, leaving unbroken only the $U(1)$ subgroup
generated by $Q$.

This already gives us enough information to work out the mass spectrum
of the vector bosons.   The kinetic energy term for $\varphi$ in the
SM Lagrangian is 
\beq
      \L =    \biggl|  D_\mu\varphi \biggr|^2
\eeqn
Replacing $\varphi$ by its vacuum value \leqn{Higgsvev}, this becomes
\beq
      \L   =  \half  \pmatrix{0 & v\cr}  (g {\sigma^a\over 2} A^a_\mu
      + g' \half B_\mu)^2 \pmatrix{0\cr v\cr} \ .
\eeqn
Multiplying this out and taking the matrix element, we find, from the
$\sigma^1$ and $\sigma^2$ terms
\beq
         {g^2v^2 \over 8}  \biggl[  (A^1_\mu)^2 + (A^2_\mu)^2 \biggr]  \ , 
\eeq{onetwoterms}
and, from the remaining terms
\beq
         {v^2 \over 8} \biggl(  - g A^3_\mu + g' B_\mu \biggr)^2 
\eeqn
So, three linear combinations of the vector fields obtain mass by
virtue of the spontaneous symmetry breaking.  This is the mechanism of
mass generation called the {\it  Higgs mechanism}~\cite{BE,Higgs,GHK}.
  The mass eigenstates
are
\beqa
     W^\pm =  (A^1\mp i A^2)/\sqrt{2} &\qquad&   m_W^2 = g^2v^2/4 \CR
    Z =  (gA^3 - g' B)/\sqrt{g^2 + g^{\prime 2}} &\qquad&   m_Z^2= 
 (g^2 + g^{\prime 2})v^2/4 \CR
    A =  (g'A^3 + gB )/\sqrt{g^2 + g^{\prime 2}} 
&\qquad&   m_A^2 =  0 
\eeqa{Vmasses}
As we will see more clearly in a moment, the massless boson $A$ is
associated with the unbroken gauge symmetry $Q$.    The combination of
local gauge symmetry and the Higgs mechanism is the only known way to
give mass to a vector boson that is consistent with Lorentz invariance
and the positivity of the theory.

The linear combinations in \leqn{Vmasses} motivate the definition of
the {\it weak mixing angle} $\theta_w$, defined by 
\beqa
     \cos\theta_w \equiv  c_w &=&   g/\sqrt{g^2 + g^{\prime 2}} \CR
     \sin\theta_w \equiv  s_w &=&   g'/\sqrt{g^2 + g^{\prime 2}} \ .
\eeqa{cwsw}
The factors $c_w$, $s_w$ will appear throughout the formulae that
appear in these lectures.   For reference, the value of the weak
mixing angle turns out to be such that
\beq
                 s_w^2 \approx  0.231
\eeq{swval}
I will describe the measurement of $s_w$ in some detail in Section 3.

An important relation that follows from \leqn{Vmasses}, \leqn{cwsw} is 
\beq
     m_W =  m_Z\  c_w  \ .
\eeq{WZrel}
This is a nontrivial consequence of the quantum number assignments for
the Higgs field, and the statement that the masses of $W$ and $Z$ come
only from the vacuum value of $\varphi$.  Using the Particle Data
Group values for the masses~\cite{PDG} and the value \leqn{swval}, we
find
\beq
     80.385~\mbox{GeV} \approx  91.188~\mbox{GeV} \cdot 0.877 =
     79.965~\mbox{GeV} \ .
\eeq{WZrelcheck}
so this prediction works well already at the leading order.   We will
see in Section 3 that, when radiative corrections are included, the
relation \leqn{WZrel} is satisfied to better than 1 part per mil.

Once we have the mass eigenstates of the vector bosons, the 
 couplings of quarks and leptons to these bosons can be worked out
 from the expresssion \leqn{covderiv} for the covariant derivative.
 The terms in \leqn{covderiv} involving $A^1_\mu$ and $A^2_\mu$
appear only for $I = \half$ particles and can be recast as
\beq
    - i {g\over \sqrt{2}} (W^+_\mu \sigma^+ + W^-_\mu \sigma^-) \ , 
\eeq{Wcoupling}
The $W$ bosons couple only to $SU(2)$ doublets, with universal
strength $g$.

The terms with $A^3_\mu$ and $B_\mu$ can similarly be recast in terms
of $Z_\mu$ and $A_\mu$,
\beqa
    - i g A^3_\mu - i g' B_\mu Y &=&  -i \sqrt{g^2 + g^{\prime 2}}
    \biggl[ c_w (c_w Z_\mu + s_w A_\mu) I^3 + s_w (-s_w Z_\mu + c_w
    A_\mu) \biggr] \CR
&= & -i \sqrt{g^2 + g^{\prime 2}}
    \biggl[ s_w c_w A_\mu( I^3 +Y) + Z_\mu (c_w^2 I^3 - s_w^2
    Y)\biggr] \CR
&= & -i \sqrt{g^2 + g^{\prime 2}}
    \biggl[ s_w c_w A_\mu( I^3 +Y) + Z_\mu (I^3 - s_w^2(I^3 +
    Y)\biggr] \ .
\eeqa{ZAcouple}
We now see explicitly that the massless gauge boson 
$A_\mu$ couples to $Q = (I^3 +Y)$, as we
had anticipated.   Its coupling constant is 
\beq  
    e =  s_w c_w \sqrt{g^2 + g^{\prime 2}} = {gg'\over \sqrt{g^2 +
        g^{\prime 2}} }\ .
\eeqn
We can then identify this boson  with the photon and the
coupling constant $e$ with the strength of electric charge.   The quantity
$Q$ is the (numerical) electric charge of each given fermion or boson
species.  The expression \leqn{ZAcouple} then simplifies to 
\beq
      - i e A_\mu Q - i {e\over s_w c_s}  Z_\mu  Q_Z \ , 
\eeq{AZcouple}
where the $Z$ charge is
\beq
    Q_Z =   (I^3 -  s_w^2 Q) \ .
\eeq{QZ}

To  complete the specification of the SM, we assign
the $SU(2)\times U(1)$ quantum numbers to the 
quarks and leptons in each generation.   As I will explain below, each
quark or lepton is build up from fields of left- and  right-handed
chirality, associated with massless left- and right-handed particles
and massless right- and left-handed antiparticles.  For the
applications developed  in Sections~3--5, it will almost
always be appropriate to ignore the masses of quarks and leptons,
so these quantum
number assignments will apply literally.
The generation of masses for quarks and leptons is part of the physics
of the Higgs field, which we will discuss beginning in Section~6.

In the SM, the left-handed fields are assigned $I=\half$,
and the right-handed fields are assigned $I = 0$.   It is not so easy
to understand how these assignments come down from fundamental
theory.  They are requred by experiment, as I will explain in later in
this section.

With this understanding, we can assign quantum numbers to the quarks
and leptons as
\beqa
  \nu_L  \ : \  I^3 = +\half, \ Y = -\half, \ Q = 0  &\qquad &   
\nu_R  \ : \  I^3 = 0, \ Y = 0, \ Q = 0 \CR
  e_L  \ : \  I^3 = -\half, \ Y = -\half, \ Q = -1  &\qquad &   
e_R  \ : \  I^3 = 0, \ Y = -1, \ Q = -1\CR
 u_L  \ : \  I^3 = +\half, \ Y = {1\over 6}, \ Q =\tthird  &\qquad &   
u_R  \ : \  I^3 = 0, \ Y = {2\over 3}, \ Q = \tthird  \CR
  d_L  \ : \  I^3 = +\half, \ Y = {1\over 6}, \ Q = -\third  &\qquad &   
d_R  \ : \  I^3 = 0, \ Y = -\third, \ Q =-\third \CR
\eeqa{QNos}
The $\nu_L$ and $e_L$, and the $u_L$ and $d_L$, belong to the same
$SU(2)$
multiplet, so they must be assigned the same hypercharge $Y$.
Note that \leqn{QNos}  gives the correct electric charge assignments for all
quarks and leptons.   The $\nu_R$ do not couple to the SM gauge
fields and will play no role in the results reviewed in these lectures.

\subsection{Massless fermions}

The idea that massless fermions can be separated into left- and
right-handed components will play a major role throughout these
lectures.
In this sentence, I introduce some notation that makes it especially
easy to apply this idea.   

To begin, write the the 4-component Dirac spinor  and the Dirac
matrices 
as 
\beq
      \Psi = \pmatrix{\psi_L\cr \psi_R\cr} \qquad   \gamma^\mu =
      \pmatrix{0 & \sigma^\mu \cr \bar\sigma^\mu & 0 \cr} \ , 
\eeq{fbasis}
with
\beq
    \sigma^\mu = (1,\vec\sigma)^\mu \qquad  
  \bar \sigma^\mu = (1,- \vec\sigma)^\mu \ .
\eeqn
In this representation, the vector current takes the form
\beq
   j^\mu =   \bar\Psi \gamma^\mu \Psi = \psi_L^\dagger \bar \sigma^\mu
   \psi_L  + \psi^\dagger_R \sigma^\mu \psi_R 
\eeqn
and splits neatly into pieces that involve only the L or R fields.
The L and R fields are mixed by the fermion mass term.  In
circumstances in which we can ignore the fermion masses, the L and R
fermion numbers are separately conserved.  We can treat $\psi_L$ and
$\psi_R$ as completely independent species and assign them different
quantum numbers, as we have already in \leqn{QNos}.   The label L, R
is called {\it chirality}.   For massless fermions, the chirality of
the fields and the helicity of the particles are identical.  For
massive fermions, there is a change of basis from the chirality states
to the helicity eigenstates.

The spinors for massless fermions are very simple.   In the basis
\leqn{fbasis},
we can write these spinors as
\beq
      U(p) = \pmatrix{u_L\cr u_R\cr } \qquad    V(p) = \pmatrix{v_R\cr
        v_L\cr } \ .
\eeqn
For massless fermions, where the helicity and chirality states are
identical, the spinors for a fermion with left-haned spin have $u_R =
0$ and the spinors for an antifermion with right-handed spin have 
$v_L = 0$; the opposite is true for a right-handed fermion and a
left-handed antifermion.   The nonzero spinor compoments for a
massless 
fermion
of energy $E$ take the form
\beqa
  u_L(p) =  \sqrt{2E}\  \xi_L   &\qquad &    v_R(p)   = \sqrt{2E} \  \xi_L \CR
  u_R(p) =  \sqrt{2E}\  \xi_R   &\qquad &    v_L(p)   =  \sqrt{2E} \  \xi_R
\eeqa{spinors}
where $\xi_R$ is the spin-up and $\xi_L$ is the spin-down 2-component
spinor along the direction of motion.  For example, for a fermion or
antifermion moving in the $\hat 3$ direction,
\beq
   \xi_L = \pmatrix{0\cr 1\cr}       \qquad   \xi_R = \pmatrix{1\cr
     0\cr} \ . 
\eeq{threeaxis}
Spinors for other directions are obtained by rotating these according
to the usual formulae for spin $\half$.   The reversal for
antifermions can be thought of by viewing right-handed (for example) 
antifermions as holes in the Dirac sea of left-handed fermions.
For a massive fermion moving in the $\hat 3$ direction, with 
\beq
       p^\mu = (E, 0, 0 , p)^\mu \ , 
\eeqn
 the solutions to the Dirac equation are
\beqa
   U_L(p)  = \pmatrix{\sqrt{E+p} \ \xi_L\cr \sqrt{E-p}\  \xi_L} &\qquad&
 V_R(p)  = \pmatrix{\sqrt{E+p} \ \xi_L\cr - \sqrt{E-p} \ \xi_L} \CR
   U_R(p)  = \pmatrix{\sqrt{E-p} \  \xi_R\cr \sqrt{E+p}\  \xi_R} &\qquad&
 V_L(p)  = \pmatrix{\sqrt{E-p}\  \xi_R\cr - \sqrt{E+p}\  \xi_R} \ ,
\eeqa{massivespinors}
with $\xi_L$, $\xi_R$ given by \leqn{threeaxis}. 
These formulae go over to \leqn{spinors} in the zero mass limit.

The matrix elements for creation or annihilation of a massless fermion
pair will appear very often in these lectures.   For annihilation of a
fermion pair colliding along the $\hat 3$ axis, 
\beqa
    \bra{0} j^\mu\ket{e^-_Le^+_R} &=&   v^\dagger_R \bar\sigma^\mu
    u_L\CR
&=& \sqrt{2E} \pmatrix{-1 & 0\cr} \ (1, -\sigma^1, -\sigma^2,
-\sigma^3) \ \sqrt{2E} \pmatrix{0\cr 1\cr} \ ,
\eeqan
Note that I have rotated the $e^+$ spinor appropriately by
180$^\circ$.  This gives
\beq
  \bra{0} j^\mu\ket{e^-_Le^+_R} = 2 E\  (0,1, -i, 0)^\mu \ . 
\eeqn
It is illuminating to write this as 
\beq
  \bra{0} j^\mu\ket{e^-_Le^+_R} = 2\sqrt{2}  E\  \eps_-^\mu \ , 
\eeq{leftann}
where 
\beq
   \eps_+^\mu  = {1\over \sqrt{2}}   (0,1, +i, 0)^\mu \qquad
     \eps_-^\mu  = {1\over \sqrt{2}}   (0,1, -i, 0)^\mu 
\eeq{epses}
are the vectors of $J^3 = \pm 1$ along the $\hat 3$ axis.   The total
spin angular momentum of the annihilating fermions ($J = 1$) is
transfered to the current and, eventually, to the final state.

More generally, we find
\beqa
  \bra{0} j^\mu\ket{e^-_Re^+_L}  &=& 2\sqrt{2} E \ \eps_+^\mu\CR
 \bra{0} j^\mu\ket{e^-_Le^+_R}  &=& 2\sqrt{2} E \ \eps_-^\mu\CR
 \bra{e^-_Re^+_L} j^\mu\ket{0}  &=& 2\sqrt{2} E \ \eps_+^{*\mu}\CR
 \bra{e^-_Le^+_R} j^\mu\ket{0}  &=& 2\sqrt{2} E \ \eps_-^{*\mu} \ .
\eeqa{anncreate}

For an annihilation process such as $e^-_Le^+_R\to \mu^-_L\mu^+_R$
with annihilation by a current and creation by another current, the
spinors appear as
\beq 
     (u^\dagger_L \bar \sigma^\mu v_R) (v^\dagger_R \bar \sigma_\mu
     u_L)  =  2 \,(2E)^2\, \eps^{\prime *}_- \cdot  \eps_-  \ .
\eeq{emurxn}
To evaluate this, rotate the $\eps_-$ vector for the muons into the
muon direction.  If the muons come off at polar  angle $\theta$, this
gives
\beq
  \eps^{\prime *}_- = {1\over \sqrt{2} } (0, \cos\theta, -i,
    -\sin\theta) \ . 
\eeqn
Then \leqn{emurxn} becomes
\beq
   2  (2E)^2 \, \eps^{\prime *}_- \cdot \eps_- =  s (1 + \cos\theta)   = -2 u  \ , 
\eeqn
in terms of the usual kinematic invariants $s$, $t$, $u$.    Another
way to write this is
\beq
  | (u^\dagger_L \bar \sigma^\mu v_R) (v^\dagger_R \bar \sigma_\mu
     u_L) |^2 = 4 \, (2 p_{e^-}\cdot p_{\mu^+})(2 p_{e^+}\cdot
     p_{\mu_-}) \ . 
\eeq{emusquare}
Similarly, for  $e^-_Le^+_R\to \mu^-_R\mu^+_L$, 
\beq
  | (u^\dagger_R \bar \sigma^\mu v_L) (v^\dagger_R \bar \sigma_\mu
     u_L) |^2 = 4 \, (2 p_{e^+}\cdot p_{\mu^+})(2 p_{e^-}\cdot
     p_{\mu_-}) \ . 
\eeq{emusquareLR}
It is a nice exercise to check these answers using the usual trace
theorems.  The trace theorems are more automatic, but the helicity
formalism gives more physical insight.

\section{Tests of the \VmA\ Interaction}

The property  that the $W$ boson only couples to
fermions of left-handed chirality is a crucial property of the
SM.  It is responsible for many of the surprising features
of the weak interactions, both the most attractive and the most
puzzling ones.  It is therefore important to understand that this
feature is extremely well supported experimentally.  In this section,
I review the most convincing experimental tests of this property.

\subsection{Polarization in $\beta$ decay}

The first applications discussed in this section involve exchange of $W$
bosons at low energy.  In this limit, we can simplify the $W$
propagator to  a pointlike interaction
\beq
      {-i \over q^2 - m_W^2} \to {i\over  m_W^2} \ . 
\eeqn
In this limit, the $W$ exchange can be represented by the product of
currents
\beq
     \Delta\L =  {g^2 \over 2m_W^2} J_\mu^+ J^{-\mu} \ , 
\eeq{VmAL}
 where
\beqa 
     J^+_\mu &=&     \nu^\dagger_{eL} \bar \sigma_\mu e_L + u^\dagger_L
     \bar\sigma_\mu d_L + \cdots \CR
    J^-_\mu &=&     e^\dagger_L \bar \sigma_\mu\nu_{eL} +d^\dagger_L
    \bar\sigma_\mu u_L + \cdots \ .
\eeqa{currents}
 Here and henceforth in these lectures, I replace the label $\psi$
 with a label that gives the flavor quantum numbers of the field.
 In \leqn{currents}, I write explicitly the terms associated with the
 first generation quarks and leptons; the omitted terms are those for
 the higher generations.  I ignore Cabibbo mixing, a reasonable
 approximation for the topics discussed in these lectures.  I will
 also ignore the masses of the neutrinos.

The theory \leqn{VmAL} is called the \VmA\ interaction, since
\beq
         u^\dagger_L \bar \sigma^\mu d_L =  \bar U \gamma^\mu
         {1-\gamma^5\over 2} D \ ,
\eeqn
the difference of a vector and an axial vector current.
   The
coefficient in \leqn{VmAL} 
 is conventionally represented by the Fermi constant $G_F$,
\beq 
      { g^2\over 2 m_W^2}   =   {4 G_F\over \sqrt{2}} \ .
\eeq{GF}
This interaction has {\it maximal parity violation} in charge-changing
weak
interactions.

\begin{figure}
\begin{center}
\includegraphics[width=0.50\hsize]{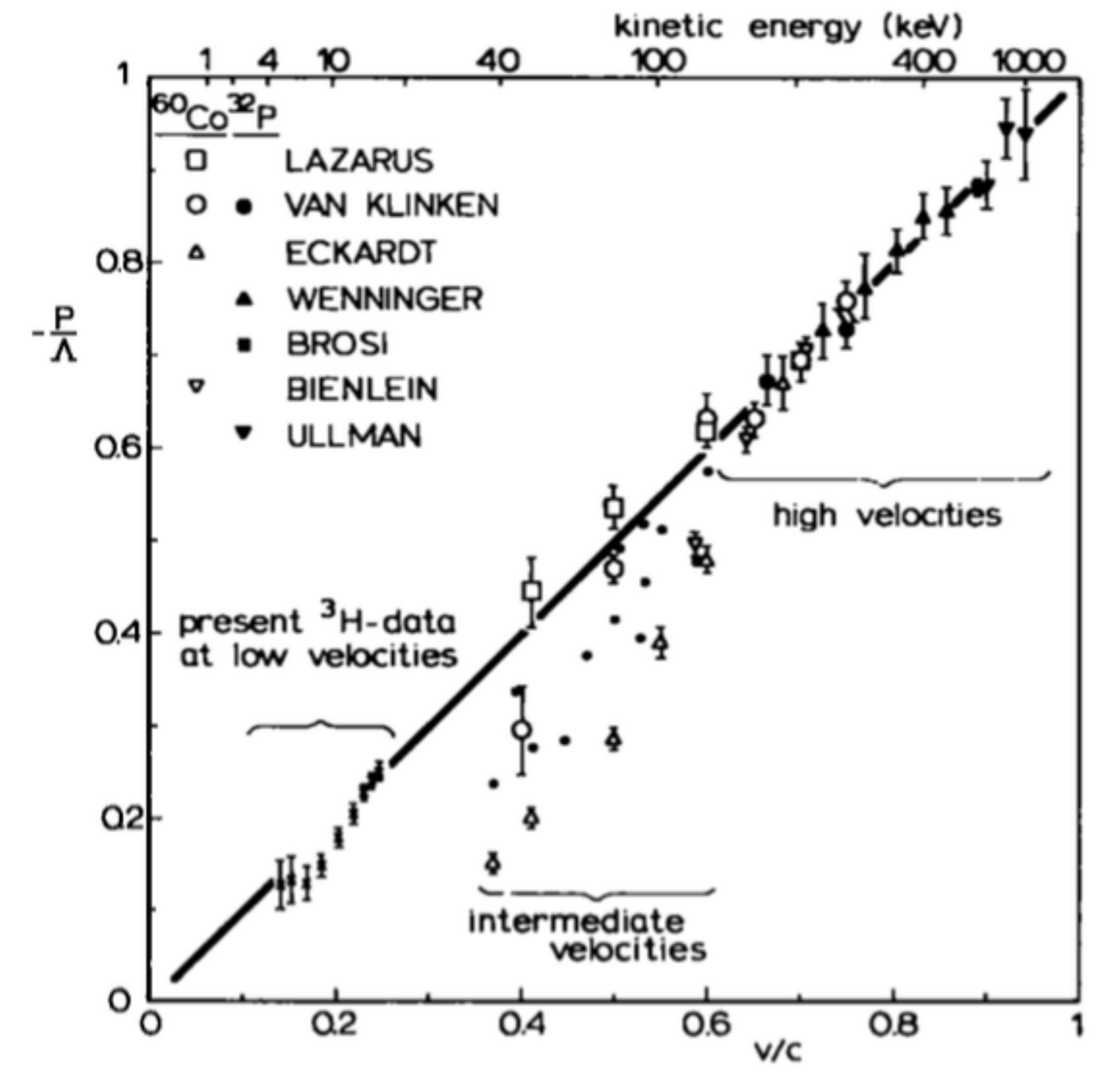}
\end{center}
\caption{Polarization of the electron emitted in $\beta$ decay for a
  variety of $\beta$ decay transitions in different nuclei, from
  \cite{KandvK}.}
\label{fig:betapol}
\end{figure}

The simplest consequence of \VmA\ is that electrons emitted in $\beta$
decay should be preferentially left-handed polarized.   Since the
energies of electrons in $\beta$ decay are of order 1~MeV, it is
typically not a good approximation to ignore the electron mass.
However, since in \VmA\ the electron is produced in the L chirality
eigenstate,  we can work out the polarization from the relative magnitude
of the $u_L$ terms in the left- and right-handed helicity massive
spinors  given in \leqn{massivespinors}.   The electron polarization, in the
left-handed
sense, is then given by 
\beq
     \mbox{Pol}(e^-) = { (\sqrt{E+p} )^2 -  (\sqrt{E-p} )^2 \over 
(\sqrt{E+p} )^2 +  (\sqrt{E-p} )^2 } = {p\over E} =  {v\over c} \ . 
\eeq{betaepol}
A data compilation is shown in Fig.~\ref{fig:betapol}~\cite{KandvK}.
Careful experiments both at high and low electron  energies verify the
regularity \leqn{betaepol}.

\subsection{Muon decay}

The \VmA\ interaction also has striking consequences for the electron
energy and polarization in muon decay. 

  It is not difficult to work
out the basic formulae for muon decay.  In \VmA\ theory, and
ignoring the electron mass,  muon
decay has a massive muon at rest decaying to $\nu_{\mu L} e^-_L\bar
\nu_{eR}$.
For the muon at rest, averaged over polarizations, we find, instead of \leqn{emusquare},
\beq
  | (u^\dagger_L \bar \sigma^\mu v_R) (v^\dagger_R \bar \sigma_\mu
     u_L) |^2 = 2 \, (2 p_{e^-}\cdot p_{\nu})(2 p_{\bar \nu}\cdot
     p_{\mu_-}) \ . 
\eeq{emunusquare}
To integrate this over phase space, let
\beq
      x_i = {2p_i\cdot p_\mu\over p_\mu^2} \ , 
\eeq{xidef}
where $i = e, \nu, \bar \nu$.   Conservation of energy-momentum $p_\mu
= p_e + p_\nu + p_{\bar\nu}$ implies
\beq
      x_e + x_\nu + x_{\bar \nu} = 2 \ . 
\eeqn
Each $x_i$ takes the maximum value 1 when that massless particle recoils
against the other two massless particles.
Note also that
\beq
     2 p_e \cdot p_\nu = (p_e+p_\nu)^2 = (p_\mu - p_{\bar \nu})^2 =
     m_\mu^2 (1 - x_{\bar \nu}) \ .
\eeqn
Three-body phase space takes a simple form in the $x_i$ variables,
\beq
    \int d\Pi_3 = {m_\mu^2\over 128 \pi^3 } 
\int  dx_e dx_{\bar \nu} \  . 
\eeqn

\begin{figure}
\begin{center}
\includegraphics[width=0.70\hsize]{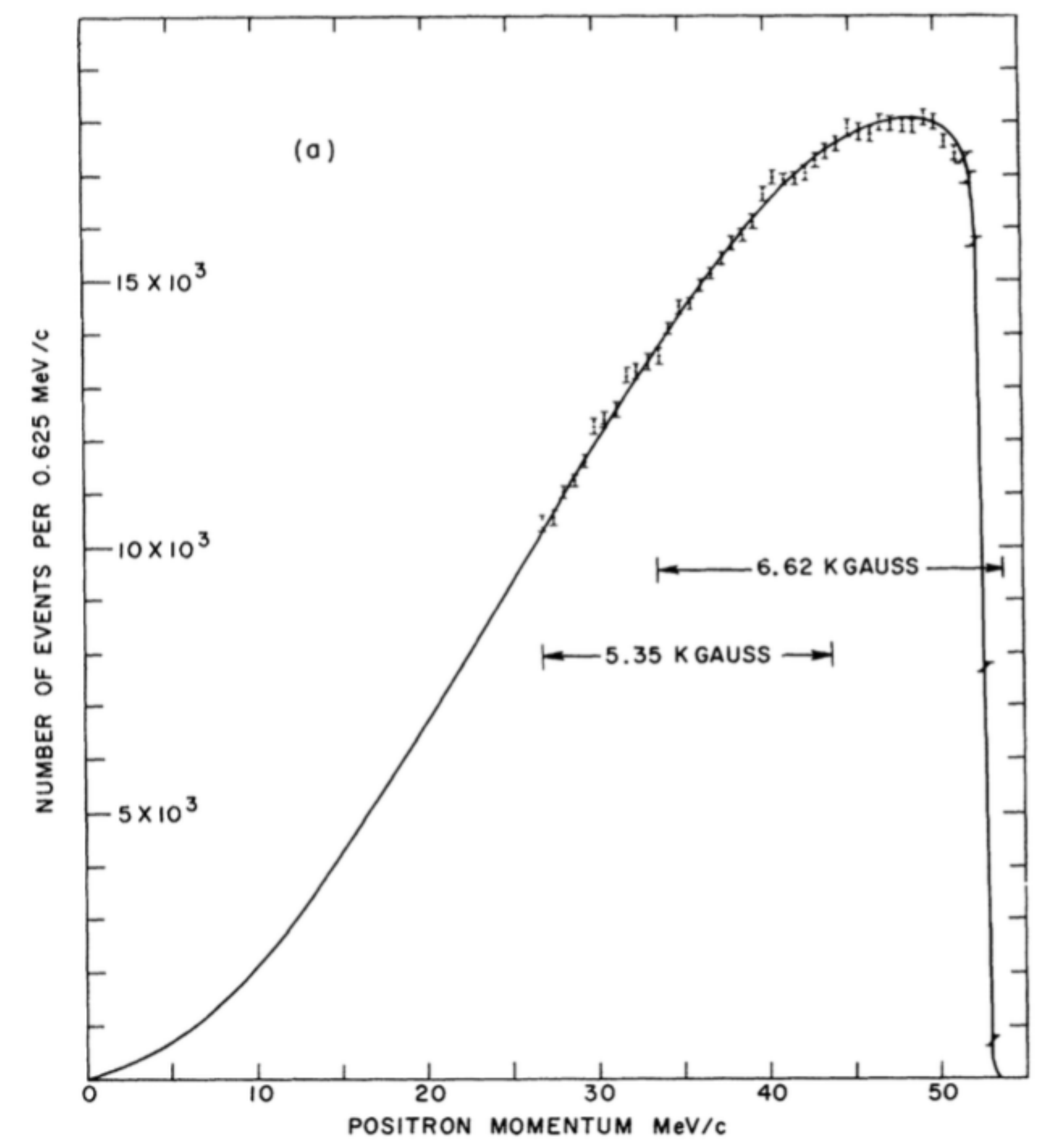}
\end{center}
\caption{Energy spectrum of $e^+$ in $\mu^+$ decay at rest, 
 from \cite{Bardon}.}
\label{fig:muond}
\end{figure}

Assembling the pieces, the muon decay rate is predicted to be
\beq
     \Gamma = {1\over 2m_\mu }\biggl( {4G_F\over \sqrt{2}}\biggr)^2
     {m_\mu^2\over 128 \pi^3}
     \int dx_e dx_{\bar\nu} \  2 m_\mu^4 x_{\bar\nu}(1-x_{\bar \nu}) \ . 
\eeqn
The integral over $x_{\bar \nu}$ is 
\beq
   \int^1_{1-x_e}  dx_{\bar\nu} \ x_{\bar \nu} (1 - x_{\bar \nu}) =
   \half x_e^2 - \third x_e^3\ .
\eeqn
Then finally we find for the electron energy distribution
\beq
   { d\Gamma\over dx_e} =  {G_F^2 m_\mu^5\over 16\pi^3  } \biggl(
     {x_e^2\over 2 } - {x_e^3\over 3} \biggr) \ .
\eeqn
This shape of this distribution is quite characteristic, with a double
zero 
at $E_e = 0$ and zero slope at the endpoint at $E_e = m_\mu/2$.   Both
effects are slightly rounded by radiative corrections, but, with these
taken into account, the prediction agrees with the measured spectrum
to high precision, as shown in Fig.~\ref{fig:muond}~\cite{Bardon}.

\subsection{Pion decay}

Charged pion decay is mediated by the \VmA\ interaction
\beq
    {4G_F\over \sqrt{2}} (d^\dagger_L \bar \sigma^\mu u_L) \biggl(
    \nu^\dagger_{eL} \bar \sigma_\mu e_L
        +  \nu^\dagger_{\mu L} \bar  \sigma_\mu \mu_L \biggr)
\eeq{piplusL}
At first sight, it might seem that the $\pi^+$ must decay equally
often to $e^+$ and $\mu^+$.   Experimentally, almost all pion decays
are to $\mu^+$.   Can this be reconciled with \VmA?

The pion matrix element is 
\beq
   \bra{0} d^\dagger_L \bar \sigma^\mu u_L \ket{\pi^+(p)} = -i \half
   F_\pi p^\mu \ , 
\eeqn
where $F_\pi$ is the pion decay constant, equal to 135~MeV.   The
matrix element of \leqn{piplusL} then evaluates to 
\beq
     {4G_F\over \sqrt{2}} \cdot (-{i\over 2} F_\pi) \  p^\mu
     U^\dagger_{\nu L} \bar\sigma_\mu V_{\ell^+} \ . 
\eeq{pimat}
The pion is at rest, so 
\beq
   p^\mu \bar \sigma_\mu =   m_\pi \cdot 1 \ .
\eeqn
The neutrino is (essentially) massless and therefore must be
left-handed.  The pion has spin 0, so angular momentum requires that
the $\ell^+$ is also left-handed.  But, from \leqn{massivespinors},
the lepton spinor is then
\beq
             V_L = \pmatrix{\sqrt{E-p} \ \xi_R\cr \times } 
\eeq{VLforpi}
The matrix element \leqn{pimat} reduces to 
\beq
   i  {4G_F\over \sqrt{2}} \cdot (\half  F_\pi) \ \sqrt{2E_\nu}
     m_\pi \sqrt{E_\ell - p_\ell} \ .
\eeq{pimattwo}
Two-body kinematics gives $E_\nu =p_\nu =  p_\ell = (m_\pi^2 -
m_\ell^2)/2m_\pi$.   Then  $(E_\ell - p_\ell) = m_\ell^2/m_\pi^2$.  Phase
space includes the factor $2p_\ell/m_\pi$, which brings another factor
of  $(E_\ell - p_\ell) $.  Finally we find
\beq
   \Gamma(\pi^+\to \ell^+\nu) =  { G_F^2 m_\pi^3 F_\pi^2 \over 8 \pi }
{m_\ell^2\over m_\pi^2} \bigl(1- {m_\ell^2\over m_\pi^2}\bigr)^2 \ .
\eeq{Gampi}
The overall factor $m_\ell^2/m_\pi^2$ comes from the matrix element
 \leqn{VLforpi}.  Angular momentum conservation 
requires the $\ell^+$ to have the 
wrong helicity with respect to \VmA, accounting for this suppression factor.

The result \leqn{Gampi} leads to the ratio of branching fractions
\beq 
     {BR(\pi^+ \to e^+ \nu_e)\over BR(\pi^+ \to \mu^+ \nu_\mu)} =
     {m_e^2\over m_\mu^2} \biggl({m_\pi^2 - m_e^2\over m_\pi^2 -
     m_\mu^2} \biggr)^2 = 1.28\times 10^{-4} \ , 
\eeqn
in good agreement with the observed value $1.23 \times 10^{-4}$.

\subsection{Neutrino deep inelastic scattering}

\begin{figure}
\begin{center}
\includegraphics[width=0.70\hsize]{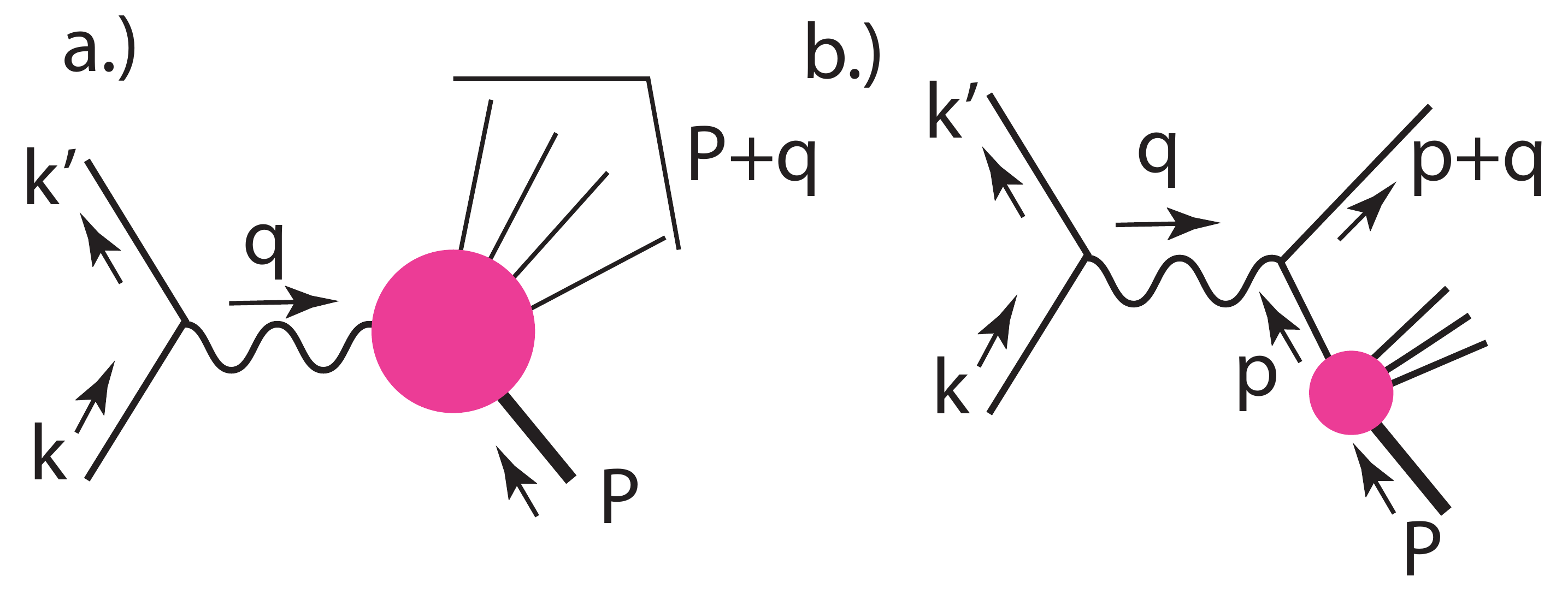}
\end{center}
\caption{Kinematics of neutrino deep inelastic scattering: (a) for
  neutrino scattering from a proton or heavy nucleus, (b) for neutrino
scattering from a quark in the parton model description.}
\label{fig:DIS}
\end{figure}

The helicity structure of the \VmA \ interaction is also seen in the
energy distributions in deep inelastic neutrino scattering.    For
electrons,
deep inelastic scattering is the scattering from a proton or nuclear
target in which the momentum transfer is large and the target is
disrupted to a high mass hadronic state.   The kinematics is shown in
Fig.~\ref{fig:DIS}(a).   In the leading order of QCD, deep inelastic
scattering is described by the scattering for the electron from a
single quark in the parton distribution of the target.   This
kinematics is shown in Fig.~\ref{fig:DIS}(b).

Neutrino deep inelastic scattering experiments are done in the
following way:  One first creates a high-energy pion beam by
scattering protons from a target.  Then the pions are allowed to
decay, producing a beam of neutrinos and muons.  The beam is made to
pass through a long path length of absorber to remove the muons and
residual pions and other hadrons.   Finally, the neutrinos are allowed
to interact with a large-volume detector.
 A charged-current neutrino reaction then leads to a scattering
event  whose result is a $\mu^\pm$, depending on the
charge of the decaying pion, and a high-multiplicity hadronic system.

If $k$ is the initial momentum of the neutrino, $k'$ is the
final momentum of the muon, and $P$ is the initial momentum of the
target proton, we let $q = (k-k')$ and define the Lorentz invariants 
\beqa
    s = (k+P)^2   & \qquad&   Q^2 = -q^2 \CR
 x = {Q^2 \over 2P\cdot q} &\qquad&   y  = {2P\cdot q\over 2P\cdot k} 
\eeqan
We are interested in the deep inelastic limit $Q^2 \gg P^2 = m_p^2$.  Then $s \approx
2p\cdot k$ and $Q^2 = xy s$.    In the lab frame $P = (m_p, \vec 0)$,
so $y = q^0/k^0$, the fraction of the initial neutrino energy
transfered to the proton.   To the extent that the initial neutrino
energy $k^0$ is known, all of the invariants $x$, $y$, and $Q^2$ can be
determined by measurement of the final muon momentum.

At leading order in QCD, a deep inelastic reaction is an essentially
elastic lepton-quark scattering, for example, $\nu + d\to \mu^- + u$.
Using Feynman's parton model, which is also the basis for QCD
predictions at hadron colliders, we model the proton or nuclear target as a collection of quarks
and antiquarks that move collinearly and  share the total momentum of the proton.  Let $p$
be the momentum of the initial quark, and approximate
\beq
       p = \xi P \ , 
\eeqn
where $0 < \xi  < 1$.  The quarks might also have transverse momentum
relative to the proton, but this is ignorable if the momentum transfer
$Q^2$ from the neutrino scattering is large.     

The final momentum of the quark is then  $p + q$.   The condition that
this quark is on-shell is
\beq
  0 = (p+q)^2 =   2p\cdot q + q^2 =   2 \xi P\cdot q - Q^2 \ .
\eeqn
Then
\beq 
     \xi =  {Q^2 \over 2 P\cdot q} = x   \ .
\eeqn
This is a remarkable result, also  due to Feynman:  To the leading order in
QCD, deep inelastic scattering events at a given value of the
invariant $x$ arise from scattering from quarks or antiquarks in the proton with
momentum fraction  $\xi = x$.

We can now evaluate the kinematic invariants for a neutrino-quark
scattering event.  I call these $\hat s$, $\hat t$, $\hat u$ to
distinguish  them from the invariants of neutrino-proton scattering.  First,
\beq
   \hat s =  (p + k)^2 =   2p\cdot k = 2 \xi P\cdot k = x \ s   \ .
\eeq{mys}
The momentum transfer can be evaluated from the lepton side, so 
\beq
    \hat t = q^2 = -Q^2 \ .
\eeqn
Finally, for scattering of approximately massless particles, $s+t+u =
0$, so
\beq
       \hat u = x s -  Q^2 =    x s (1-y) \ . 
\eeq{myu}

The aspect of the deep inelastic scattering cross section that is most
important for the subject of this lecture is the distribution in $y$.
To begin, consider the deep inelastic scattering of a $\nu_\mu$.  The
quark-level reaction is 
\beq
       \nu + d \to \mu^- + u
\eeqn
In the \VmA\ theory, the $\nu $ and the $d$ must be  left-handed. 
Similarly to  \leqn{emusquareLR},  
\beq
  | (u^\dagger_L (\mu^-)\bar \sigma^\mu u_L(\nu))
 (u^\dagger_L(u) \bar \sigma_\mu
     u_L(d) ) |^2 = 4 \, (2 p_{\mu^-}\cdot p_{u})(2 p_{\nu}\cdot
     p_{d}) =  4 \hat s^2 \ . 
\eeq{nuampsq}
On the other hand, antineutrino scattering from a quark, which
proceeds  by the reaction
\beq
       \bar\nu + u \to \mu^+ + d \ , 
\eeqn
is, in \VmA\ theory, the scattering of a right-handed $\bar \nu$ and a
left-handed $u$.  Then 
\beq
  | (v^\dagger_R (\mu^-)\bar \sigma^\mu v_R(\bar\nu))
 (u^\dagger_L(u) \bar \sigma_\mu
     u_L(d) ) |^2 = 4 \, (2 p_{\mu^+}\cdot p_{u})(2 p_{\bar\nu}\cdot
     p_{d}) =  4 \hat u^2 \ . 
\eeq{nubarampsq}

\begin{figure}
\begin{center}
\includegraphics[width=0.40\hsize]{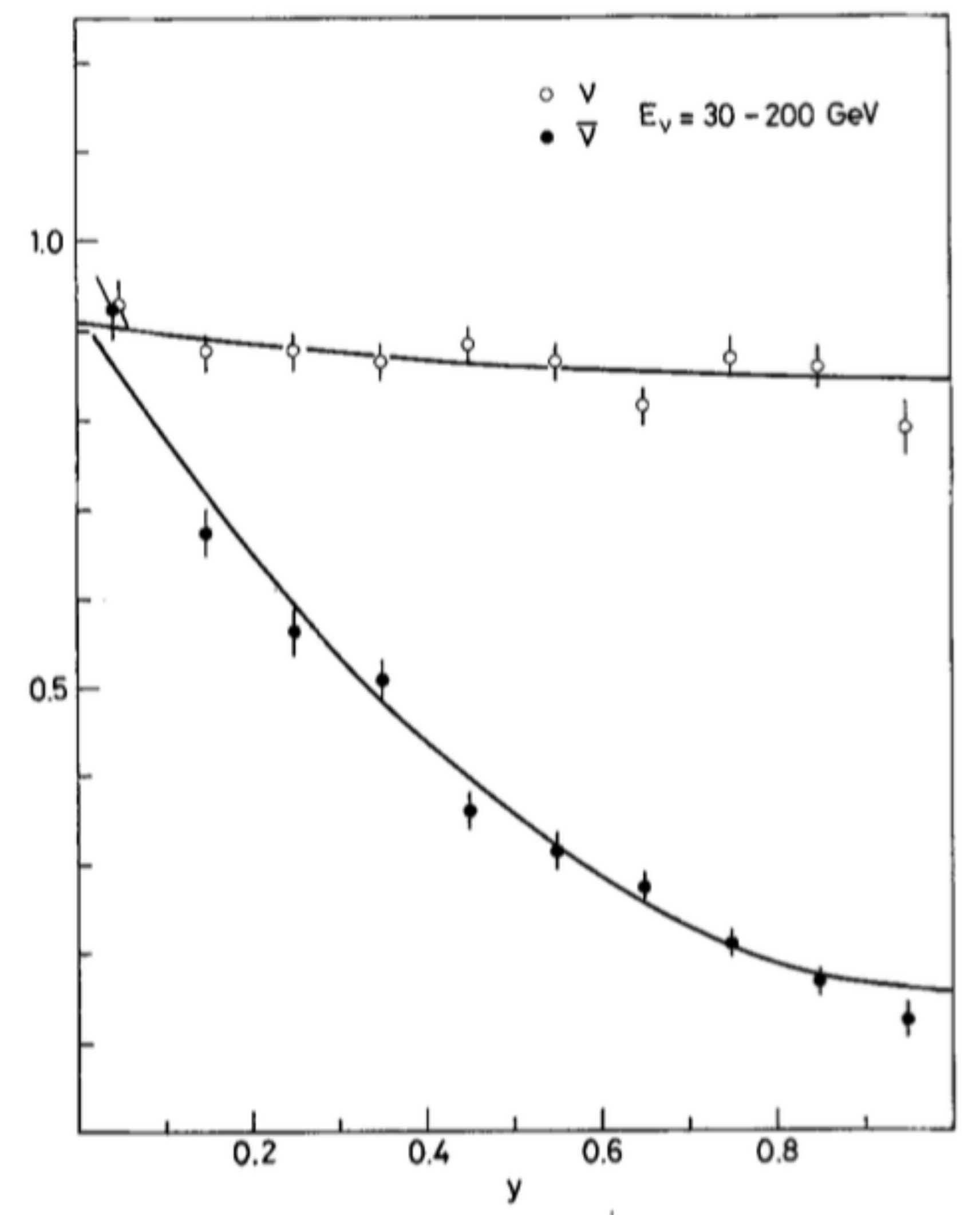}
\end{center}
\caption{Dependence on the variable $y$ of the cross sections for
  neutrino and antineutrino scattering on an iron target, 
 from \cite{CDHS}.}
\label{fig:CDHS}
\end{figure}

Inserting \leqn{mys}, \leqn{myu}, we see that the dependence of the
deep inelastics scattering cross section on $y$ should be 
\beqa
   {  d\sigma\over d y}(\nu p \to \mu^- X) &\sim&  \hat s^2 \sim  1   \CR
  {  d\sigma\over d y}(\bar\nu p \to \mu^+ X) &\sim& \hat u^2 \sim  ( 1 -y)^2    \ . 
\eeqan
These results, which I have derived for a proton target, hold for any
nuclear  target under the assumption that we consider only scattering
from quarks and not from antiquarks.    For scattering from
antiquarks, the dependence on $y$ is reversed, with a $(1-y)^2$
dependence for neutrino scattering.
The experimental result, from the CDHS experiment, a CERN neutrino
experiment of the1980's, is shown in Fig.~\ref{fig:CDHS}~\cite{CDHS}.   The $y$
distribution
for neutrino scattering is indeed almost flat, and that for
antineutrino scattering is close to $(1-y)^2$.   The deviations from
these ideal results are consistent with arising from the antiquark
content of 
the proton and neutron.

The same regularity can be seen in collider physics.  For example, the
Standard
Model predicts that, in quark-antiquark annihilation to a $W$ boson,
\beqa
    {d\sigma\over d \cos\theta}(d\bar u \to W^-\to \mu^-\bar\nu) \sim
& u^2& \sim   (1 + \cos\theta)^2\CR
   {d\sigma\over d \cos\theta}(u
\bar d \to W^+\to \mu^+\nu) \sim
&t^2& \sim   (1 -\cos\theta)^2\ ,
\eeqan
and these distributions are well verified at the LHC~\cite{ATLASW,CMSW}.

\subsection{$\ee$ annihilation at high energy}

The angular distributions in annihilation through the neutral current
are more complex, first, because of photon-$Z$ interference, and,
second, 
because the weak neutral current couples to both
left- and right-handed quarks and leptons.  

\begin{figure}
\begin{center}
\includegraphics[width=0.70\hsize]{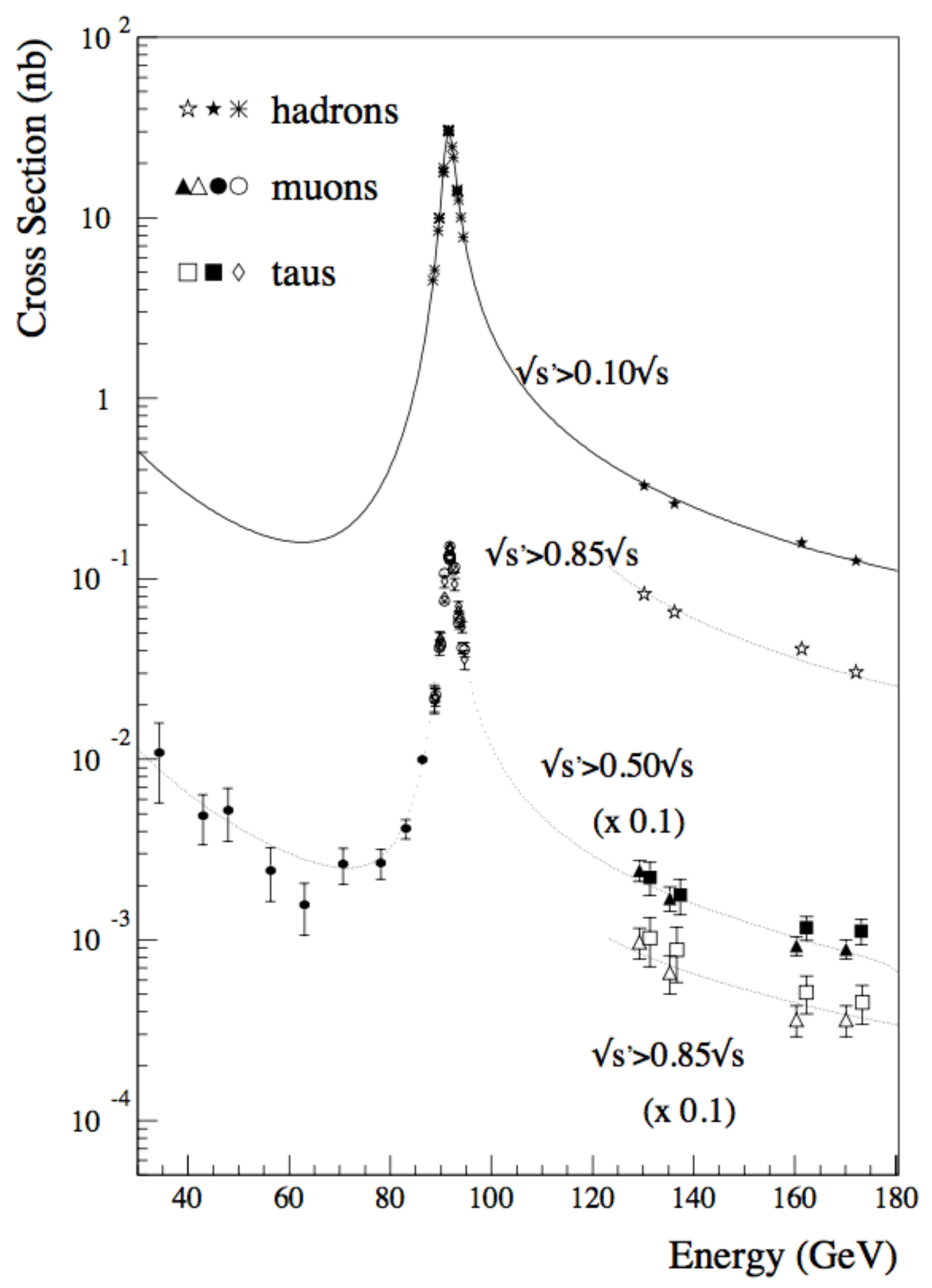}
\end{center}
\caption{Total cross section for $\ee\to$ hadrons, 
$\ee\to \mu^+\mu^-$, and $\ee\to \tau^+\tau^-$,  as a function of
center of mass energy, as measured by the DELPHI
experiment at the collider LEP~\cite{DELPHI}.  The continuous lines
are the predictions of the SM.}
\label{fig:DELPHItot}
\end{figure}

 To write formulae for the
cross sections in $\ee$ annihilation to a fermion pair, it is simplest
to begin with the cross sections for polarized initial and final
states.    Using the same principles for evaluating spinor products
as before, it is not difficult to work these out.   The general form
of the differential cross sections is
\beqa
{d\sigma\over d \cos\theta}(e^-_Le^+_R\to f_L\bar f_R) &=& {\pi
  \alpha^2\over 2s} \ | s\,F_{LL}(s)|^2 \ (1+\cos\theta)^2 \CR
{d\sigma\over d \cos\theta}(e^-_Re^+_L\to f_L\bar f_R) &=& {\pi
  \alpha^2\over 2s} \ | s\, F_{RL}(s)|^2 \ (1-\cos\theta)^2 \CR
{d\sigma\over d \cos\theta}(e^-_Le^+_R\to f_R\bar f_L) &=& {\pi
  \alpha^2\over 2s} \ |  s\, F_{LR}(s) |^2 \ (1-\cos\theta)^2 \CR
{d\sigma\over d \cos\theta}(e^-_Re^+_L\to f_R\bar f_L) &=& {\pi
  \alpha^2\over 2s} \ |  s\, F_{RR}(s)|^2 \ (1+\cos\theta)^2 \ .
\eeqa{eeforms}
The form factors $F_{IJ}(s)$ reflect photon$\gamma$--$Z$ interference,
with the p$\gamma$ charges $Q$ and the $Z$ charges $Q_Z$ in
\leqn{QZ}.  Using the subscript $f$ to denote the flavor and chirality
of the fermion, 
\beqa
  F_{LL}(s) &=&   {Q_f\over s}  + {(1/2 - s_w^2)(I^3_f - s_w^2 Q_f)\over 
 s_w c_w} {1 \over s - m_Z^2} \CR 
  F_{RL}(s) &=&   {Q_f\over s}  + {( - s_w^2)( - s_w^2 Q_f)\over 
 s_w c_w} {1 \over s - m_Z^2} \CR 
  F_{LR}(s) &=&   {Q_f\over s}  + {1/2 - s_w^2)(I^3_f - s_w^2 Q_f)\over 
 s_w c_w} {1 \over s - m_Z^2} \CR 
  F_{RR}(s) &=&   {Q_f\over s}  + { (- s_w^2)(- s_w^2 Q_f)\over 
 s_w c_w} {1 \over s - m_Z^2} \ .
\eeqa{eefactors}
The total cross sections predicted from these formulae for $\ee\to$
hadrons,
 $\ee\to \mu^+\mu^-$, and $\ee\to \tau^+\tau^-$ are shown in 
Fig.~\ref{fig:DELPHItot} and compared to data from the DELPHI
experiment at the CERN $\ee$ collider LEP.   The resonance at the center
of mass energy of 91~GeV is of course the $Z$ boson.

\begin{figure}
\begin{center}
\includegraphics[width=0.80\hsize]{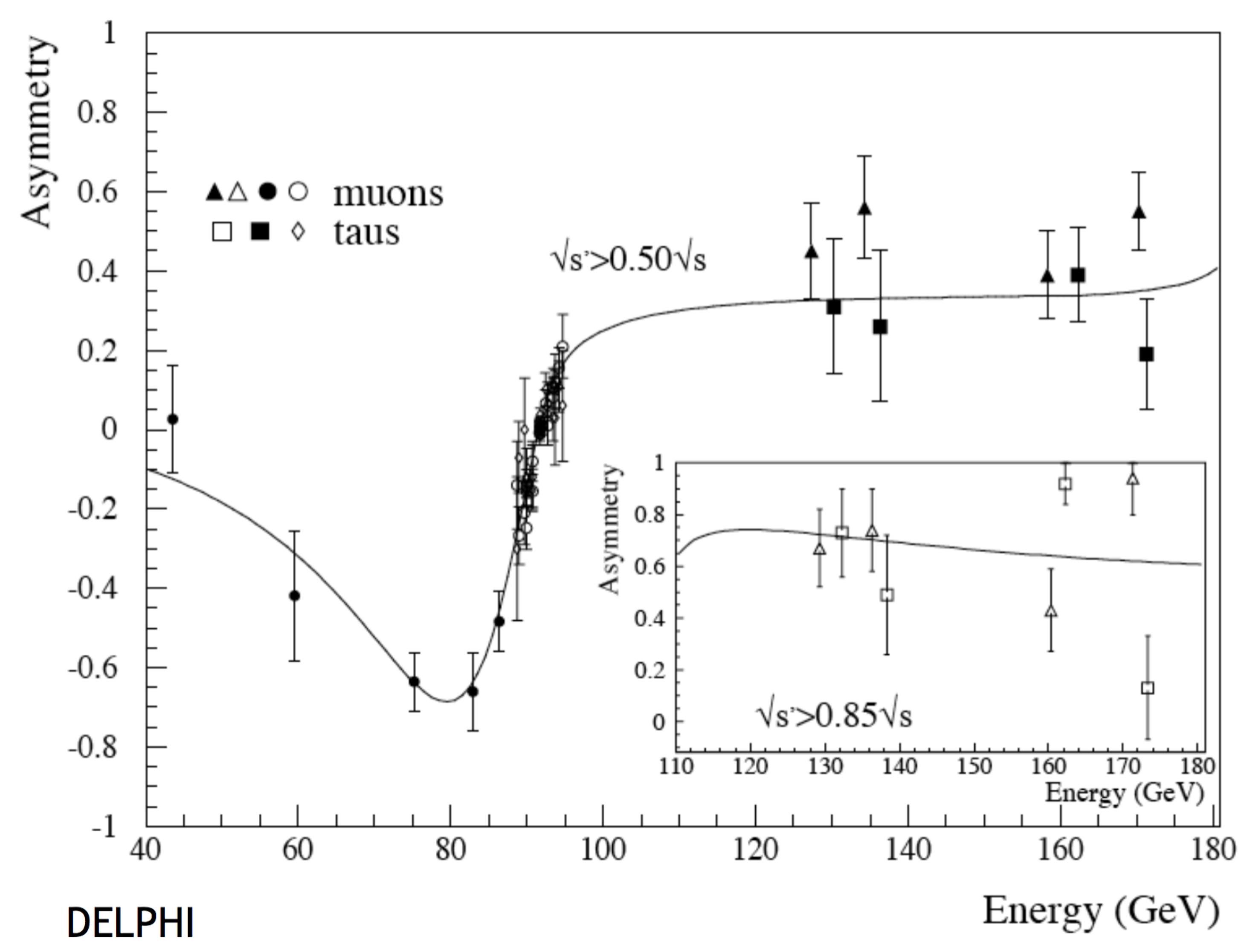}
\end{center}
\caption{Forward-backward asymmetry in the reactions 
$\ee\to \mu^+\mu^-$ and $\ee\to \tau^+\tau^-$,  as a function of
center of mass energy, as measured by the DELPHI
experiment at the collider LEP~\cite{DELPHI}. .}
\label{fig:DELPHI}
\end{figure}

Notice that, for $s > m_Z^2$, we have constructive interference in the
LL and RR polarization states and destructive interference for RL and
LR.   Then in an experiment with unpolarized beams (as in the program
of $\ee$ experiments at LEP), the LL and RR modes
should 
dominate and  produce a positive forward-backward asymmetry in the
angular distribution.   This behavior is actually seen in the data.
Figure~\ref{fig:DELPHI} shows the forward-backward asymmetry in
$\ee\to \mu^+\mu^-$ and $\ee\to \tau^+\tau^-$ measured by the DELPHI
experiment at LEP~\cite{DELPHI}.   The solid line is the prediction of
the SM.

It is interesting to explore the high-energy limits of the expressions
\leqn{eefactors}.  Begin with $F_{RL}(s)$, corresponding to
$e^-_Re^+_L\to f_L\bar f_R$.   In the limit $s \gg
m_Z^2$ and inserting $Q = I^3_f + Y$, this becomes
\beqa
    F_{RL} &\to&  {s_w^2 c_w^2 (I^3_f + Y_f) - s_w^2 I^3_f + s_w^4(I^3_f
      + Y_f)\over s_w^2  c_w^2  \ s }\CR
 & & ={ s_w^2 Y_f \over s_w^2 c_w^2 \ s}  \CR
 & & = {1\over e^2} \biggl( {g^{\prime 2}Y_{eR} Y_f\over s}\biggr) \ .
\eeqan
The expression in parentheses is exactly the amplitude for $s$-channel
exchange of the $U(1)$ boson $B$ in the situation in which the
original $SU(2)\times U(1)$ symmetry was not spontaneously broken.  So
we see that the full gauge symmetry is restored at high energies.

Here is the same analysis for $F_{LL}(s)$: 
 \beqa
    F_{RL} &\to&  {s_w^2 c_w^2 (I^3_f + Y_f) +(1/2 -  s_w^2)( I^3_f - s_w^2(I^3_f
      + Y_f))\over s_w^2  c_w^2  \ s }\CR
 & & ={ (1/2) c_w^2 I^3_f + (1/2) s_w^2 Y_f \over s_w^2 c_w^2 \ s}  \CR
 & & = {1\over e^2} \biggl({g^2 I^3_{eL}I^3_f\over s} + 
 {g^{\prime 2}Y_{eR} Y_f\over s}\biggr) \ .
\eeqan
Now the result is a coherent sum of $A^3$ and $B$ exchanges in the
$s$-channel.   Again, this is the result expected in a theory of
unbroken $SU(2)\times U(1)$.

\begin{figure}
\begin{center}
\includegraphics[width=0.70\hsize]{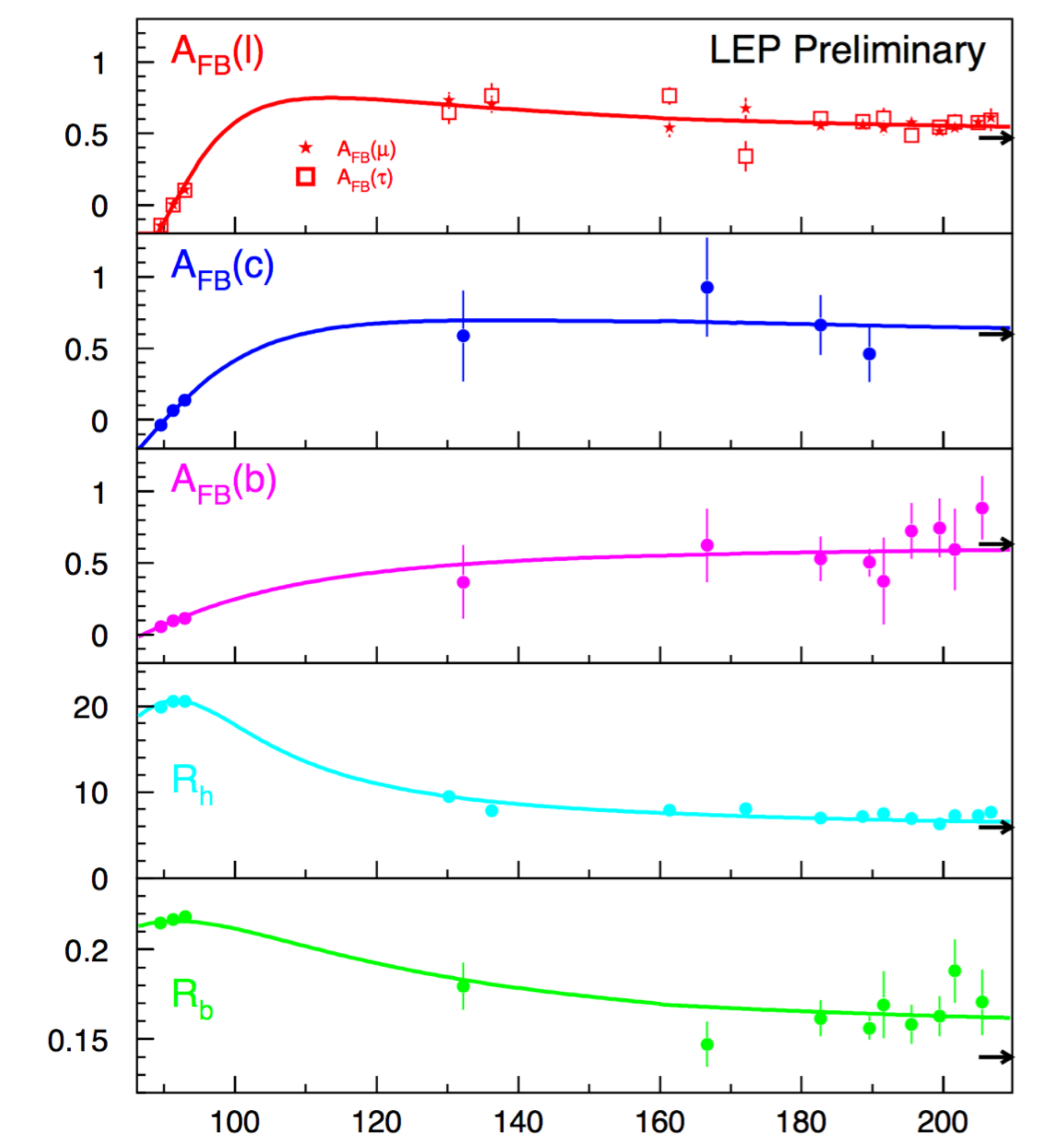}
\end{center}
\caption{Compilation of preliminary LEP
  measurements of the forward-backward asymmetry in lepton, $c$, and
  $b$ pair production, the hadron to lepton ratio $R_h$ and the $b$ to
  all hadron ratio $R_b$~\cite{Hildreth}.  The solid curves show the SM
  prediction.  The arrows at the right are the predictions of unbroken
  $SU(2)\times U(1)$.}
\label{fig:Hildreth}
\end{figure}

It is interesting to compare the values of ratios and asymmetries
measured at LEP to the asymptotic values predicted by unbroken
$SU(2)\times U(1)$.   This comparison is shown in
Fig.~\ref{fig:Hildreth} from a compilation of preliminary LEP
results~\cite{Hildreth}; final LEP results on 2-fermion processes are
collected in \cite{finalLEP}.   The arrows at the extreme right show
the values for restored $SU(2)\times U(1)$.  The calculation of $R_b$
involves a top quark box diagram that does not yet reach its
asymptotic limit at 200~GeV.   It is remarkable that,
for allother  observables, the LEP measurements at center of mass
energies of 200~GeV  are already close
to the asymptotic values predicted at high energy.

\section{Precision electroweak measurements
 at the $Z$ resonance}

It is possible to test the SM theory of the weak interactions  more
incisively by focusing more tightly on the properties of the $Z$
boson.  The $Z$ boson appears as a resonance in $\ee$ annihilation.
In the 1990's, the accelerators LEP at CERN and SLC at SLAC tuned
their energies to the $Z$ boson resonance to produce large numbers of
$Z$ bosons at rest in the lab, in an appropriate setting for precision
measurements.  In this section, I review the results of these
precision measurements, which continue to provide important
constraints on the SM and its generalizations.

\subsection{Properties of the $Z$ boson in the Standard Model}

My discussion will be based on the leading order matrix elements for
$Z$
decay to $f_L \bar f_R$ and $f_R \bar f_L$.  It is straightforward to
work these out based on the spinor matrix elements computed in Section
2.2.
The leading order matrix element for $Z$ decay to $f_L\bar f_R$ is
\beq
   \M(Z\to f_L \bar f_R) = i {g\over c_w} Q_{Zf} \  u_L^\dagger
   \bar\sigma^\mu v_R \   \eps_{Z\mu} \ ,
\eeqn
with 
\beq
   Q_Z = I^3 - s_w^2 Q \ ,
\eeq{QZdef}
as in \leqn{QZ}.    Using \leqn{anncreate} for the spinor matrix
element, this becomes
\beq
    \M = i {g\over c_w} \sqrt{2} m_Z \eps_{-}^* \cdot \eps_Z \ . 
\eeqn
Square this and average over the direction of the fermion, or,
equivalently, average over three orthogonal directions for the $Z$
polarization vector.  The result is 
\beq 
     \VEV{|\M|^2} = {2\over 3} {g^2\over c_w^2} Q_{Zf}^2 m_Z^2 \ . 
\eeqn
Then, since 
\beq
     \Gamma(Z \to f_L \bar f_R) =  {1\over 2m_Z} {1\over 8\pi}
     \VEV{|\M|^2} \ ,
\eeqn
we find
\beq
\Gamma(Z \to f_L\bar f_R) =  {\alpha_w m_Z\over 6 c_w^2}  Q^2_{Zf} N_f \
, 
\eeq{GZf}
where 
\beq
   \alpha_w = {g^2\over 4\pi} 
\eeqn
and
\beq
   N_f = \cases{1 & lepton \cr 
        3 (1 + \alpha_s/\pi + \cdots) & quark\cr} 
\eeqn
accounts the number of color states and the QCD correction.   The same
formula holds for the $Z$ width to $f_R\bar f_L$.

To evaluate this formula, we need values of the weak interaction
coupling constants.   The electromagnetic coupling $\alpha$ is
famously close to  1/137.  However, in quantum field theory, $\alpha$
is a running coupling constant that becomes larger at smalll distanct
scales.  At a scale of $Q = m_Z$,  $\alpha(Q) = 1/129$.   Later in the
lecture, I will defend a value of the weak mixing angle
\beq
          s_w^2 = 0.231 \ . 
\eeq{refsstw}
Then the $SU(2)$ and $U(1)$ couplings take the values
\beq
          \alpha_w = {g^2\over 4\pi} = {1\over 29.8} \qquad 
\alpha' = {g^{\prime 2}\over
            4\pi} = { 1\over 99.1}
\eeq{refggprime}
It is interesting to compare these values to other fundamental
SM couplings taken at the same scale $Q = m_Z$,
\beq
          \alpha_s = {1\over 8.5} \qquad  \alpha_t = {y_t^2\over 4\pi}
          = {1\over 12.7} \ . 
\eeq{refalphat}
All of these SM couplings are roughly of the same order of 
magnitude.

Using \leqn{refsstw} or \leqn{refggprime}, we can tabulate the values
of the $Z$ couplings to left- and right-handed fermions,
\beq
\begin{tabular}{lllcc}
species &  $Q_{ZL}$ & $Q_{ZR}$ & $S_f$ & $A_f$ \\   \hline
$\nu$ & $+\half$ & - &  0.250 & 1.00 \\
$e $ & $-\half + s_w^2 $ &  $+s_W^2$ & 0.126  &  0.15  \\
$u $ & $+\half -\tthird s_w^2 $ &  $-\tthird s_W^2$ & 0.143 &  0.67  \\
$d $ & $-\half + \third s_w^2 $ &  $+\third s_W^2$ & 0.185 &  0.94
\end{tabular}
\eeq{Zvalues}
In this table, the quantities evalated numerically are
\beq
S_f = Q_{ZL}^2 + Q_{ZR}^2 \qquad  A_f = { Q_{ZL}^2 -Q_{ZR}^2 \over 
Q_{ZL}^2 + Q_{ZR}^2 }\ .
\eeq{SandA}
The quantity $S_f$ gives the contribution of the species $f$ to the total
decay rate of the $Z$ boson.  The quantity $A_f$ gives the
polarization asymmetry for $f$, that is, the
preponderance of $f_L$ over $f_R$, in $Z$ decays,

\subsection{Measurements of the $Z$ properties}

It is possible to measure many of the total rates and polarization
asymmetries for individual species in a very direct way through experiments on the $Z$
resonance.   This subject is reviewed in great detail in the report
\cite{LEPEWWG}.   Values of the $Z$ observables given below are taken
from this reference unless it is stated otherwise.

The $S_f$ are tested by the measurement of the $Z$ resonance width and
its branching ratios.  Using \leqn{GZf}, we find for the total width
of the $Z$
\beqa
\Gamma_Z& =&  {\alpha_w m_Z\over 6 c_w^2} \biggl[   3\cdot 0.25 +
3\cdot 0.126 \CR
   & & \hskip 0.1in + 2 \cdot (3.1) \cdot 0.144 + 3 \cdot (3.1) \cdot
   0.185 \biggr] \ .
\eeqa{Zwidth}
The four terms denote the contributions from 3 generations of
$\nu$, $e$, $u$, and $d$, minus the top quark, which is too heavy to
appear in $Z$ decays. The numerical prediction is
\beq
      \Gamma_Z = 2.49~\mbox{GeV} 
\eeq{measZwidth}
  The separate terms in \leqn{Zwidth} give the branching ratios
\beqa
      BR(\nu_e\bar\nu_e) = 6.7\%  & \qquad&   BR(\ee) = 3.3\% \CR
     BR(u\bar u ) = 11.9\% &\qquad& BR(d\bar d) = 15.3\%
\eeqa{ZBRs}

The measured value of the total width, whose extraction I will discuss
in a moment, is 
\beq
    \Gamma_Z = 2.4952 \pm 0.0023~\mbox{GeV}\ .
\eeqn
This is in very good agreement with \leqn{measZwidth}, with accuracy
such that a valid comparison with theory requires the inclusion of
electroweak radiative corrections, with typically are of order 1\%.
The measurements of branching ratios and polarization asymmetry that I
review later in this section are also of sub-\% accuracy.   At the end
of this section, I will present a more complete comparison of
theory and experiment, including radiative corrections to the
theoretical predictions.

To begin our review of the experimental measurements, we should
discuss the measurement of the $Z$ resonance mass and width in more
detail.  Ideally, the $Z$ is a Breit-Wigner resonance, with cross
section shape
\beq
    \sigma \sim \biggl|  {1\over s - m_Z^2 + i m_Z \Gamma_Z} \biggr|^2
    \ .
\eeq{simpleBW}
At first sight, it seems that we can simply read off the $Z$ mass as
the maximum of the resonance and the width as the observed width at
half maximum.  However, we must take into account that the resonance
is  distorted by initial-state radiation.   As the electron and
positron collide and annihilate into a $Z$, they can radiate hard
collinear 
photons.  Because of this, the resonance is pushed over to higher
energies, an effect that shifts the peak and creates a long tail on
above the resonance.   The magnitude of the photon radiation is given
by the parameter
\beq
   \beta  =  {2\alpha\over \pi} (\log{s\over m_e^2} - 1) =  0.108
   \quad \mbox{at} \ s = m_Z^2 
\eeqn
In addition, since the $Z$ is narrow, the effect of this radiation  is
magnified, since even a relatively soft photon can push the center of
mass energy off of the resonance.   The size of the correction can be
roughly 
estimated as 
\beq
           - \beta \cdot \log{m_Z\over \Gamma_Z} = 40\% \ . 
\eeqn

To make a proper accounting of this effect, we need to include
arbitrary numbers of radiated collinear photons.  Fadin and Kuraev
introduced the idea of viewing the radiated photons and the final
annihilating electron as partons in the electron in the same way that
quarks and gluons are treated as partons in the proton~\cite{FK}.
For the proton, the parton distribution is generated by
non-perturbative effects, but for the electron  the parton
distributions are generated only by QED, so that they can be
calculated as a function of $\alpha$.  The result for the parton
distribution of the electron in the electron, to order $\alpha$, is
\beq
   f_e(z,s) = {\beta\over 2} (1-z)^{\beta/2-1} (1 + {3\over 8} \beta)
   - {1\over 4} \beta (1 + z) + \cdots \ , 
\eeq{fez}
where $z$ is the momentum fraction of the original electron carried
into the $\ee$ annihilation to a $Z$ boson.   The cross section for
producing a $Z$ boson would then be a convolution of the Breit-Wigner
cross section \leqn{simpleBW} with the parton distribution \leqn{fez}
and the corresponding distribution for the positron.  For the LEP
experiments, this theory was extended to include two orders of
subleading logarithms and finite corrections of order
$\alpha^2$~\cite{fullZwidth}.

The experimental aspects of the measurement of the $Z$ resonance
lineshape were also very challenging; see Section~2.2 of \cite{LEPEWWG}.  Careful control was needed for
point-to-point normalization errors across the $Z$ resonance.  The
absolute energy of the LEP ring was calibrated using resonant
depolarization of a single electron beam and then corrected for
two-beam effects.   This calibration was found to depend on the season
and the time of day.  Some contributing effects were the changes in
the size of the LEP tunnel due to the annual change in the water level
in Lake Geneva and current surges in the LEP magnets due to the
passage to the TGV leaving Geneva for Paris.

\begin{figure}
\begin{center}
\includegraphics[width=0.70\hsize]{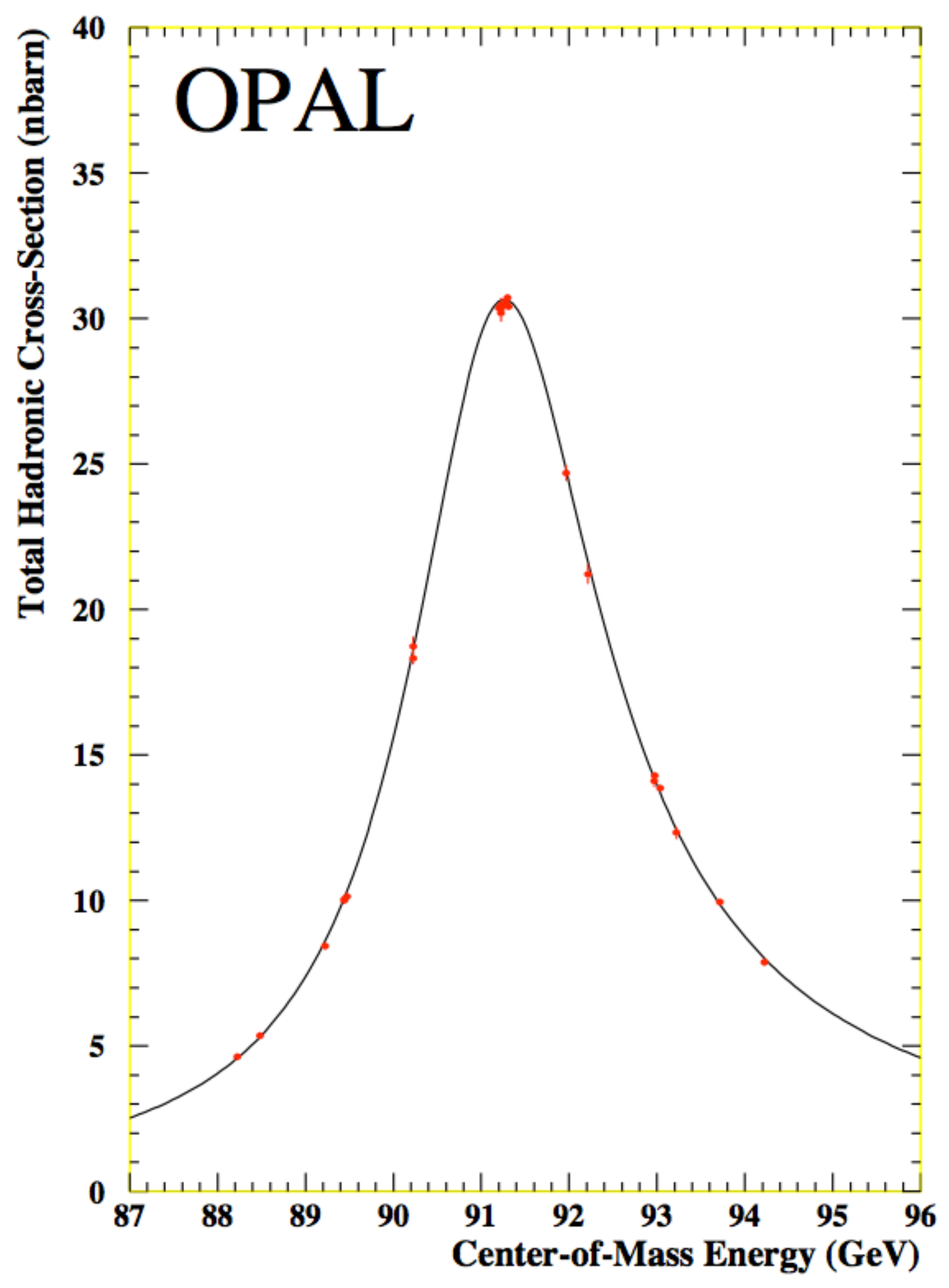}
\end{center}
\caption{Resonance line shape of the $Z$ in $\ee$ annihilation, as 
measured by the OPAL experiment~\cite{OPALZ}.}
\label{fig:OPALZ}
\end{figure}

\begin{figure}
\begin{center}
\includegraphics[width=0.50\hsize]{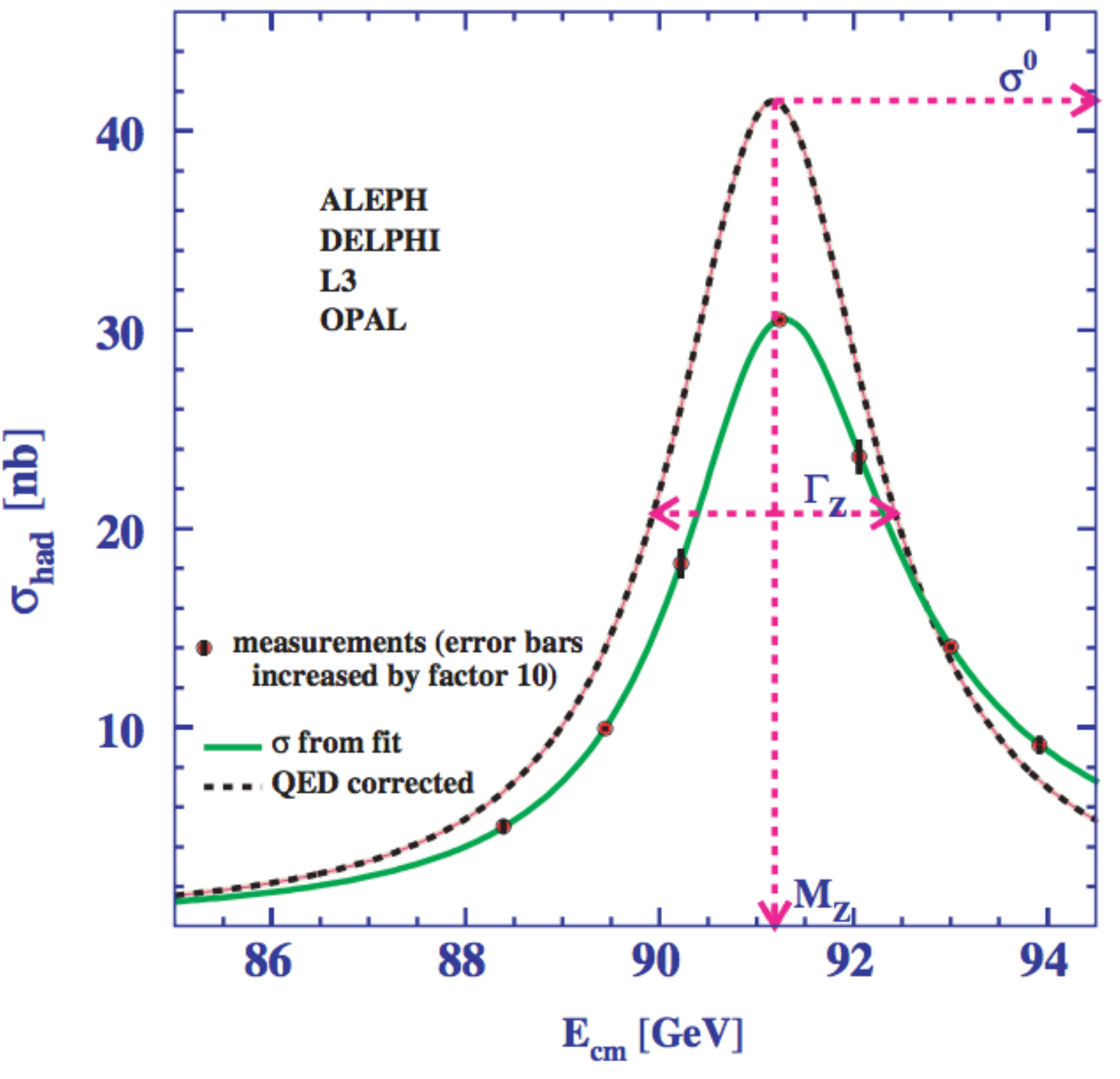}
\end{center}
\caption{Resonance line shape of the $Z$ in $\ee$ annihilation, as
  measured by the four LEP experiments, from \cite{LEPEWWG}.  The
  dotted curve shows the zeroth-order resonance line shape of the $Z$
  resonance.  The solid line shows the Standard Model prediction
  including initial-state radiative corrections.}
\label{fig:LEPZ}
\end{figure}

Some final results for the resonance line shape measurement are shown
in Figs.~\ref{fig:OPALZ}, \ref{fig:LEPZ}.   The first of these figures
shows the measurements by the OPAL experiment over the resonance and
the detaied agreement of the shape between theory and experiment~\cite{OPALZ}.
The second shows the combination of the resonance height and width
measurements from the four LEP experiments ALEPH, DELPHI, L3, and
OPAL~\cite{LEPEWWG}.   In this figure, the lower curve is the
radiatively corrected result; the higher curve is the inferred
Breit-Wigner distribution excluding the effects of radiative corrections.

\begin{figure}
\begin{center}
\includegraphics[width=0.60\hsize]{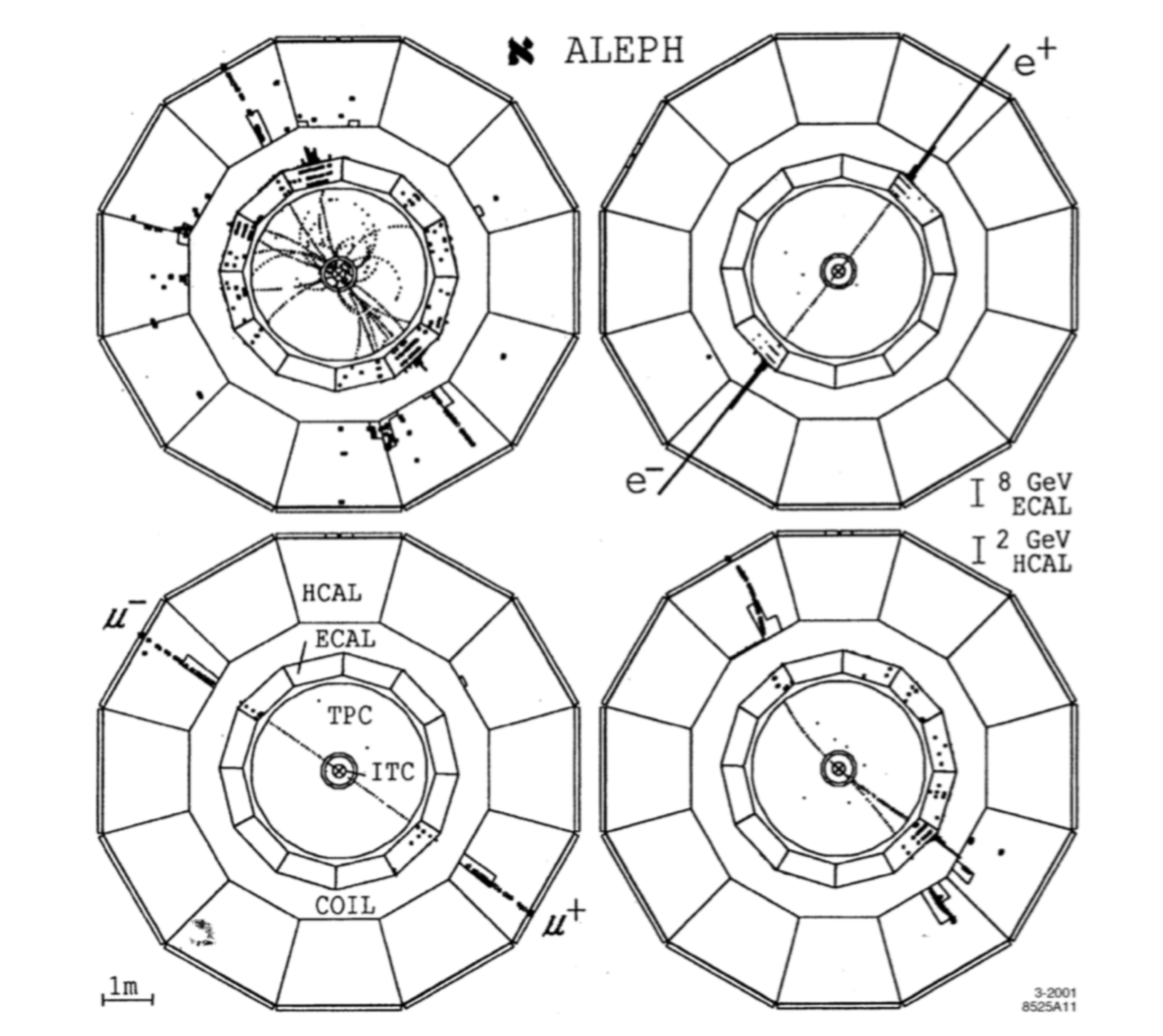}
\end{center}
\caption{Typical $\ee\to Z$ events corresponding to the $Z$ decays to
  hadrons, to $\ee$, to $\mu^+\mu^-$, and to $\tau^+\tau^-$, from 
\cite{ALEPHtypes}. }
\label{fig:ALEPHtypes}
\end{figure}

The measurement of branching ratios is more straightforward.  It is
necessary only to collect $Z$ decay events and sort them into
categories.   The various types of leptonic and hadronic decay modes
have very different, characteristic forms.  Typical events are shown in 
Fig.~\ref{fig:ALEPHtypes} for hadronic, $\ee$, $\mu^+\mu^-$, and
$\tau^+\tau^-$ decays~\cite{ALEPHtypes}.   The major backgrounds are
from Bhabha scattering and 2-photon events.  These do not resemble $Z$
decay events and are rather straightforwardly separated.   Nonresonant $\ee$
annihilations are also a small effect, generally providing backgrounds at only
the level of parts per mil. An exception is the $Z$ decay to
$\tau^+\tau^-$, which can be faked by hadronic
$\ee$ annihilations with radiation to provide a
background level of a few percent.   Still, these high signal to
background ratios are completely different from the situationn at the
LHC and enable measurements of very high precision.

Two particular branching ratios merit special attention.   First,
consider $Z$ decays to invisible final states.   The SM includes $Z$
decays to 3 species of neutrino, with a total branching ratio of
20\%.   Even though these decays are not seen in the detector, 
 the presence of invisible final states affects the resonance
lineshape  by increasing the $Z$ width and decreasing the $Z$ peak
height to visible modes such as hadrons.  Measurement of the resonance
parameters then effectively gives the number of light neutrinos into
which the $Z$ can decay.   The result is
\beq
    n_\nu = 2.9840 \pm  0.0082 \ ,
\eeqn
strongly constraining extra neutrinos or more exotic neutral
particles.

\begin{figure}
\begin{center}
\includegraphics[width=0.40\hsize]{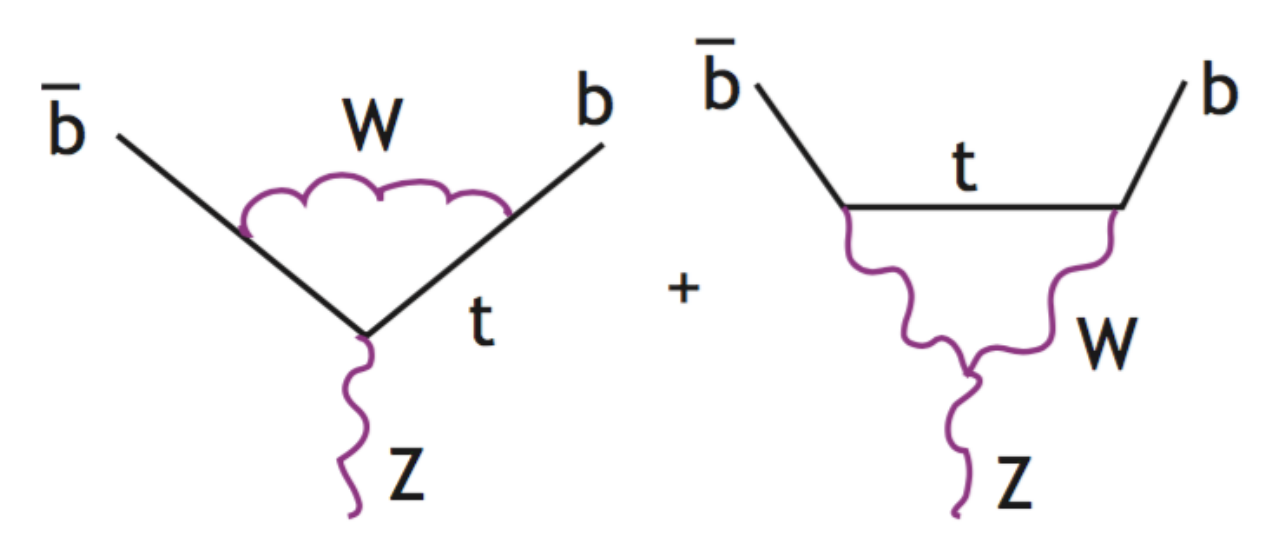}
\end{center}
\caption{Diagrams containing the top quark which give a relatively
  large
correction to the partial width for $Z\to b\bar b$.}
\label{fig:Ztoploops}
\end{figure}

Second, the $Z$ branching ratio to  $b$ quarks is of special interest,
for two reasons.  First, the $b$ belongs to the same $SU(2)\times
U(1)$ multiplet as the top quark, and, even in the SM, there is a
relatively large radiative correction due to top quark loops, from the
diagrams shown in Fig.~\ref{fig:Ztoploops}.   These produce 
\beq
    Q_{ZbL} = - \biggl( \half - \third s_w^2 - {\alpha\over 16\pi
    s_w^2}{m_t^2\over m_W^2} \biggr) \ , 
\eeq{shiftZb}
a shift of about $-2$\%.  More generally,  the $b$ is a
third-generation particle that might have a nontrivial coupling to
new, heavier, particles.

\begin{figure}
\begin{center}
\includegraphics[width=0.83\hsize]{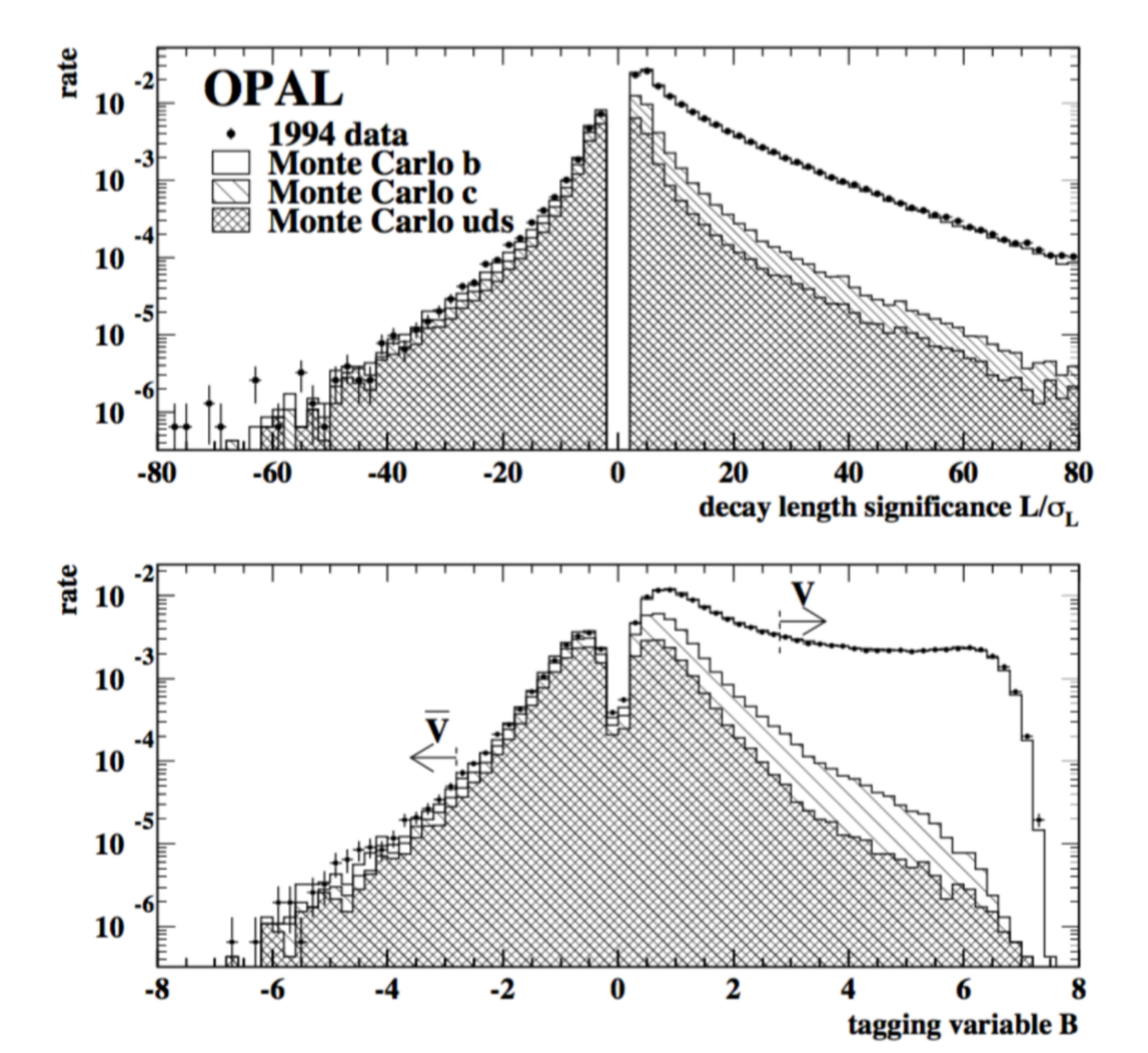}\\
\includegraphics[width=0.60\hsize]{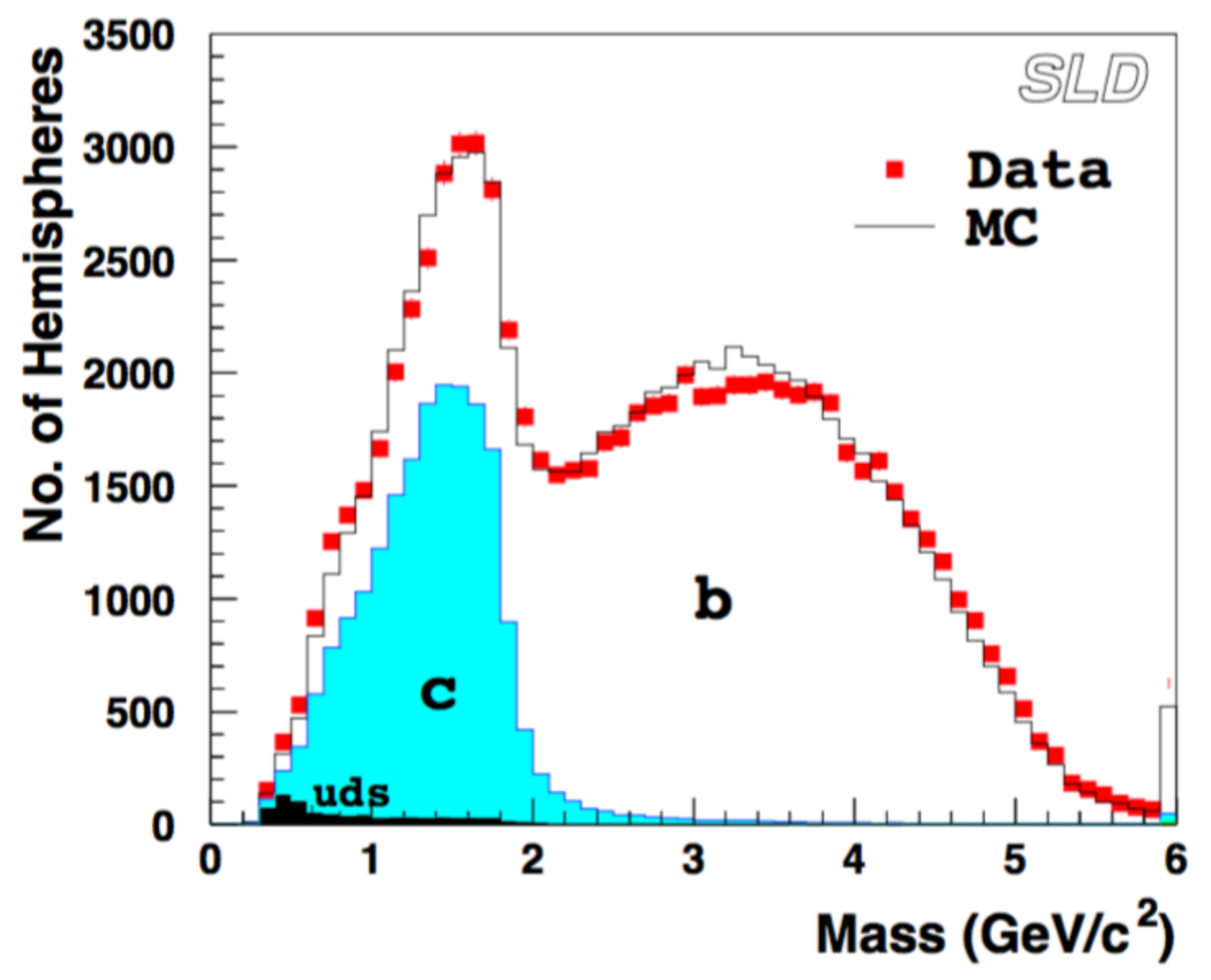}
\end{center}
\caption{Measurements of the $b$ and $c$ branching fractions of the
  $Z$.  Top:  Distributions in decay length significance and the $b$
  quark tagging variable,  from the OPAL experiment,
  showing the relative contributions of light quarks, $c$, and $b$,
  from \cite{OPALb}.  Right: Vertex mass distribution from the SLD
  experiment, showing the contributions from $c$ and $b$ meson decays,
  from \cite{SLDbc}. }
\label{fig:bvertex}
\end{figure}

An observable that specifically tracks this effect is 
\beq
          R_b =   {\Gamma(Z\to b\bar b)\over \Gamma(Z\to
            \mbox{hadrons})} \ . 
\eeqn
At leading order, we predict $R_b = 0.22$, but in the full SM this
value should be reduced according to \leqn{shiftZb}.   $Z$ decays to
$b\bar b$ could be identified by vertex tags.    The SLD detector at
SLAC included a pixel vertex detector capable of separating decays to  $b$ and
$c$ by vertex mass and by the presence of tertiary charm decay
vertices in $b$ jets.   Fig.~\ref{fig:bvertex}(a)  shows the signal
and background separation in the OPAL experiment~\cite{OPALb}.
Fig.~\ref{fig:bvertex}(b) shows a corresponding result from SLD,
in which the observed vertex mass was used to discriminate between the
$c$ and $b$ contributions~\cite{SLDbc}.  The final LEP
and SLC results gave
\beqa
         R_b &=&   0.21629\pm 0.00066  \CR
         R_c &=&   0.1721 \pm 0.0030    \ , 
\eeqan
confirming the shift predicted by \leqn{shiftZb} and demonstrating
consistency with the SM also for $Z\to c\bar c$. 

While the total rates for the $Z$ decay to the various species have
similar values, the asymmetries listed in \leqn{Zvalues} vary over a
wide range, from 15\% for the charged leptons to almost maximal for
the $d$-type quarks.   The SM predicts these disparate values from a
common value of $s_w^2$.

There are three very different methods to measure the lepton
asymmetries  $A_e$.  First, the $A_e$ can be found from the 
forward-backward asymmetry for $\ee\to f\bar f$ at the $Z$. 
 Second, $A_e$ can be determined from
the final-state polarization effects in the decays of $\tau^+\tau^-$
produced at the $Z$.   Finally,
$A_e$ can be measured directly from the rate for $Z$ production
from polarized electron beams. 

For unpolarized beams, the angular distribution for $\ee\to f\bar f$
can be found from \leqn{eeforms}.  On the $Z$ resonance, the
distribution takes the form
\beqa
{d\sigma\over d\cos\theta} &=&  \bigl({1+A_e\over 2}\bigr)\bigl({1 +
A_f\over 2}\bigr)  (1+\cos\theta)^2 +  \bigl({1-A_e\over 2}\bigr)\bigl({1 +
A_f\over 2}\bigr)  (1-\cos\theta)^2 \CR
&  & + \bigl({1+A_e\over 2}\bigr)\bigl({1 -
A_f\over 2}\bigr)  (1-\cos\theta)^2 +  \bigl({1-A_e\over 2}\bigr)\bigl({1 -
A_f\over 2}\bigr)  (1+\cos\theta)^2 \ .
\eeqan
The forward-backward asymmetry predicted by this expression is 
\beq
    A_{FB} = {3\over 4} A_e A_f
\eeqn
Especially for $b$ quarks, which have an almost maximal asymmetry, the
dependence of this quantity on $s_w^2$ is mainly through $A_e$.

\begin{figure}
\begin{center}
\includegraphics[width=0.60\hsize]{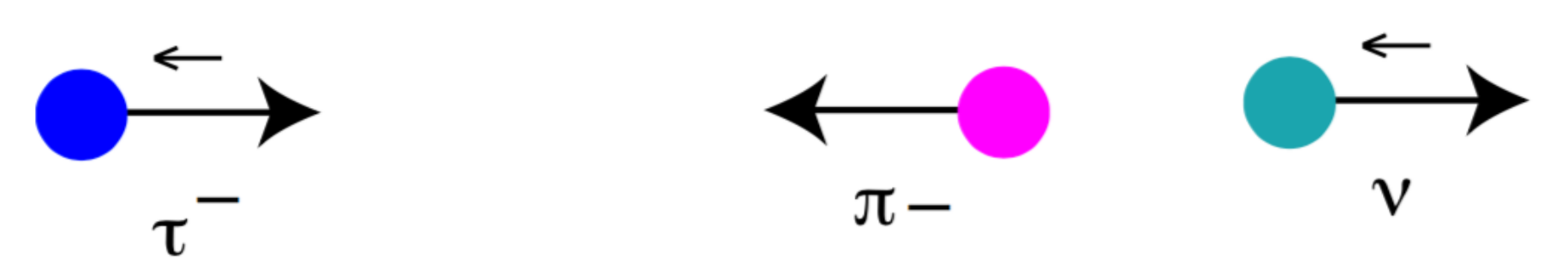}
\end{center}
\caption{Kinematics of $\tau\to \nu\pi$ decay.}
\label{fig:taunupi}
\end{figure}

\begin{figure}
\begin{center}
\includegraphics[width=0.60\hsize]{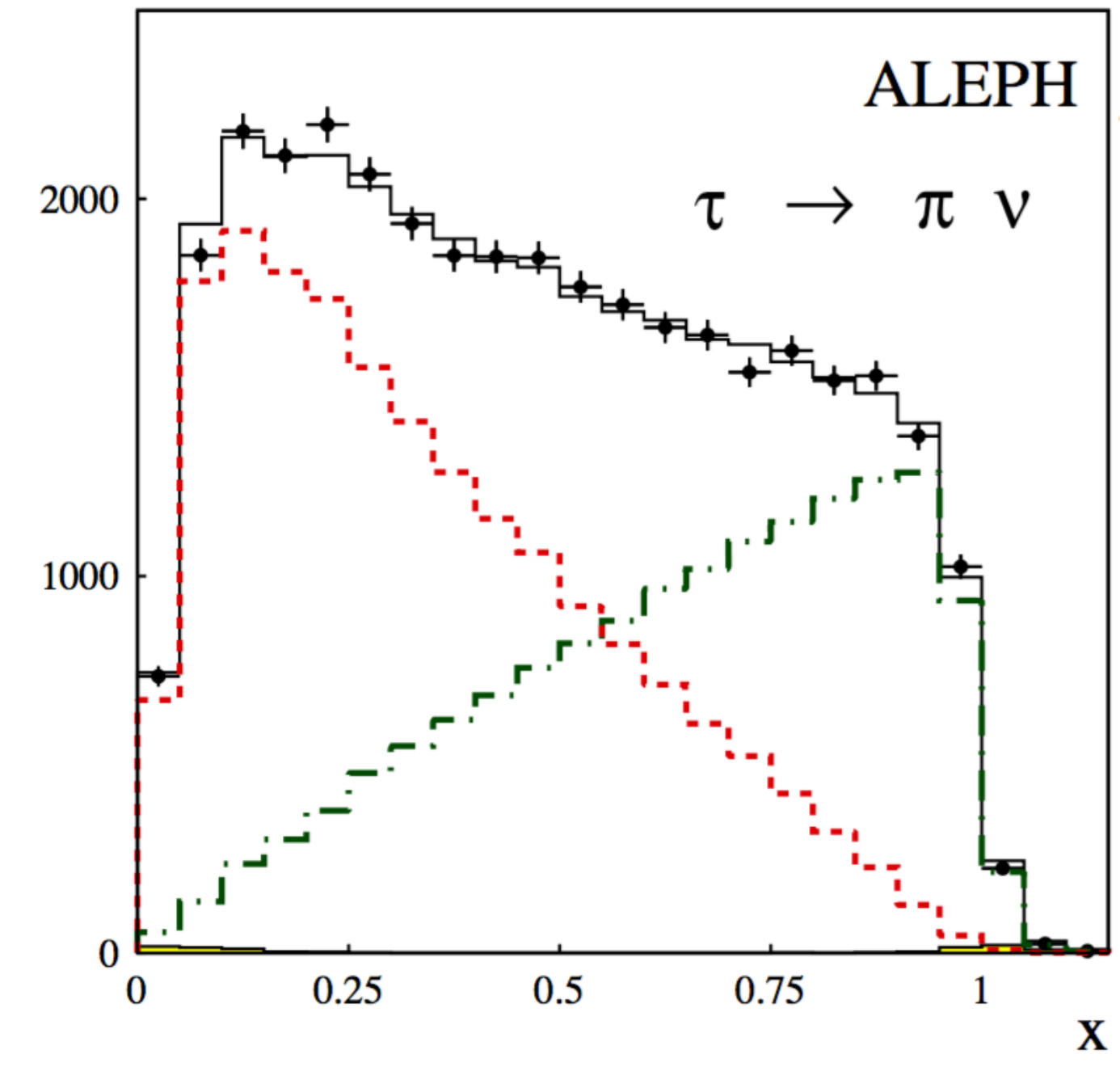}
\end{center}
\caption{Pion energy spectrum in $\tau\to \nu \pi$ decays at the $Z$
  resonance, from \cite{ALEPHtau}.  The ordinate $x =2 E_\pi/m_Z$.
  The separate contributions from $\tau_L$ and $\tau_R$ decays are indicated.}
\label{fig:ALEPHtau}
\end{figure}

The value of $A_e$ determines the polarization of $\tau$ leptons
produced in $Z$ decays, and this polarization becomes visible through
the \VmA\ structure of the $\tau$ decays.   The easiest case to
understand is the decay $\tau^-\to \nu_\tau \pi^-$.   Since the
neutrino is always left-handed and the pion has zero spin, a $\tau^-$
at rest with $S^3 = - \half$ will decay to a forward neutrino and a
backward $\pi^-$, as shown in Fig.~\ref{fig:taunupi}.   When the
$\tau^-$ is boosted, a left-handed $\tau$ will decay to a high-energy
neutrino and a slow pion.  A right-handed $\tau$ will decay to a
low-energy neutrino and a fast pion.   More generally,  if $x$ is
the fraction of the $\tau$ momentum carried by the $\pi^-$,
\beq
       \tau_L \ : \ {d\Gamma\over dx} \sim (1-x) \qquad \tau_R \ : \
       {d\Gamma\over dx} \sim x \ .
\eeqn
Similar asymmetries appear in the other $\tau$ decay modes.  
 Fig.~\ref{fig:ALEPHtau}
shows the distributions measured by the ALEPH experiment for $\tau\to
\pi \nu$,  compared to the expected
distributions from $\tau_L$ and $\tau_R$.  The 15\% asymmetry is
apparent.   The SM also predicts a correlation
between polarization and $\cos\theta$ that can be used to improve the
$s_w^2$ measurement.

The SLC produced $\ee\to Z$ events using linear acceleration of the
electrons.  This technique allowed the preservation of electron
polarization from the source to the collisions.  The experiment was
conducted by flipping the 
the electron polarization in each bunch randomly, and measuring the
correlation between the polarization orientation and the total $Z$
production rate---measured 4~km downstream of the
source.  This gave a  direct
 measurement~\cite{SLDpol}
\beq
      A_e = 0.1516 \pm 0.0021 
\eeqn

\begin{figure}
\begin{center}
\includegraphics[width=0.60\hsize]{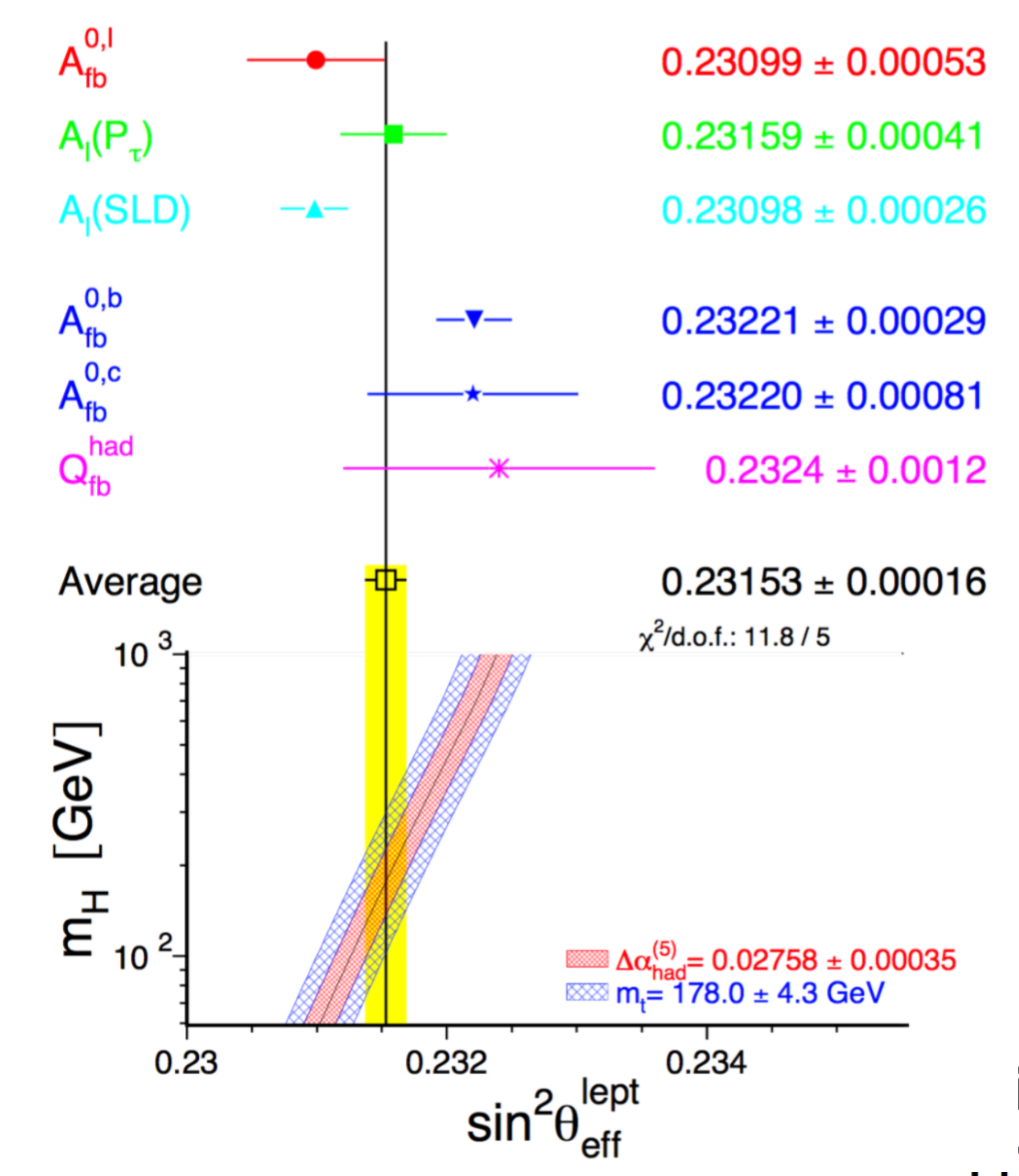}
\end{center}
\caption{Summary of $A_\ell$ measurements at the $Z$ resonance from
  different observables, from~\cite{LEPEWWG}.}
\label{fig:Aesummary}
\end{figure}

Figure~\ref{fig:Aesummary} shows the summary of the various
determinations of $s_w^2$ from the leptonic asymmetries~\cite{LEPEWWG}.   The
measurements are statistically consistent and lead to a very precise
value.

\begin{figure}
\begin{center}
\includegraphics[width=0.80\hsize]{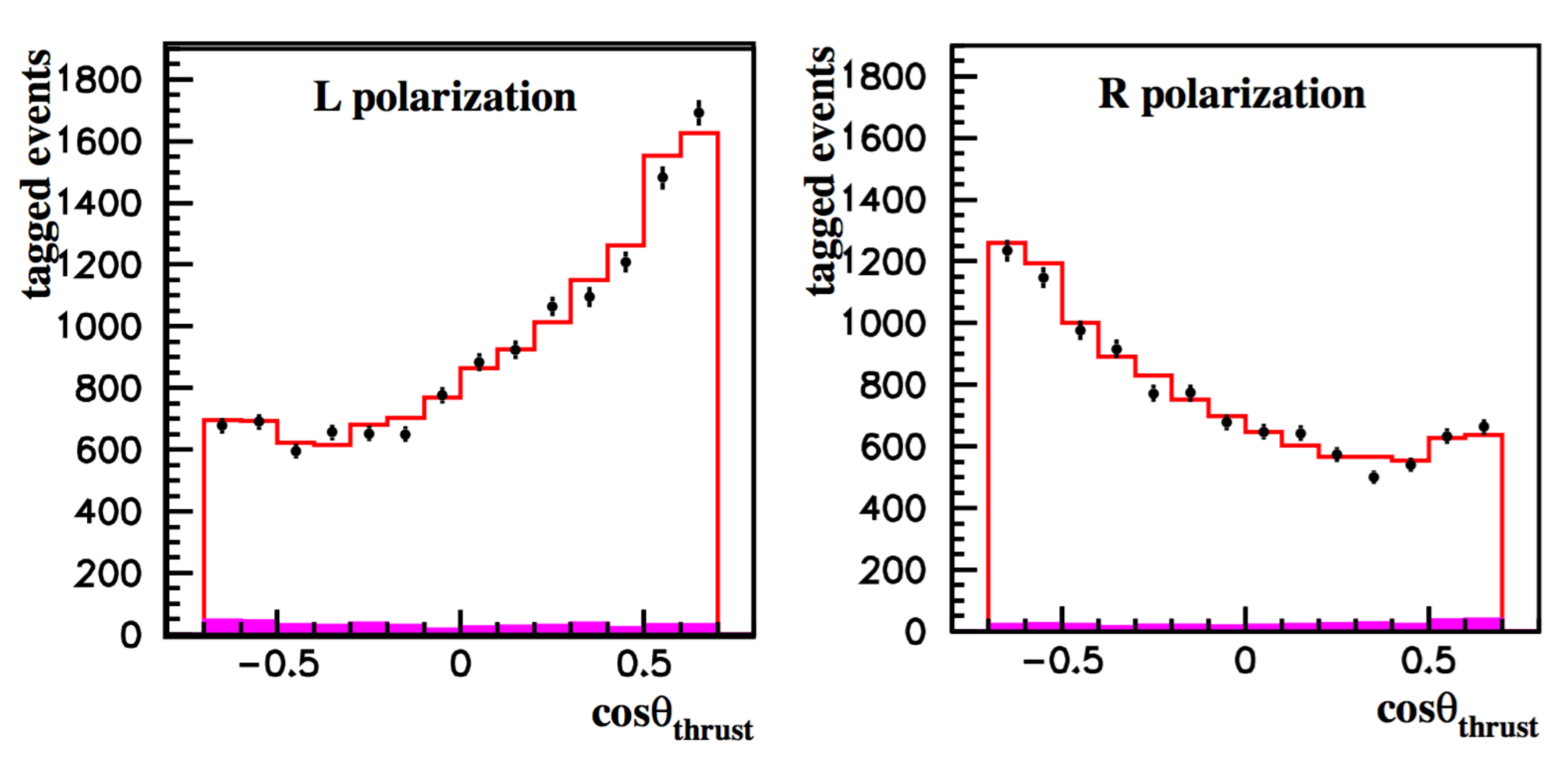}
\end{center}
\caption{Angular distribution of $\ee\to Z\to b\bar b$ events measured
  by the SLD experiment for left- and right-handed polarized beams,
  from~\cite{SLDbpol}. }
\label{fig:bdist}
\end{figure}

The prediction that the $b$ asymmetry is close to maximal implies that
the angular distribution of $\ee\to b\bar b$ at the $Z$ should show a
large dependence on beam polarization.  The distribution should be
close to $(1+\cos\theta)^2$ for a left-handed polarized beam and close
to $(1-\cos\theta)^2$ for a right-handed polarized beam.   The
distributions measured by the SLD experiment at the SLC  for left- and
right-handed beams are shown in
Fig.~\ref{fig:bdist}.   Allowing for the expected  confusion in
separating $b$ and $\bar b$ jets, the results are consistent with a
high $b$ polarization in $Z$ decays.  The difference in normalization
of the two distributions reflects the 15\% asymmetry in the production
cross section.

\begin{figure}
\begin{center}
\includegraphics[width=0.80\hsize]{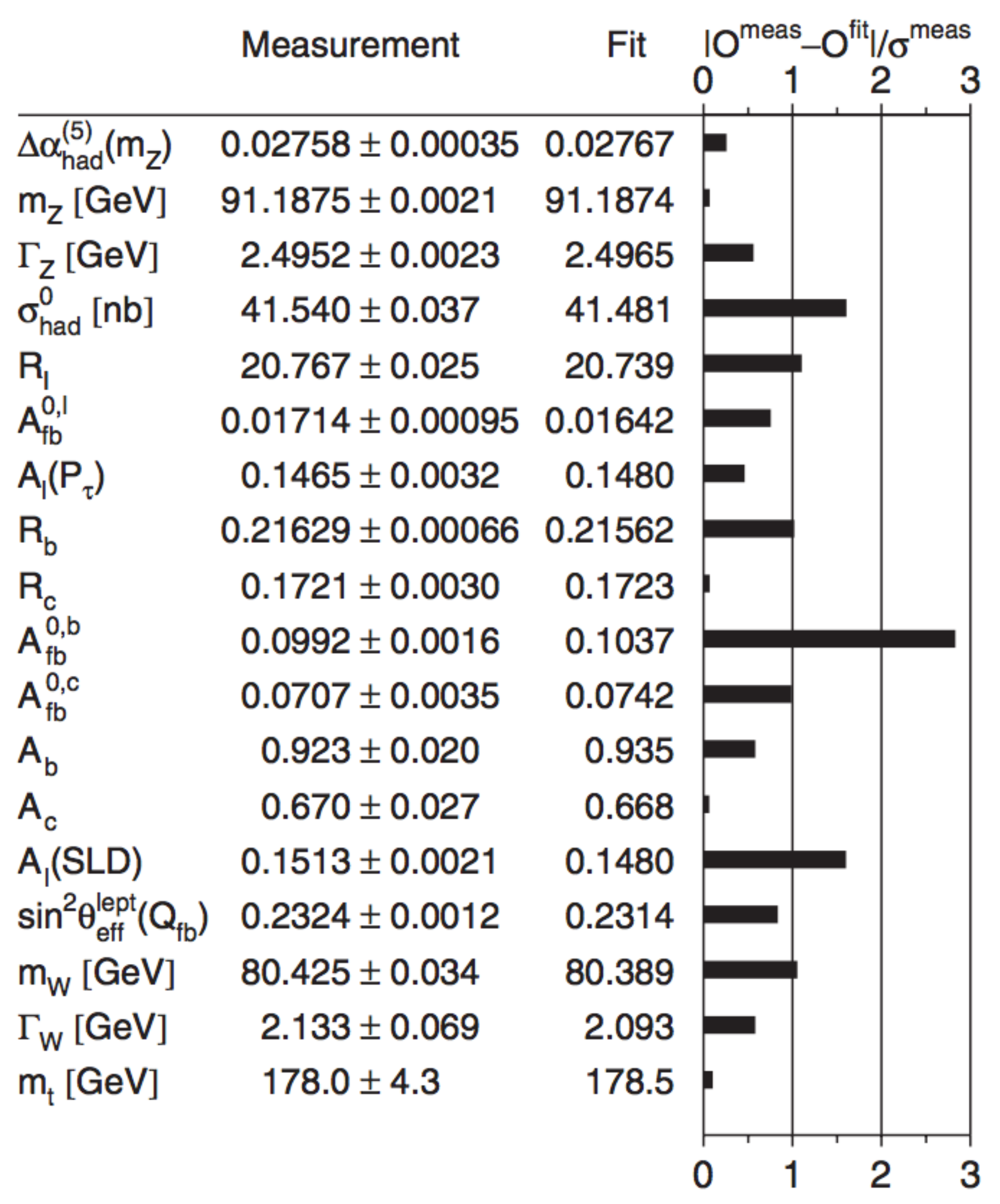}
\end{center}
\caption{Summary of precision electroweak measurements at the $Z$
  resonance, from \cite{LEPEWWG}.}
\label{fig:Zsummary}
\end{figure}

Figure~\ref{fig:Zsummary} shows a summary of the precision
measurements of the properties of the $Z$ boson~\cite{LEPEWWG}.
   The measured values 
listed in the first column are compared to the values from the best
fit to the SM, including one-loop radiative corrections.   The bars
show the deviations from the SM prediction, in units of the $\sigma$
of the measurement.  This is an impressive confirmation of the
$SU(2)\times U(1)$ weak interaction model.

\subsection{Constraints on oblique radiative corrections}

From the excellent agreement of the $Z$ measurements with the SM, it
is possible to put general constraints on possible new particles
coupling to the weak interactions.  

To explain this, we should first discuss the properties of one-loop
corrections to the $SU(2)\times U(1)$ predictions in more detail.
The SM contains a large number of parameters.  However,
the predictions discussed in this Section depend, at the three level,
only on the three parameters
\beq
      g \ , \ g'\ , v  \  .
\eeqn
The loop corrections will include divergences, including quadratically
divergent corrections to $v^2$.   However, because the $SU(2)\times
U(1)$ theory is renormalizable, once these three parameters are fixed,
all of the 1-loop corrections must be  finite.   Then each specific
reaction aquires a finite prediction, which is a testable consequence
of the SM.

DIfferent schemes are used to fix the three underlying divergent
amplitudes.  Each gives different expressions for the measurable cross
sections.  Three common schemes are
\begin{itemize}
\item  applying $\msb$ subtraction, as in QCD
\item  fixing $\alpha(m_Z)$, $m_Z$, $m_W$ to their measured values
  (Marciano-Sirlin scheme)~\cite{MarcianoSirlin}
\item  fixing $\alpha(m_Z)$, $m_Z$, $G_F$   to their measured values
  (on shell $Z$ scheme)
\end{itemize}
In the $\msb$ scheme, used by the Particle Data Group, the
$\msb$ parameters $g$, $g'$, and $v$ are unphysical but can be
defined as the values that give the best fit to the corpus of 
SM measurements~\cite{PDGweak}.  

The various schemes for renormalizing the $SU(2)\times U(1)$ model
lead to different definitions of $s_w^2$ that are found in the
literature.  In the Marciano-Sirlin scheme, we define $\theta_w$ by 
\beq
    c_w \equiv  m_W/m_Z  \ . 
\eeq{defcw}
This leads to 
\beq
    s_w^2 = 0.22290 \pm 0.00008 \ .
\eeq{marcianosstw}
We will see in Section 4 that the relation \leqn{defcw} is often needed
to insure the correct behavior in high-energy reactions of $W$ and
$Z$, so it is useful that this relation is insured at the tree level.
 Thus, the Marciano-Sirlin definition of $\theta_w$ is the most common
 one used in event generators for LHC.    However, one should note
 that the value \leqn{marcianosstw} is significantly different from
 the value \leqn{refsstw} that best represents the sizes of the $Z$
 cross sections and asymmetries.

In the on-shell $Z$ scheme, $\theta_w$ is defined by 
\beq
   sin^2 2\theta_w  = (2 c_w s_w)^2 \equiv   {4\pi \alpha(m_Z)\over
     \sqrt{2} G_F m_Z^2}\ , \ 
\eeqn
leading to 
\beq
    s_w^2 = 0.231079 \pm 0.000036 \ .
\eeqn
 This defintion gives at  tree level a value that is much closer to
 \leqn{refsstw}.   All three values of  $\sstw$
 lead to the same predictions for the relation of observables to
 observables after the (scheme-dependent)  finite 1-loop corrections are
 included.

One particular class of radiative corrections is especially simple to
analyze.  If new particles have no direct coupling to light fermions,
they can apprear in radiative corrections to the $Z$ observables only
through vector boson vacuum polarization amplitudes.  Effects of this
type are called {\it oblique} radiative corrections.   These effects
can be analyzed in a quite general way.

\begin{figure}
\begin{center}
\includegraphics[width=0.80\hsize]{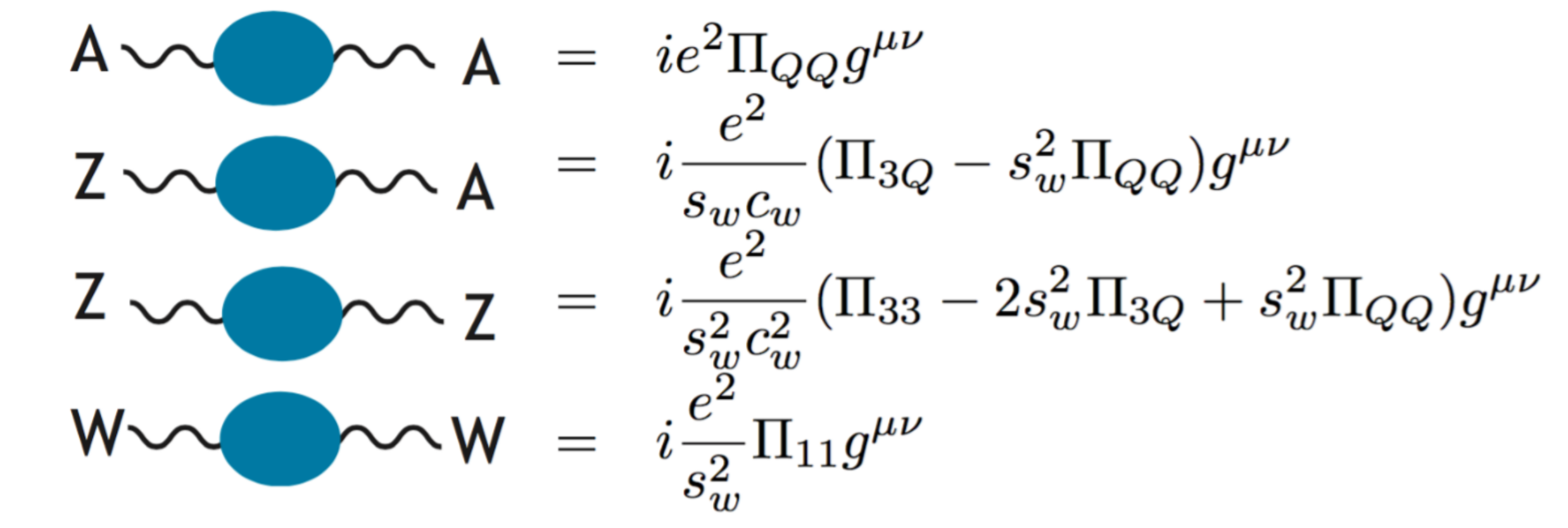}
\end{center}
\caption{Vector boson vacuum polarization diagrams..}
\label{fig:VPs}
\end{figure}

There are four electroweak vacuum polarization amplitudes  $\Pi_{AB}(q^2)$.  I will
notate them as shown in Fig.~\ref{fig:VPs}.  The subscripts $1,3$
refer to the weak isospin currents $j^{\mu a}$,  $a = 1,3$; the
subscript $Q$ refers to the electromagnetic current.   The $Z$ vacuum
polarizations
are found from these elements using \leqn{QZdef}.  If the particles in the loop
have large masses $M$, we can Taylor expand the vacuum polarization amplitudes in
powers of $q^2/M^2$.  Up to order $q^2/M^2$, we find
\beqa
       \Pi_{QQ} (q^2) &= & A q^2 + \cdots  \CR
      \Pi_{3Q} (q^2) &=& B q^2 + \cdots \CR
         \Pi_{33} (q^2) &= & C + D q^2 + \cdots  \CR
      \Pi_{11} (q^2) &=& E  + F q^2 + \cdots 
\eeqan

There are six constants in this set of formulae.   Three of them are
fixed by the renormalizations of $g$, $g'$, $v$.   This leaves 3
finite combinations of vacuum polarization amplitudes will be
predicted in any new physics model.   These combinations are 
canonically defined as~\cite{PT}
\beqa
  S &=&   {16\pi\over m_Z^2} \biggl[ \Pi_{33}(m_Z^2) - \Pi_{33}(0) -
  \Pi_{3Q}(m_Z^2) \biggr] \CR
T &=&   {4\pi\over s_w^2 m_W^2} \biggl[ \Pi_{11}(0) - \Pi_{33}(0) \biggr] \CR
U &=&   {16\pi\over m_Z^2} \biggl[ \Pi_{11}(m_Z^2) - \Pi_{11}(0) -
  \Pi_{33}(m_Z^2) + \Pi_{33}(0)\biggr] 
\eeqa{STUdef}
In \cite{PT}, the amplitudes appearing in \leqn{STUdef} are the new
physics contributions only, but other analyses, for example,
\cite{PDGweak}, use different conventions.   The three parameters in
\leqn{STUdef} have clear physical interpretations.  $T$ parametrizes
the size of 
weak isospin violating corrections to the relation $m_W = m_Z
c_w$.  $S$ parametrizes the $q^2/M^2$ corrections.  $U$ requires both
effects and is predicted to be very small in most new physics models.

The leading oblique corrections to electroweak observables can then be
expressed as linear shifts proportional to $S$ and $T$.  For example,
\beqa
     {m_W^2 \over m_Z^2} - c_0^2 &=  &  {\alpha c_w^2 \over c_w^2 -
       s_w^2} \biggl( - \half  S + c_w^2 T\biggr) \CR
   s_*^2 - s_0^2 &=  &  {\alpha \over c_w^2 -
       s_w^2} \biggl( - \half {1\over 4}  S - s_w^2 c_w^2 T\biggr)  \
     , 
\eeqa{STpredict}
where $s_0$, $c_0$ are the values of $s_w$ and $c_w$ in the on-shell
$Z$ scheme and $s_*$ is the value of $s_w$ used to evaluate the $Z$
asymmetries $A_f$.   By fitting to the formulae such as
\leqn{STpredict}, we can obtain general constraints that can be
applied to a large class of new physics models.

Some guidance about the expected sizes of $S$ and $T$ is given by the
result for one new heavy  electroweak doublet,
\beq
    S = {1\over 6 \pi} \qquad  T =  {|m_U^2 - m_D^2|\over m_Z^2} \ . 
\eeqn
A complete heavy fourth generation gives $S  = 0.2$.  The effects of
the SM top quark and Higgs boson can also be expressed approximately
in the $S$, $T$ framework,
\beqa
     \mbox{top\ :} \qquad  S = {1\over 6\pi} \log{m_t^2\over m_Z^2}
     &\qquad&
           T = {3\over 16\pi s_w^2 c_w^2} {m_t^2\over m_Z^2} \CR
     \mbox{Higgs\ :} \qquad  S = {1\over 12\pi} \log{m_h^2\over m_Z^2}
     &\qquad&
           T = - {3\over 16\pi c_w^2} \log {m_h^2\over m_Z^2}
\eeqan
The appearance of corrections proportional to $m_t^2/m_Z^2$, which we
have already seen in \leqn{shiftZb}, will be explained in Section~5.

\begin{figure}
\begin{center}
\includegraphics[width=0.55\hsize]{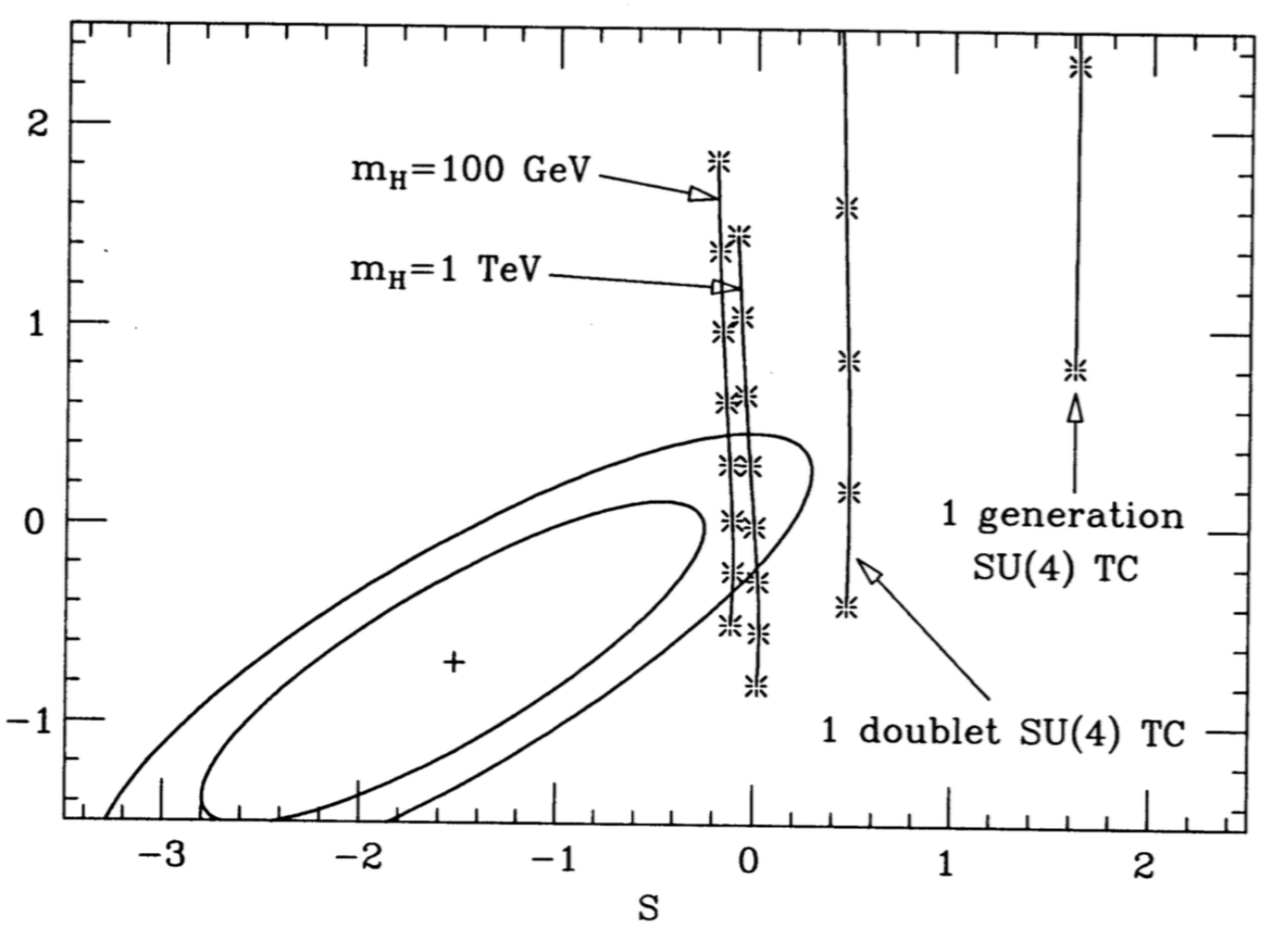} \\
\includegraphics[width=0.60\hsize]{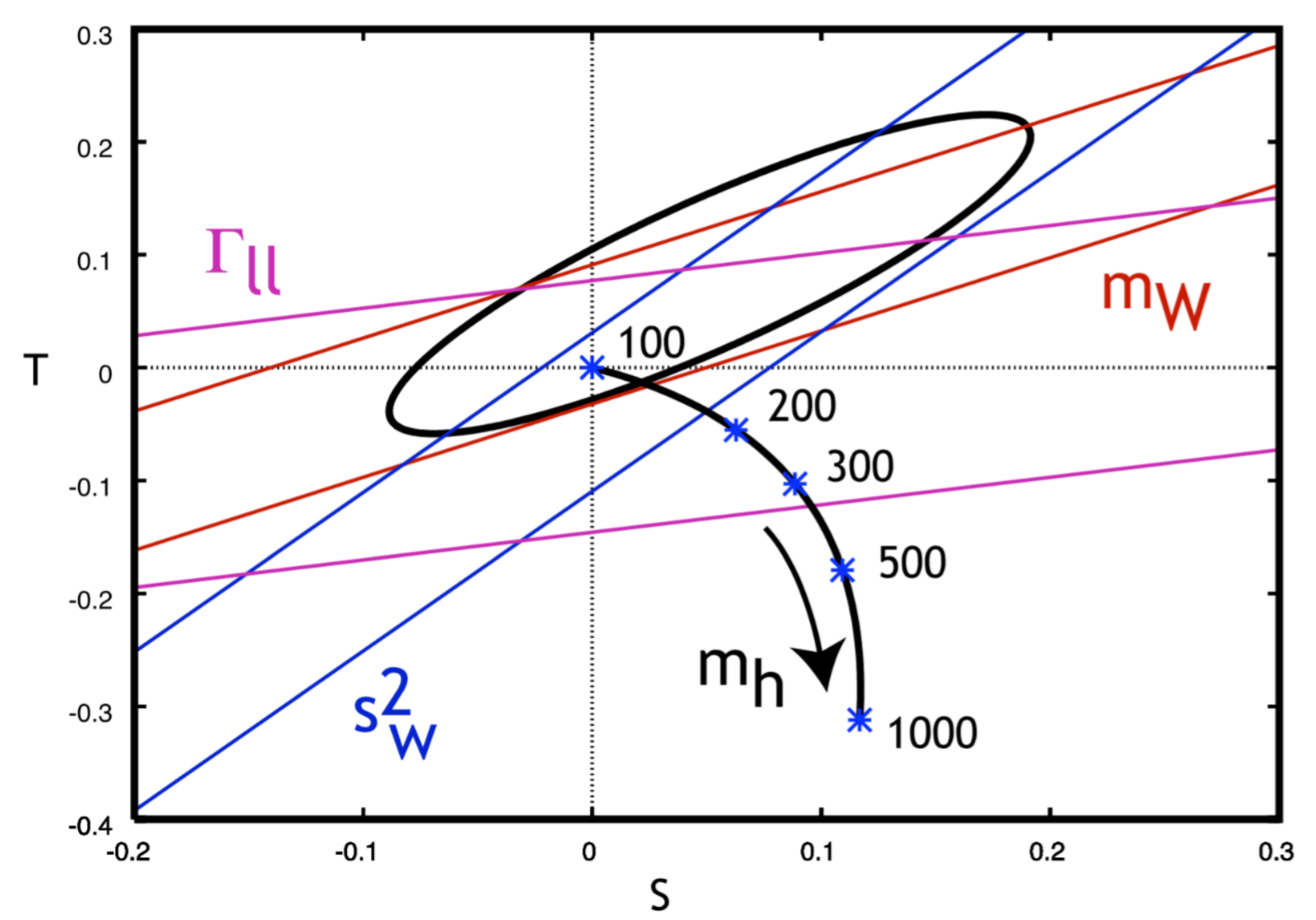} \\
\includegraphics[width=0.60\hsize]{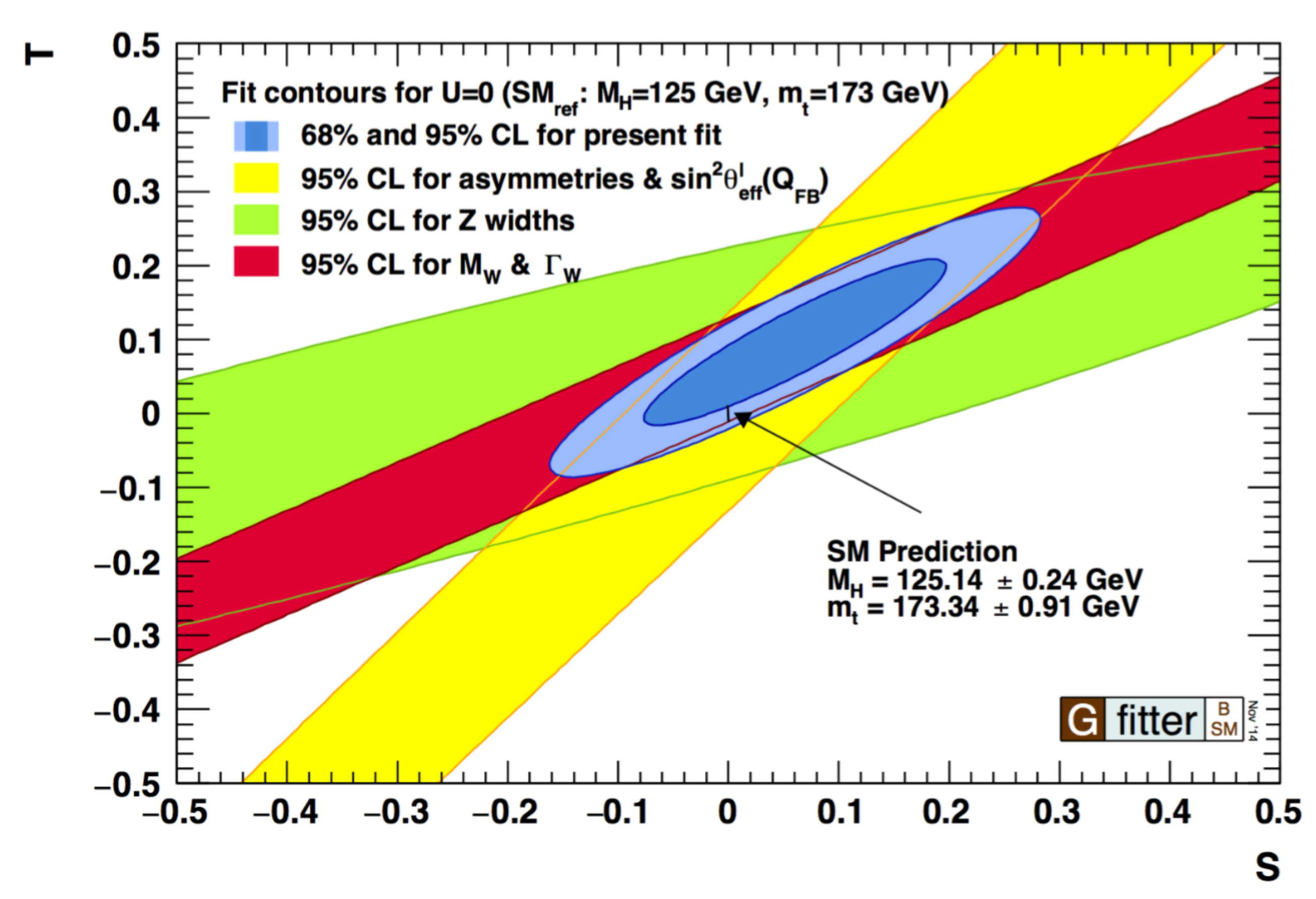} \\
\end{center}
\caption{Allowed domain for the $S,T$ parameters in three different
  eras:  in 1991, before the discovery of the top quark~\cite{PT}; in
  2008, before the discovery of the Higgs boson; today~\cite{Gfitter}.}
\label{fig:STfits}
\end{figure}

Figure~\ref{fig:STfits} shows the progress of the $S$, $T$ fit with our
improved understanding of the SM.   Figure~\ref{fig:STfits}(a)
reflects the situation in 1991, before the discovery of the top
quark~\cite{PT}.  The two vertical lines to the left are predictions of the SM
with a varying top quark mass.  Values of $m_t$ in the range of
170--180~GeV are highly favored by the precision electroweak data.
The measurement of $S$, even without the value of $m_t$, strongly
constrained the ``technicolor''  models of electroweak symmetry
breaking.  (I will describe these models at the end of Section~7.2.)
 Figure~\ref{fig:STfits}(b)  shows the $S$, $T$ fit in 2008.  The
 solid curve shows the predictions of the SM with a variable Higgs
 boson mass.  Values of the Higgs mass close to 100~GeV are strongly
 favored.  Figure~\ref{fig:STfits}(c)  shows the current $S$, $T$
 fit~\cite{Gfitter}.   The fit is in good agreement with the SM with
 the now-measured values of $m_t$ and $m_h$.  It also is in
 substantial tension with the presence of a fourth generation of
 quarks and leptons.

\section{The Goldstone Boson Equivalence Theorem}

In this section, I will describe the properties of the weak
interactions at energies much greater than $m_W$ and $m_Z$.   Some new
conceptual
issues appear here.  These affect the energy-dependence of $W$ and $Z$
boson reactions at high energy and the parametrization of possible
effects of new physics.  I will introduce a way of thinking that can
be used as a skeleton key for understanding these issues, called the
Goldstone Boson Equivalence Theorem.

\subsection{Questions about  $W$ and $Z$ bosons at high energy}

To begin this discussion, I wil raise a question, one that turns out
to be one of the more difficult questions to answer about
spontaneously broken gauge theories.

In its rest frame, with $p^\mu = (m, 0,0,0)^\mu$, 
a massive vector boson has 3 polarization states,
corresponding to the 3 orthogonal spacelike vectors
\beqa 
    \eps_+^\mu  &=& {1\over \sqrt{2}}  (0, 1 , +i ,0 )^\mu \CR
    \eps_0^\mu  &=&  (0, 0, 0 ,1)^\mu \CR
    \eps_-^\mu  &=&  {1\over \sqrt{2}} (0, 1 , -i ,0 )^\mu \ .
\eeqa{restpol}
These vectors represent the states of the vector boson with definite
angular momentum $J^3 = +1, 0, -1$. 

Now boost along the $\hat 3$ axis to high energy, $p^\mu = (E, 0,
0,p)^\mu$.    The boosts of the
polarization
vectors in \leqn{restpol} are
\beqa
   \eps_+^\mu  &=& {1\over \sqrt{2}}  (0, 1 , +i ,0 )^\mu \CR
    \eps_0^\mu  &=&  ({p\over m}, 0, 0 ,{E\over m} )^\mu \CR
    \eps_-^\mu  &=&  {1\over \sqrt{2}} (0, 1 , -i ,0 )^\mu \ .
\eeqa{boostpol}
The transverse polarization vectors $\eps_+$, $\eps_-$ are left
unchanged by the boost.  However,  for the longitudinal polarization
vector $\eps_0$, the components grow
without bound.  At very high energy
\beq
      \eps_0^\mu \to  {p^\mu \over m} \ . 
\eeq{epslimit}
Another way to understand this is to recall that the polarization sum  for a
massive vector boson is written
covariantly as 
\beq
  \sum_i  \eps_i^\mu \eps_j^\nu = - \biggl(  g^{\mu\nu} -
  {p^\mu p^\nu\over m^2 } \biggr) \ .
\eeqn
In the rest frame of the vector boson, this  is the projection onto
the 3 spacelike polarization vectors.   For a highly boosted vector
boson, however, 
the second term in parentheses in this expression has matrix elements
that grow large in the same way as \leqn{epslimit}. 

This potentially leads to very large contributions to amplitudes for
high-energy vector bosons, even threatening violation of unitarity.
An example of this problem is found in the production of a pair of
massive vector bosons in $\ee$ annihilation.   The amplitude for
production of a pair of scalar bosons in QED is 
\beq
   i\M(\ee\to \phi^+\phi^-) =  - i {e^2\over s} (2E) \sqrt{2} \eps_-
   \cdot (k_- - k_+)  \ ,
\eeqn
where $k_+$, $k_-$ are the scalar particle momenta.   In $\ee\to
W^+W^-$, we might expect that this formula generalizes to 
\beq
   i\M(\ee\to \phi^+\phi^-) =   i {e^2\over s} (2E) \sqrt{2} \eps_-
   \cdot (k_+ - k_-) \  \eps^*(k_+) \cdot \eps^*(k_-) \ .
\eeq{firsteeWW}
where $\eps(k_+)$, $\eps(k_-)$ are the $W^+$ and $W^-$ polarization vectors.
For longitudinally polarized $W$ bosons, this extra factor becomes
\beq
       {   k_+\cdot k_-\over m_W^2}   = { s - 2m_W^2\over 2m_W^2}
\eeq{WWenhancement}
at high energy.   This growth of the production amplitude really would
violate
unitarity.

This raises the question: Are the enhancements due to $\eps_0 \sim
p/m$ at high energy actually present?  Do these enhancements appear
 always, sometimes, or
never?

The answer to this question is given by the Goldstone Boson
Equivalence Theorem (GBET) of Cornwall, Levin, and Tiktopoulos and
Vayonakis~\cite{Cornwall,Vayonakis}.

When a $W$ boson or other gauge boson acquires mass through 
the Higgs mechanism, this boson must also acquire a longitudinal
polarization state that does not exist for a massless gauge boson.
The extra degree of freedom is obtained from the symmetry-breaking
Higgs field, for which a Goldstone boson is gauged away.  
When the $W$ is at rest, it is not so clear which polarization state
came from the Higgs field.  However, for a highly boosted $W$ boson,
there is a clear distinction between the transverse and longitudinal
polarization states.   The GBET states, in the limit of high energy,
the couplings of the longitudinal polarization state are precisely
those of the original Goldstone boson,
\beq
  \M(X\to Y + W^+_0(p)) = \M(X\to Y +\pi^+(p))\  \bigl(1 + {\cal
    O}({m_W\over E_W}) \bigr)
\eeq{theGBET}
The proof is too technical to give here. Some special cases are
analyzed in Chapter 21 of \cite{PeskinSchr}.   A very elegant and complete
proof, which accounts for  radiative corrections and includes the
possibility of multiple boosted vector bosons, 
 has been given by Chanowitz and Gaillard in \cite{CG}.   Both arguments
 rely  in an essential way on the underlying  gauge invariance of the theory.

In the rest of this section, I will present three examples that
illustrate the various aspects of this theorem.

\subsection{$W$ polarization in top quark decay}

The first application is the theory of the polarization of the $W$
boson emitted in top quark decay, $t \to b W^+$. 

It is straightforward to compute the rates for top quark decay to
polarized $W$ bosons.   These rates follow directly from the form of
the \VmA \ coupling.   The matrix element is 
\beq
  i \M =  i {g\over \sqrt{2}} u^\dagger_L(b) \ \bar \sigma^\mu \
  u_L(t) \ \eps_\mu^*  \ .
\eeqn
  In evauating this matrix element, I will ignore
the $b$ quark mass, a very good approximation.  I will use coordinates
in which the $t$ quark is at rest, with spin orientation given by a
2-component spinor $\xi$, and the $W^+$ is emitted in the $\hat 3$
direction.   The $b$ quark is left-handed and moves in the $-\hat
3$ direction.   Then the spinors are 
\beq
        u_L(b)  = \sqrt{2E_b} \pmatrix{-1\cr 0\cr} \qquad u_L(t) =
        \sqrt{m_t} \xi \ . 
\eeqn

For a $W^+_-$, 
\beq
     \bar \sigma\cdot \eps^*_- = {1\over\sqrt{2}}( \sigma^1 + i \sigma^2) =
       \sqrt{2}\sigma^+
\eeqn
and so the amplitude is 
\beq
   i\M = i g \sqrt{2 m_t E_b} \xi_2  \ . 
\eeqn
with, from 2-body kinematics, $E_b = (m_t^2 - m_w^2)/2m_t$. 
For a $W^+_+$, the sigma matrix structure is proportional to
$\sigma^-$ and the amplitude vanishes.  For a $W^+_0$, 
\beq
     \bar \sigma\cdot \eps^*_0 =  -  {p + E\sigma^3 \over m_W} 
\eeqn
and the amplitude is
\beq
    i \M =  ig \sqrt{2m_t E_b} \ {m_t\over m_W} \ \xi_1 \ . 
\eeq{tbWlong}
Squaring these matrix elements, averaging over the $t$ spin direction, and integrating over phase
space, we find
\beqa
    \Gamma(t\to bW^+_-) &=& {\alpha_w\over 8} m_t  \bigl( 1 -
    {m_W^2\over m_t^2} \bigr)^2 \CR
   \Gamma(t\to bW^+_+) &=&0 \CR
   \Gamma(t\to bW^+_0) &=& {\alpha_w\over 8} m_t  \bigl( 1 -
    {m_W^2\over m_t^2} \bigr)^2  \cdot {m_t^2 \over 2 m_W^2} \ . 
\eeqan
  From these formulae, we see that the fraction of longitudinally
  polarized $W$ bosons is
\beq
   {\Gamma(t\to b W^+_0)\over \Gamma(t\to b W^+)} = 
       {m_t^2/2 m_W^2\over 1 + m_t^2/2m_W^2} \approx 70\% \ . 
\eeqn

\begin{figure}
\begin{center}
\includegraphics[width=0.40\hsize]{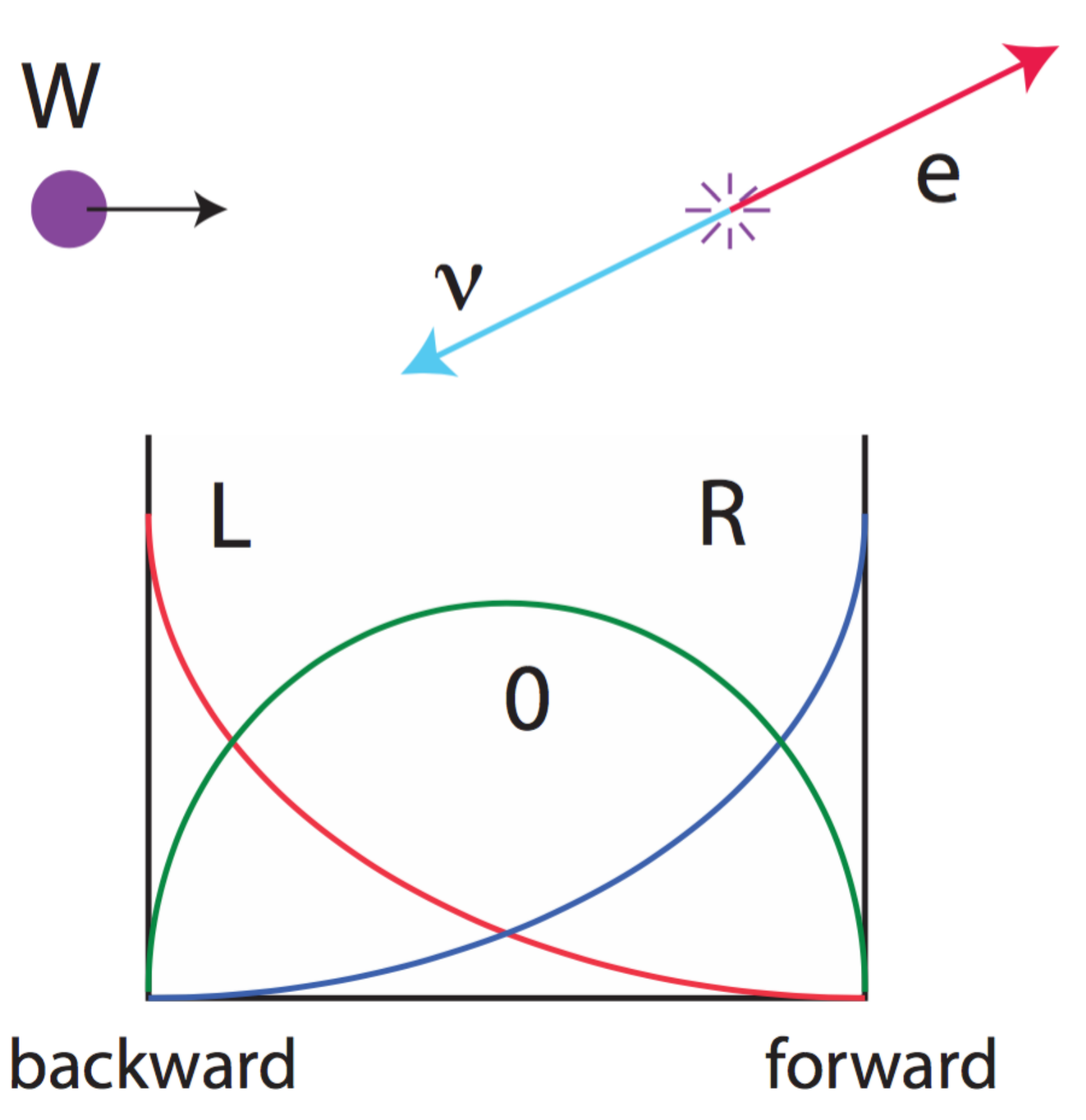}
\end{center}
\caption{Angular distributions of $\cos\theta_*$ in $W$ boson decay
  for each of the three possible polarization states.}
\label{fig:Wdecay}
\end{figure}

\begin{figure}
\begin{center}
\includegraphics[width=0.95\hsize]{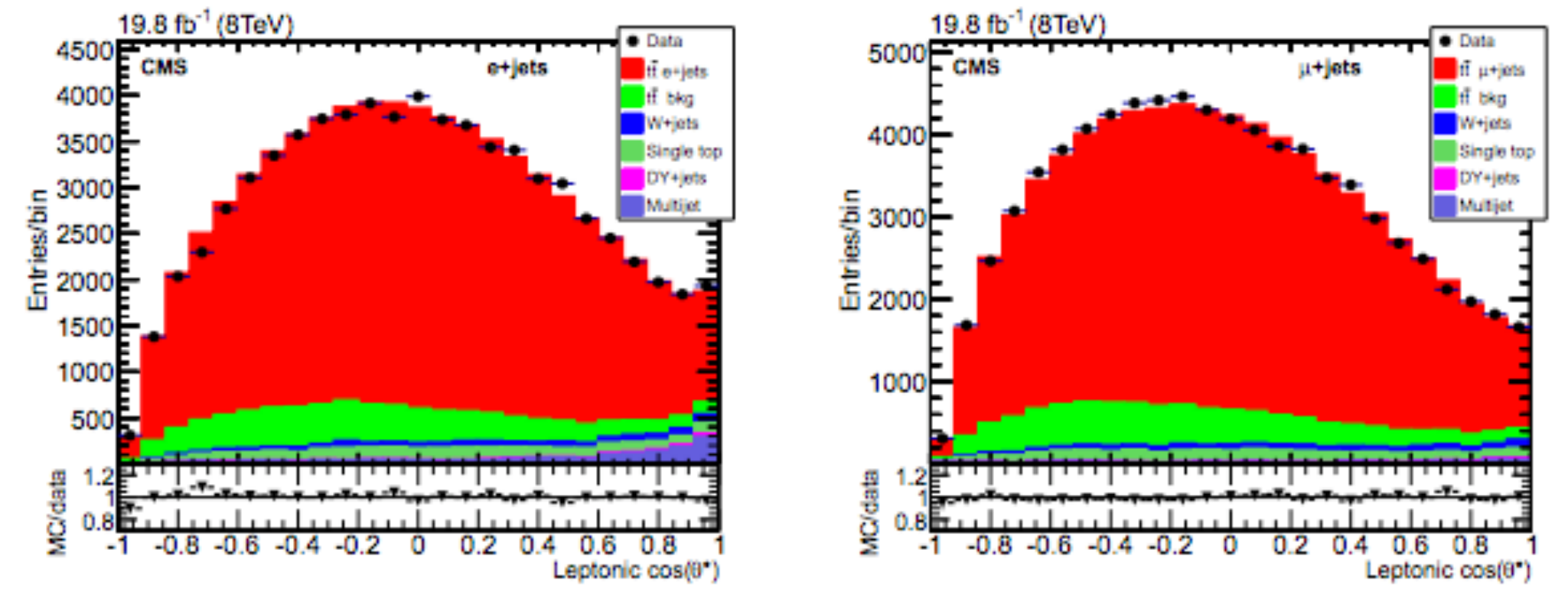}
\end{center}
\caption{CMS measurement of the $\cos\theta_*$ distribution in top
  decay, compared to a simulation that represents the SM expectation~\cite{CMStpol}.}
\label{fig:CMStW}
\end{figure}

The polarization of $W$ bosons in $t$ decay can be measured by
reconstructing full
$pp\to t\bar t\to \ell \nu + 4$~jet events.  Beginning in the 
$t$ rest frame, we boost the leptonically decaying $W$ to rest.   The angular
distribution of the decay lepton in the $W$ frame is then given by for
the three polarization states by 
\beq
    {d\Gamma\over d \cos \theta_*} \sim \cases{  (1+ \cos\theta_*)^2 &
      $+$ 
      \cr \sin^2\theta_*/2    & $0$ \cr  (1 - cos\theta_*)^2 &
      $-$\cr} \ , 
\eeq{Wpolstates}
where $\theta_*$ is the angle between the boost direction and the
lepton direction.    These angular distributions, which are also a
consequence of \VmA, are illustrated in Fig.~\ref{fig:Wdecay}.    The
actual distributions measured in hadron collisions are distorted from
the idealized ones, since leptons with $\cos\theta_*$ near $-1$, which
implies low lab-frame energy, have low acceptance.
Figure~\ref{fig:CMStW} shows the $\cos\theta_*$ distribution measured
by the CMS experiment at the LHC and indicates an excellent agreement with
the SM prediction~\cite{CMStpol}.

An interesting feature of this prediction is the form of the amplitude
\leqn{tbWlong}.   This amplitude is enhanced by a factor $m_t/m_W$,
just as we might have expected from \leqn{epslimit}.  This behavior
can be understood using the GBET.  According to the GBET, we should
find
\beq
   i\M (t\to b W^+_0) \to i\M(t\to b \pi^+) \ . 
\eeqn
The amplitude for emission of a Higgs boson should be proportional to
the top quark Yukawa coupling $y_t$, given by
\beq
     m_t =   {y_t v \over \sqrt{2}} \ . 
\eeqn
So the GBET predicts that the rate for $t$ decay to a longitudinal $W$
should be larger than the rate to a transverse $W$ by the factor
\beq
     {y_t^2\over g^2} =  { 2m_t^2/v^2 \over 4 m_W^2 /v^2}  =
   {  m_t^2\over 2 m_W^2} \ , 
\eeqn
and this is exactly what we found in the explicit calculation.

\subsection{High energy behavior in $\ee\to W^+W^-$}

The next example to study is the high energy behavior of the reaction
\beq
    \ee \to W^+W^-  \ .
\eeqn
I argued earlier that the amplitude for this process cannot show the
enhancement \leqn{epslimit}, at least in the most straightforward way,
since this would lead to an amplitude that violates unitarity.
Indeed, the prediction of the GBET is that
\beq
    \M(\ee\to W^+_0W^-_0) \to \M(\ee\to \pi^+\pi^-) \ .
\eeq{WWlimit}
Using \leqn{anncreate}, the high-energy limit of $SU(2)\times U(1)$,
and  the quantum
numbers of the Higgs field  $(I,Y) = (\half, \half)$, we can readily
work out that the right-hand side of \leqn{WWlimit} is, for an
$e^-_Re^+_L$ initial state,
\beq
     i\M =  - i (2E) \sqrt{2} \ \eps_+ \cdot (k_- - k_+) \cdot
     {e^2\over 2 c_w^2} \ {1\over s} \ , 
\eeq{RLanswer}
and for an $e^-_Le^+_R$ initial state,
\beq
     i\M =  - i (2E) \sqrt{2} \ \eps_-\cdot (k_- - k_+) \cdot \biggl(
     {e^2\over 4 c_w^2} \ {1\over s}  +    {e^2\over 4 c_w^2} \
     {1\over s}  \biggr) \ , 
\eeq{LRanswer}
where $k_-$ and $k_+$ are the final-state momenta.   So it must be
that the expression we guessed in \leqn{firsteeWW} is either incorrect
or is cancelled by other factors.

\begin{figure}
\begin{center}
\includegraphics[width=0.60\hsize]{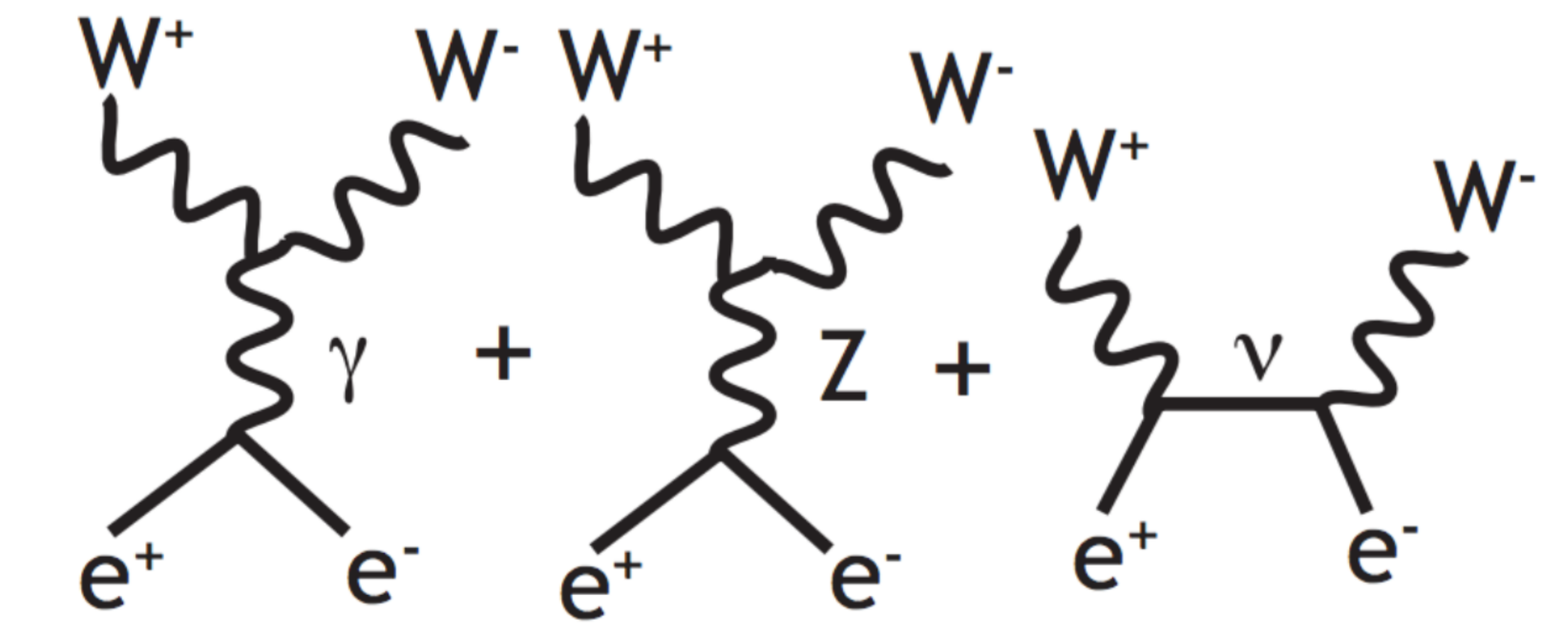}
\end{center}
\caption{Feynman diagrams for the process  $\ee\to W^+W^-$.}
\label{fig:eeWW}
\end{figure}

In the SM, the complete tree level amplitude for $\ee\to W^+W^-$ is given by a
sum of three diagrams, shown in Fig.~\ref{fig:eeWW}.  It will be
instructive to work out the sum of diagrams in a careful way.   I will
do this first for the initial state $e^-_Re^+_L$, for which the
neutrino diagram does not appear.

The full matrix element involves the Yang-Mills vertex for the
$WW\gamma$ and $WWZ$ interactions.  It is 
\beqa
i\M &=&  (-ie)(ie) 2E \sqrt{2} \ \eps_{+\mu}  \biggl[ {-i \over s} + 
    {-s_w^2\over s_w c_w}{c_w\over s_w} { -i \over s-m_Z^2}\biggr]
     \CR
& &\hskip -0.2in \cdot \biggl[ \eps^*(-) \eps^*(+) (k_- - k_+)^\mu +
(-q-k_-)\eps^*(+) \eps^{*\mu}(-) + (k_+ +q) \eps^*(-) \eps^{*\mu}(+)
\biggr] \ ,\CR
\eeqan
where $q = k_-+k_+$  and, in the second line, $\eps^*(-)$ and $\eps^*(+)$
are the $W$ polarizations.   To evaluate the high-energy limit for
longitudinally polarized $W$ bosons, send
\beq
      \eps^*(-) \to {k_-\over m_W} \qquad  \eps^*(+) \to {k_+\over
        m_W} \ .
\eeqn
Then the second term in brackets becomes
\beqa
   & &   {1\over m_W^2} \biggl[ k_-k_+ (k_--k_+)^\mu  - 2 k_-k_+
   k^\mu_- + 2k_+ k_- k^\mu_+ \biggr]  \CR
  & & \hskip 0.2in =   -{ k_-k_+\over m_W^2}
 (k_- - k_+)^\mu =  - {s - 2m_W^2\over 2m_W^2} (k_--k_+)^\mu \ . 
\eeqan
This expression has the enhancement \leqn{WWenhancement}.   However, 
there is a nice cancellation in the first term in brackets,
\beq
    \biggl[ {-i\over s} - {-i\over s- m_Z^2} \biggr] =    {i\ 
      m_Z^2\over s(s-m_Z^2)} \ . 
\eeqn
Assembling the pieces and using   $m_W^2 = m_Z^2 c_w^2$, we find
\beq
i\M =  i e^2 \ 2E \sqrt{2} \ \eps_{+\mu} (k_--k_+)^\mu \biggl(  -{ s -
2m_W^2\over 2c_w^2 s (s-m_Z^2) } \biggr)\ ,
\eeqn
which indeed agrees with \leqn{RLanswer} in the high energy limit.

For the $e^-_Le^+_R$ case, the $\gamma$ and $Z$ diagrams do not
cancel, and so the neutrino diagram is needed.  The first two diagrams
contribute
\beqa
i\M &=&  (-ie)(ie) 2E \sqrt{2} \ \eps_{+\mu}  \biggl[ {-i \over s} + 
    {(1/2-s_w^2)\over s_w c_w}{c_w\over s_w} { -i \over s-m_Z^2}\biggr]
     \CR
& &\hskip -0.2in \cdot \biggl[ \eps^*(-) \eps^*(+) (k_- - k_+)^\mu +
(-q-k_-)\eps^*(+) \eps^{*\mu}(-) + (k_+ +q) \eps^*(-) \eps^{*\mu}(+)
\biggr] \ ,\CR
\eeqan
After the reductions just described, there is a term in the
high-energy behavior that does not cancel,
\beqa
i\M& = & i e^2 \ 2E \sqrt{2} \ \eps_{-\mu} (k_--k_+)^\mu\biggl[ {1\over
  2 s_W^2 s} \biggr] \biggl(  -{ s \over 2m_W^2} \biggr) \CR
& = &{ i e^2\over 4s_w^2}  \ 2E \sqrt{2} \ \eps_{-\mu} (k_--k_+)^\mu  {1\over
 m_W^2} \ .
\eeqa{firsttwo}
We must add to this the neutrino diagram, which contributes
\beq
  i\M = (i{g\over \sqrt{2}})^2 \   v_R(\bar p)^\dagger \, \bar\sigma\cdot
  \eps^*(+)\, { i \sigma\cdot(p - k_-)\over (p-k_-)^2} \,
  \bar\sigma\cdot \eps^*(-) \ u_L(p) \ . 
\eeqn
Substituting $\eps^*(-) \to k_-/m_W$, the second half of this formula becomes
\beq
 {i \sigma\cdot (p-k_-)\over (p-k_-)^2} \, \bar \sigma \cdot {k_-\over
   m_W} u (p) \ . 
\eeqn
Since $\bar\sigma\cdot p\  u_L(p) = 0 $, this can be written
\beq
  {i \sigma\cdot (p-k_-)\over (p-k_-)^2} \, \bar \sigma \cdot {(k_- -p) \over
   m_W} u (p)  = - {i \over m_W^2} u(p) \ . 
\eeqn
Sending $\eps^*(+)\to k_+/m_W = ((p + \bar p)/2 + (k_+ - k_-)/2)/m_W$
and 
using $ (\bar \sigma \cdot p )u_L = v_R^\dagger (\bar \sigma\cdot \bar
p) = 0 $, we finally find
\beq
i\M = -  { i e^2\over 2s_w^2}  \ 2E \sqrt{2} \ \eps_{-\mu}\half  (k_--k_+)^\mu  {1\over
 m_W^2} \ ,
\eeqn
and this indeed cancels the high-energy behavior \leqn{firsttwo} from
the $\gamma$ and $Z$ diagrams. To fully verify \leqn{LRanswer}, we
would need to carry out this calculation more exactly to pick up all
subleading terms at high energy.  It does work out
correctly, as was first shown by Alles, Boyer, and Buras~\cite{Buras}.

\begin{figure}
\begin{center}
\includegraphics[width=0.80\hsize]{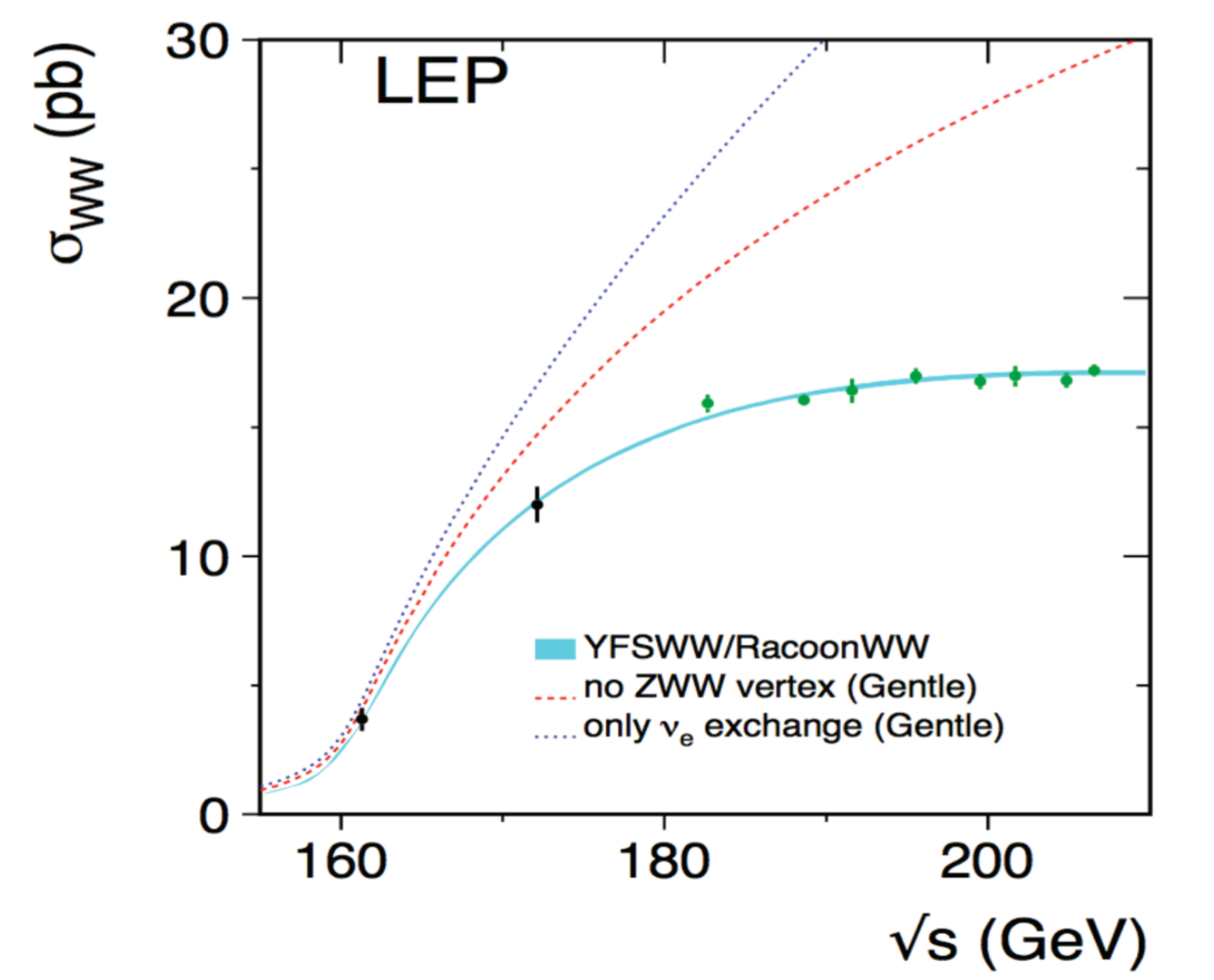}
\end{center}
\caption{Measurement of $\sigma(\ee\to W^+W^-)$ from the four LEP
  experimenta, from \cite{finalLEP}. }
\label{fig:eeWWdata}
\end{figure}

The cross section for $\ee\to W^+W^-$ was measured by the LEP
experiments.  The result is shown in
Fig.~\ref{fig:eeWWdata}~\cite{finalLEP}.   The lowest, solid line is
the prediction of the SM, including one-loop radiative corrections.
It is in excellent agreement with the measurements.   The upper curves
show the effect of omitting, first, the $Z$ diagram and, second, both
the $\gamma$ and $Z$ diagrams.  Apparently, the cancellation I have
demonstrated here is important not only at very high energy but even
in the qualitative behavior of the cross section quite close to threshold.

\subsection{Parametrizing corrections to the Yang-Mills vertex}

The cancellation described in the previous section clearly requires
the precise structure of the Yang-Mills vertex that couples three
vector bosons.  Before the LEP measurements, when the gauge boson nature of
the $W$ and $Z$ was less clear, theorists suggested that the
$WW\gamma$ and $WWZ$ vertices might be modified form the Yang-Mills
form, and that such modifications could be tested by measurements of
$W$ reactions at high energy.  

The most general  Lorentz-invariant, $CP$ conserving $WW\gamma$ vertex
in which the photon couples to a conserved current 
has the form~\cite{Hagiwara}
\beqa
\Delta\L &=& e \biggl[ i g_{1A} A_\mu (W^-\nu W^{+\mu\nu} - W^+_\nu
W^{-\mu\nu})  + i \kappa_A  A_{\mu\nu}  W^-\mu W^+_\nu \CR
& & \hskip 0.7in  + i \lambda_A {1\over m_W^2} W^-_{\lambda\mu}
W^{+\mu\nu} A_\nu{}^\lambda \biggr] \ .
\eeqan
In this formula, for each vector field, $V_{\mu\nu} = (\del_\mu V_\nu
- \del_\nu V_\mu)$.  We can write a similar generalization of the SM
$WWZ$ vertex, with parameters $g_{1Z}$, $\kappa_Z$, $\lambda_Z$ and
overall coupling $e c_w/s_w$.   The
choice
\beq
     g_{1\gamma} = g_{1Z} = \kappa_A = \kappa_Z = 1 \qquad \lambda_A =
     \lambda_Z = 0 
\eeq{SMgk}
gives the SM coupling.  If we relax the assumption of CP conservation,
several more terms can be added. 

It was quickly realized that any changes to the SM vertex produce
extra contributions to the $W$ production amplitudes that are enhanced
by the factor $s/m_W^2$.   In view of the discussion earlier in this
section, this is no surprise.   If the  additional terms violate the
gauge invariance of the theory, the GBET will not be valid, and the
cancellations it requires will not need to occur.  However, this idea
would seem to be already excluded by the strong evidence from the
precision electroweak measurements that the $W$ and $Z$ are the vector
bosons of a gauge theory.

Still, there is a way to modify the $WW\gamma$ and $WWZ$ vertices in a
way that is consistent with gauge invariance.   It is certainly
possible that there exist new heavy particles that couple to the gauge
bosons of the SM.  The quantum effects of these particles can be
described as a modification of 
the SM Lagrangian by the addition of new gauge-invariant operators.
This approach to the parametrizatoin f new physics effects has become
known as Effective Field Theory (EFT). 
The SM already contains the most general $SU(2)\times U(1)$-invariant
operators up to dimension 4, but new physics at high energy can add
higher-dimension operators, beginning with dimension 6. 

There are many dimension 6 operators that can be added to the SM.
Even for 1 generation of fermions, there are 84 independent dimension
6 operators, of which 59 are baryon-number and
CP-conserving~\cite{Warsaw}.
The theory of these operators has a complexity that I do
not have room to explain here.   It is possible to make many different
choices for the basis of these operators, using the fact that 
combinations of these operators are set to zero by the SM equations of
motion.  The theory of EFT modifications of the SM is reviewed
in \cite{dimsixreview} and, in rather more detail, in
\cite{Murayama}.  I will give only a simple example here.

Consider, then, adding to the SM the dimension-6 operators
\beq
  \Delta\L =  {c_T\over 2 v^2} \Phi^\mu \Phi_\mu  + {4gg'\over m_W^2}
  \Phi^a W^a_{\mu\nu} B^{\mu\nu} + {g^3 c_{3W}\over m_W^2} \eps^{abc}
  W^a_{\mu\nu} W^{b\nu}{}_\rho W^{c \rho\mu} \ , 
\eeq{firstdimsix}
where, in this formula, $W^a_{\mu\nu}$ and $B_{\mu\nu}$ are the
$SU(2)$ and $U(1)$ field strengths and $\Phi_\mu$, $\Phi^a$ are
bilinears in the Higgs field, 
\beq
  \Phi_\mu = \varphi^\dagger D_\mu \varphi - (D_\mu \varphi)^\dagger
  \varphi \qquad    \Phi^a =  \varphi^\dagger {\sigma^a\over 2}
  \varphi \ .
\eeqn
It can be shown that these shift the parameters of the $WW\gamma$ and
$WWZ$ couplings to
\beqa
     g_{1Z} &=&   1 + \biggl[ {c_T \over 2(c_w^2 - s_w^2) }-  {8s_w^2
         c_{WB}\over c_w^2 (c_w^2 - s_w^2)} \biggr] \CR
     \kappa_A &=&  1 - 4 c_{WB}\CR
    \lambda_A &=& - 6 g^2 c_{3W}
\eeqan
The parameter $g_{1A} = 1$ is not shifted; this is the electric charge
of the $W$ boson.   The remaining two parameters obey
\beq
     \kappa_Z = g_{1Z} - {s_w^2\over c_w^2} (\kappa_A -1) \qquad
     \lambda_Z = \lambda_A \ .
\eeq{fixedrels}
It can be shown that the relations \leqn{fixedrels} are maintained for
any set of dimension-6 perturbations of the SM.   They may be modified
by dimension-8 operators.

Dimension-6 operators also contribute to the $S$ and $T$ parameters
discussed at the end of the previous section.  From the perturbation
\leqn{firstdimsix}, 
\beqa 
   \alpha  S &=&   32  s_w^2 c_{WB}\CR
   \alpha T &=&  c_T
\eeqan 

Given that EFT is based on gauge-invariant Lagrangian, this formalism
for pa\-ra\-me\-triz\-ing new physics can be worked out explicitly in great
detail.  QCD and electroweak radiative corrections can be included.
The higher-dimension operators in the EFT must of course be
renormalized according to some scheme, and the detailed formulae will
depend on the scheme.    

A
dimension-6 operator has a coefficient with the units of
(GeV)$^{-2}$.  Thus, the effects of such operators are suppressed by
one factor of  $s/M^2$, where $M$ is then mass scale of new
particles.  Contributions from dimension-8 operators suppressed by
$(s/M^2)^2$, and similarly for operators of still higher dimension.
So, an analysis that puts constraints on dimension-6 operators,
ignoring the effects of dimension-8 operators  is
properly valid only when $s/M^2 \ll 1$.

As a corollary to this point, I call your attention to a Devil's
bargain that arises frequently in tests of the structure of $W$ and
$Z$ vertices at hadron colliders.    In $pp$ collisions, the parton
center of mass energy $\hat s$ varies over a wide range.   There is
always a region of phase space where $\hat s$ becomes extremely
large.   This is the region that has the greatest sensitivity to
higher-dimension operators.  It is tempting to apply event selections
that emphasize this region to obtain the strongest possible limits.

However, this is exactly the region where operators of dimension 8 and
higher might also be important.  In many models, these give negative
contribution.  Then  a parametrization that uses only  dimension-6
operators leads to limits on their coefficients that are stronger than
the limits that would be obtained in a more complete theory.

The question of how to interpret limits on dimension-6 EFT
coefficients is now hotly debated in the literature.   My personal
position is on one extreme, that only analyses in which $\hat s/M^2
\ll 1$ for all events included in the analysis  should be trusted.  
The authors of \cite{FalkowskiWW} advocate
for a much more aggressive approach.   Experimenters who quote such
limits should study this issue carefully.

On the other hand, the SM itself makes precise predictions in all
regions of $\hat s$. Your first priority should be to discover a
deviation from these predictions.   If you are able to demonstrate a
substantial deviation from the SM predictions in any region of phase
space, we can all have 
fun quarreling about the interpretation of this result.

\subsection{$W$ parton distributions}

As a final topic in this section, I will discuss a situation in which
the GBET might be expected to apply, but it does not.  This is
involves processes in which a $W$ boson is radiated from a quark or
lepton with
small transverse momentum relative to the fermion direction.   In  
QCD, the collinear radiation of gluons from initial quarks is
essential is creating the observed quark and gluon parton
distributions.  In Section~4.2, we saw that collinear radiation of
photons from initial electrons and positrons is also an important
effect that makes qualitative changes in the $Z$ resonance line shape.
In this section, I will present the analogous theory for collinear $W$
boson emission~\cite{Dawson}.    I will carry out the analysis for
quark initial states, but the same theory applies to electron and
positron initial states.

\begin{figure}
\begin{center}
\includegraphics[width=0.30\hsize]{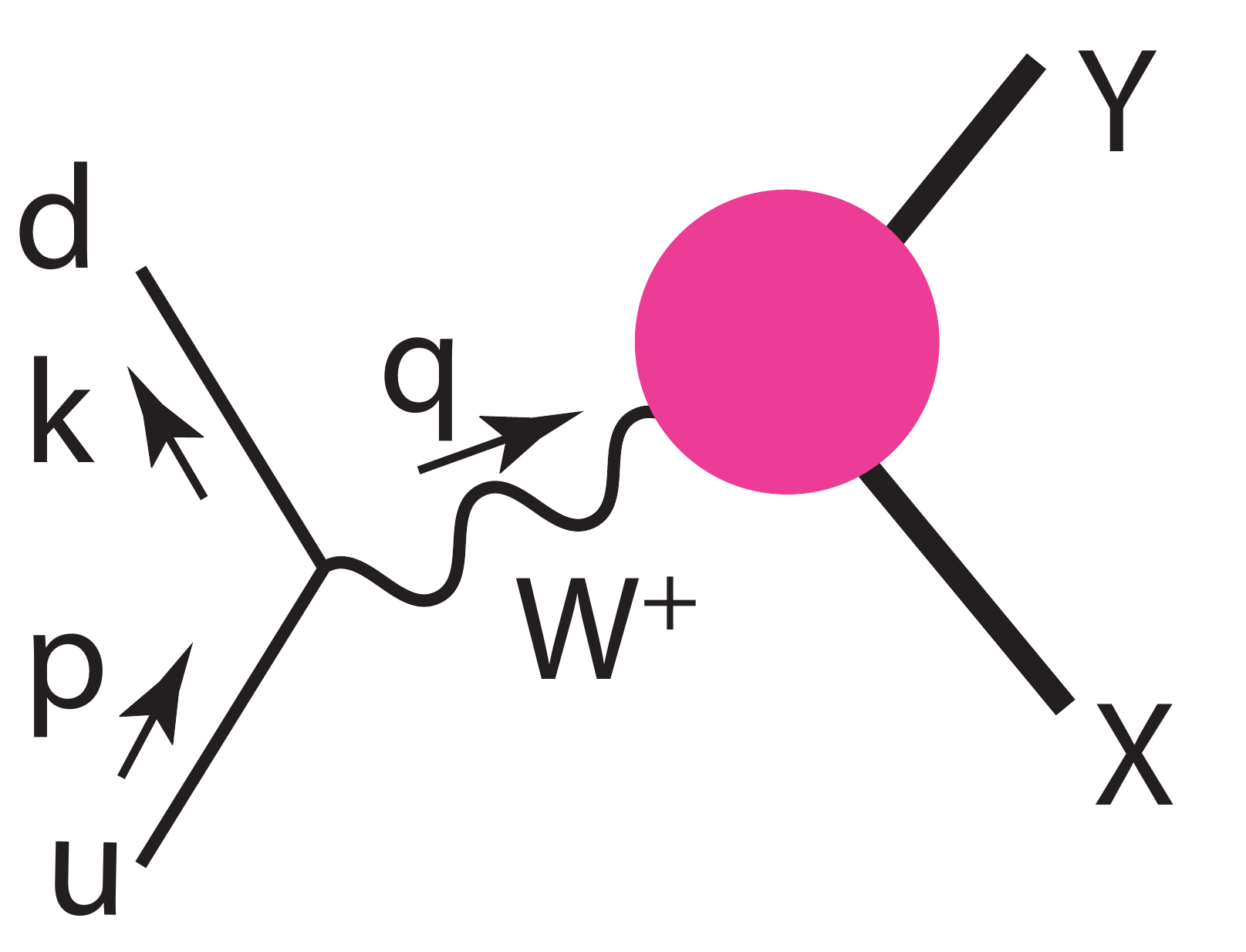}
\end{center}
\caption{Kinematics of a process in which a $W$ is emitted collinearly
  from a quark and then initiates a large-momentum-transfer reaction.}
\label{fig:Wexchange}
\end{figure}

For definiteness, consider the following setup:  An  initial $u$
quark, with momentum $p$, emits an almost collinear $W^+$ boson, with
momentum $q$, 
\beq
     u(p) \to d(k) + W^+(q)  \ . 
\eeqn
The $W$ boson must be off-shell. This emission will be part of a
process shown in Fig.~\ref{fig:Wexchange}, in which the virutal $W$
collides with a parton from the other proton to initiate a
hard-scattering reaction.   An important class of processes of this
type is $WW$ scattering, including the reaction $W^+W^-\to h$ that we
will discuss in Section~6.2.

For $W$ reactions that involve the Higgs boson, it will be important
to have $W$ bosons with longitudinal polarization.  According to the GBET, a longitudinally
polarized $W$ boson should  have a coupling equal to that of the
corresponding Goldstone boson $\pi^+$ from the Higgs sector.  Then the
study of high energy $W$ boson reactions allows us to directly measure
the strength of Higgs boson interactions.   However, it is not clear
that it is possible to radiate  longitudinally polarized $W$ bosons
from initial quarks.  A 
$\pi^+$ couples to a light fermion with its  Higgs Yukawa coupling,
that is, negligibly, the the radiation of longitudinally polarized $W$
bosons would seem to be forbidden by the GBET. 

To understand the correct story, we must compute the $u\to W d$ emission
amplitude explicitly. In this calculation, I will take the $W$ boson to
be emitted approximately collinearly with the $u$ quark.  The analysis
is very similar to calcuation of the Altarelli-Parisi splitting
functions that you will find, for example, in Chapter 17 of
\cite{PeskinSchr}.    I will assume that the $W$ has  $p_T \sim m_W
\ll  p_\parallel$.

First, I write the momentum vectors for the
quarks, taking the $u$ quark to move in the $\hat 3$ direction and the
$d$ quark to carry away an energy fraction $(1-x)$ and to have a small
transverse momentum,
\beqa
    p &=& (E,0,0,E)\CR
    k &=& ((1-z)E, -p_T, 0, (1-z)E - {p_T^2\over 2 (1-z) E} )  \ . 
\eeqan
The momentum $k$ is on-shell to order $p_T^2$.  
 The $W$ momentum vector is then determined by momentum conservation
\beq 
    q =  (zE, p_T, 0,  zE + { p_T^2\over 2 (1-z) E} ) \ .
\eeqn
The denominator of the $W$ propagator is then
\beq
     q^2 - m_W^2 =  - p_T^2 - {z\over (1-z)} p_T^2 - m_W^2 =  - \bigl(
     {p_T^2\over 1-z} + m_W^2\bigr)  \ .
\eeq{Wdenom}

Next, we compute the matrix elements for $W$ emission
\beq
    i\M =   ig \   u^\dagger_L(k) \ \bar\sigma\cdot \eps_W^* \ u_L(p)
  \eeqn
to first order in $(p_T,m_W)$.   The explicit form of the spinors is
\beq
       u_L(k) = \sqrt{2 (1-z)E} \pmatrix{ p_T/2(1-z)\cr 1 \cr} \qquad
       u_L(p) = \sqrt{2E} \pmatrix{0\cr 1\cr} \ .
\eeqn
The $W$ polarization vectors are
\beq
     \eps_\pm^{*\mu} = (0,1, \mp i , -p_T/zE)^\mu /\sqrt{2} 
\eeqn
for the transverse polarizations, and 
\beq
    \eps_0^{*\mu} =  ( q, p_T, 0, zE)^\mu / m_W 
\eeqn
for the longitudinal polarization state.   In this formula
\beq
     q =  [ (zE)^2 - m_W^2]^{1/2} =  zE - {m_W^2\over 2zE} 
\eeqn
Then
\beqa
     \bar\sigma \cdot \eps_+^* = {1\over \sqrt{2}} \pmatrix{ - p_T/zE
       & 0 \cr 2 & p_T/zE \cr } \CR
      \bar\sigma \cdot \eps_-^* = {1\over \sqrt{2}} \pmatrix{ - p_T/zE
       & 2 \cr 0& p_T/zE \cr } \CR
     \bar\sigma \cdot \eps_0^* = {1\over m_W} \pmatrix{ q+zE
       &p_T \cr p_T & q-zE \cr } 
\eeqan
With these ingredients, it is straightfoward to work out the matrix
elements for the three $W$ polarization states,
\beq
   i\M (u\to dW^+) =   ig \cdot \cases{    \sqrt{1-z}\ p_T / z    &     $+$ \cr
     \sqrt{1-z}\ p_T / z(1-z)     &     $-$ \cr
 - \sqrt{1-z} \ m_W /\sqrt{2}  z    &     $0$ \cr}    \ .
\eeq{Wemiss}

We can convert these expressions to cross sections for complete
$W$-induced processes.   The cross section for a
process
$u X \to d Y$, in the approximation in which the $W$ is almost on
shell, is given by 
\beqa
    \sigma &=& {1\over 2s} \int {d^3k\over (2\pi)^3 2k} \int d\Pi_Y \
    (2\pi)^4 \delta^{(4)}(p + p_X - k - p_Y) \CR
& & \hskip 0.4in   \biggl|  \M(u\to dW^+) {1\over q^2 - m_W^2}
\M(W^+X\to Y) \biggr|^2 
\eeqa{firstWcs}
In the collinear kinematics, with $\hat s = zs$
\beq
    {1\over 2s} \int {d^3k\over (2\pi)^3 2k}  =  {1\over 2 \hat s/z} 
  \int {dz E d^2p_T \over 16 \pi^3 E (1-z)} = {1\over 2\hat s} \int
  {dz dp_T^2 \pi\over 16\pi^3}  {z\over (1-z)} 
\eeqn
Then, also using \leqn{Wdenom},  \leqn{firstWcs} simplifies to 
\beqa
  \sigma &=& \int dz \int {dp_T^2\over (4\pi)^2} {z\over (1-z)}
  \bigl|\M(u\to dW^+)\bigr|^2 {1\over p^2_T/(1-z) + m_W^2)^2}\CR
 && \hskip 0.3in  \cdot {1\over 2\hat s} \int d\Pi_Y (2\pi)^4 
\delta^{(4)}(q+ p_X - p_Y) \bigl|\M(W^+ X \to Y)\bigr|^2 
\eeqa{nextWcs}

The last line of  \leqn{nextWcs} is $\sigma(W^+(q) + X \to Y)$.   Then
\leqn{nextWcs} has the form of a parton model cross section
\beq
    \sigma(uX\to dY)  = \int dz   f_{W\leftarrow u}(z) \
    \sigma(W^+X\to Y)
\eeqn
where $ f_{W\leftarrow u}(z)$ is the  parton distribution for a $W$
  boson in the $u$ quark,
\beq
   f_{W\leftarrow u}(z)  = \int {dp_T^2\over (4\pi)^2} {z\over (1-z)}
   { (1-z)^2 \over (p_T^2 + (1-z) m_W^2)^2}   \bigl|\M(u\to
   dW^+)\bigr|^2 \ .
\eeqn
We can evaluate this parton distribution for each $W$ polarization
state by  using the formula \leqn{Wemiss}.  The result is 
\beqa
     f_{W-}(z) = {\alpha_2\over 4\pi} \int  { dp_T^2 \ p_T^2 \over 
   (p_T^2 + (1-z) m_W^2)^2 }\  {1\over z} \CR
     f_{W+}(z) = {\alpha_2\over 4\pi} \int  { dp_T^2 \ p_T^2 \over 
   (p_T^2 + (1-z) m_W^2)^2 }\  {(1-z)^2 \over z} \CR
     f_{W0}(z) = {\alpha_2\over 8\pi} \int  { dp_T^2 \ m_W^2 \over 
   (p_T^2 + (1-z) m_W^2)^2 }\  {(1-z)^2\over z} 
\eeqa{Wparton}

For the transverse polarizations, we find a resut very similar to the 
Altarelli-Parisi splitting function for collinear gluon emission,
\beq
   f_{WT}(z) =   {\alpha_w\over 4\pi} { 1 + (1-z)^2\over z} \cdot
   \log{Q^2\over m_W^2} \ , 
\eeqn
where $Q^2$ is the upper limit of the $p_T^2$ integral, which is set
by the momentum transfer in the hard reaction.

For the longitudinal $W$ polarization, the story is different.  The
integral over $p_T$ is convergent, so that the $p_T$ is restricted to
the region $p_T\sim m_W$. In this regime, as we see explicitly, 
 longitudinal $W$ bosons can be produced with coupling strength $g$.
Apparently, in this process, the error term in the GBET is actually ${\cal
  O}(m_W/p_T)$, which is consistent with \leqn{theGBET} but, still,
  larger than we might expect.   The reduction of the longitudinal $W$
  boson to a Higgs boson then is not accurate in the region $p_T\sim 
  m_W$, though it does apply---and cuts off the amplitude---when $p_T
  \gg m_W$.

When we perform the convergent integral over $p_T$, we find that 
 the parton distribution for $W_0$ is substantial~\cite{Dawson},
\beq
      f_{W0}(z) = {\alpha_w\over 8\pi} \ {1-z\over z} \ .
\eeqn
Then the proton does contain longitudinal $W$ bosons, which can induce 
Higgs sector reactions when this proton collides with another proton at
high energy.  The collinear longitudinal  $W$ bosons have $p_T\sim m_W$ but not
higher, a kinematic feature that can be used to suppress backgrounds
from reactions involving transversely-polarized $W$ bosons.

\section{The Standard Model theory of Higgs boson decays}

There remains one heavy particle of the SM that we have not yet
discussed, the Higgs boson.  The Higgs boson has a central role in the
structure of the weak interactions.   Its field is the agent that
breaks the $SU(2)\times U(1)$ symmetry and generates the masses of all
quarks, leptons, and vector bosons.   This at the same time forms a
unified picture of the electroweak interactions as we have studied
them so far and also points to new mysteries whose explanations are
still to be found.

The best way to enter a discussion of the Higgs boson is to understand
thoroughly the predictions for the properties of this particle given
by the SM.   The Higgs sector involves one more parameter of the SM
beyond those we have discussed already, the Higgs field self-coupling
$\lambda$.  However, this coupling is fixed by the measurement of the
Higgs
boson mass.  Thus, the SM makes precise predictions for all of the
Higgs boson cross sections and branching fractions.   These
predictions provide a starting point for any discussion of the
properties of the Higgs boson in model that generalize the SM. An
excellent reference on the theory of the Higgs boson in the Standard
Model is \cite{HHGuide}.  The best current calculations of the Higgs
boson properties are compiled in \cite{HiggsHandbook}.

\subsection{Decay modes of the Higgs boson}

\begin{figure}
\begin{center}
\includegraphics[width=0.70\hsize]{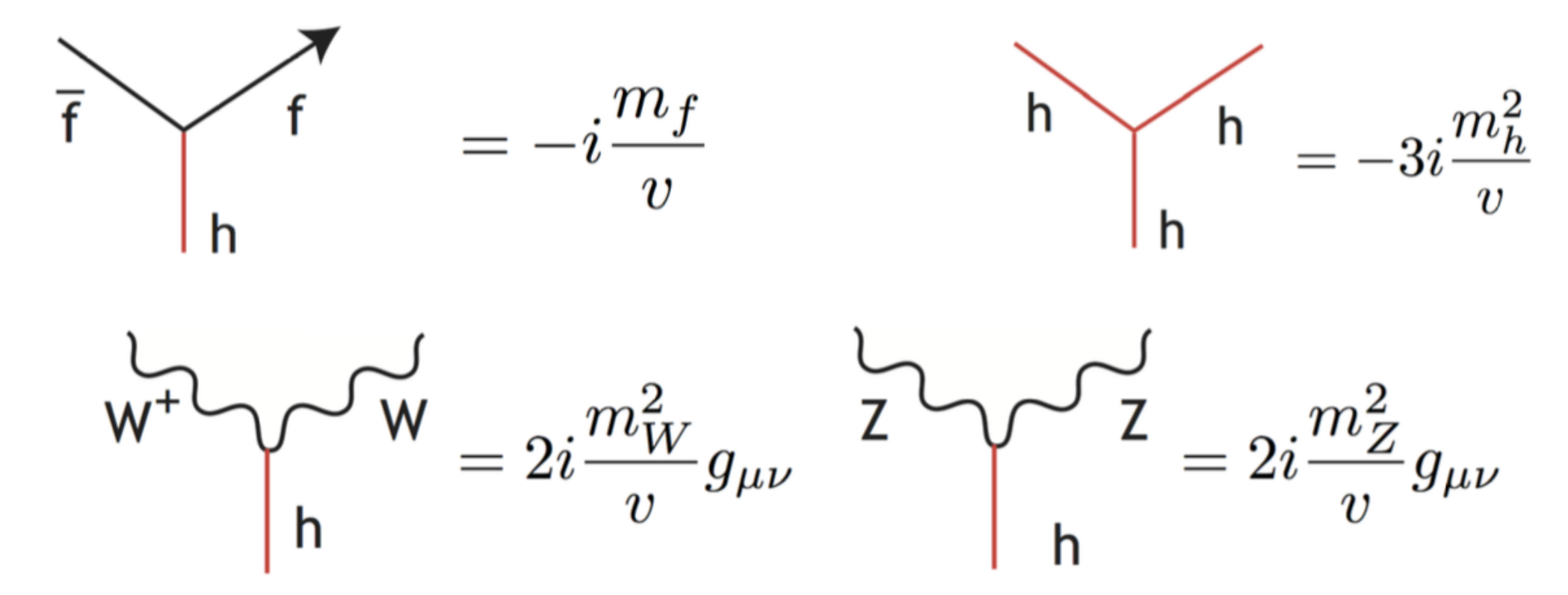}
\end{center}
\caption{Feynman rules for couplings of the Higgs boson.}
\label{fig:hcouple}
\end{figure}

The basic elements of the SM description of the Higgs boson are
extremely simple.  A general configuration of the Higgs field can be 
written in the form of an $SU(2)$ gauge transformation acting on a 
simple scalar field
\beq
\varphi(x) =   \exp[ -i\alpha^a(x) \sigma^a/2] \pmatrix{0\cr (v +
  h(x))/\sqrt{2}\cr}
\ .
\eeqn
We can remove the prefactor by a choice of gauge.  Then the Higgs
field reduces to a vacuum expectation value $v$ and the dynamical
scalar field $h(x)$.   The values of $m_W$ and $g$ give
\beq
           v = 246~\mbox{GeV} \ . 
\eeqn
The vertices of $h(x)$ are given by shifting $v$ everywhere it appears
in the SM
\beq
    v \to v + h(x) \ . 
\eeq{hrescale}
This gives rise to the Feynman rules shown in Fig.~\ref{fig:hcouple}.
Within the SM, there is no freedom to change these vertices.

The couplings in Fig.~\ref{fig:hcouple} imply that a heavy Higgs boson
would  decay dominantly into pairs of the other heavy particles of the
SM,
\beq
    h\to W^+W^- \ , \ h\to ZZ\ , \  h \to t\bar t
\eeqn
However, it has been found at the LHC that there is no heavy resonance
that decays to these final states.  On the other hand, a narrow
resonance with the properties of the Higgs boson has been found at the
LHC at a mass of 125~GeV.   At this mass value, the otherwise dominant
decay modes of the Higgs boson are kinematically forbidden.  The
actual decay modes of the Higgs are all suppressed in some way, by
factors
\beq
         {m_f^2\over m_W^2} ,\quad   {\alpha_w\over 4\pi} \ , \quad \mbox{or} \quad 
             \bigl( {\alpha_s\over 4\pi} \bigr)^2  \ .
\eeqn
This means that the decay pattern of the Higgs boson will be more
complex that might have been expected, but also that it should be very
rich, with a large number of decay modes accessible to observation.

To describe these decays, I begin with the decays to fermions.   The
matrix element for Higgs decay to a light fermion is
\beq
   i \M(h\to f_R \bar f_R)   = -i {m_f\over v} u^\dagger_R v_R = - i
   {m_f\over v} (2E) \ .
\eeqn
and similarly for decay to $f_L\bar f_L$.   The total decay rate is 
\beq
     \Gamma ( h \to f\bar f) = {1\over 2 m_h} {1\over 8\pi} {m_f^2
       m_h^2\over v^2} \cdot 2 \ ,
\eeqn
or, using $v^2 = 4 m_W^2/ g^2$, 
\beq
   \Gamma(h\to f\bar f) =  {\alpha_w\over 8} \ m_h\ {m_f^2\over m_W^2} \
   .
\eeq{hff}

For final-state leptons, we can immediately evaluate this,
\beq
     \Gamma(h\to \tau^+\tau^-) = 260~\mbox{keV} \qquad
      \Gamma(h\to \mu^+\mu^-) = 9~\mbox{keV}  
\eeqn
for $m_h = 125$~GeV.

For decays to quarks, a few more details must be added.   The quark
mass must be defined by some renormalization convention.   An
appropriate choice that absorbs large logarithms is to set the quark
mass in \leqn{hff} equal to the $\msb$ quark mass evaluated at $Q =
m_h$.  This is related to the quark mass as usually quoted by
\beq
  m_f(m_h) =  m_f(m_f) \biggl[ {\alpha_s(m_h)\over
    \alpha_s(m_f)}\biggr]^{4/b_0} \bigl(1 + {\cal O}(\alpha_s) \bigr) \ ,
\eeqn
where $b_0$ is the first coefficient of the QCD $\beta$ function,
equal to 23/3 for 5 light quark flavors.   This means that the values
of the quark masses appropriate to the calculation of Higgs boson
branching ratios are
\beq
\begin{tabular}{ccccc}
  $m_u$ & $m_d$ &  $m_s$ & $m_c$ &  $m_b$\\
  1.5    &     3     &   60   &   700  &   2800 \\
\end{tabular}
\eeqn
with all values in MeV.  The formula \leqn{hff} must also be
multiplied by the color factor of 3 and a substantial QCD correction
\beq
             3 \cdot   \bigl(1 + {17\over 3\pi} \alpha_s(m_h)+ \cdots
             \bigr) =    3 \cdot 1.24 \ .
\eeqn
Then, for example, 
\beq
     \Gamma(h\to b\bar b) = {\alpha_w m_h\over 8} \bigl( {2.8\over
       m_W}\bigr)^2 \cdot 3 \cdot 1.24 =   2.4~\mbox{MeV} \  .
\eeqn
After we compute the other major Higgs boson decay rates, this will
correspond to a branching fractionn of 58\%.   Then the total width of
the Higgs boson is predicted to be about 4.1~MeV, and the other
fermion
branching fractions should be 
\beq 
\begin{tabular}{cccc}
  $\tau^+\tau^-$ & $c\bar c $ &  $s\bar s$ & $\mu^+\mu^-$ \\
  6.3\%   &  3\%    &  0.03\%   &   0.02\% \\
\end{tabular}
\eeqn
It is somewhat surprising the that the branching ratio for
$\tau^+\tau^-$ is larger than that for $c\bar c$, despite the presence
of the color factor of 3.

For a heavy Higgs boson that can decay to on-shell $W$ and $Z$ bosons,
the decay amplitudes would be 
\beqa
     i\M(h\to W^+W^-) &=&  i {2m_W^2\over v} \eps^*(+) \cdot \eps^*(-)
     \CR
    i\M(h\to ZZ) &=&  i {2m_Z^2\over v} \eps^*(1) \cdot \eps^*(2) \ .
 \eeqan
For a very heavy Higgs boson, there is a further enhancement for the
longitudinal polarization states, 
\beq 
    \eps_0^*(1) \cdot \eps^*_0(2)  \sim  {k_1\cdot k_2\over m_Z^2}
    \sim    {m_h^2 \over 2 m_Z^2} \ . 
\eeqn
This factor is just
\beq
               {  \lambda  \over (g^2 + g^{\prime 2})  } \ .
\eeqn
so the longitudinal $Z$ and $W$ couple to the Higgs boson as
Higgs boson rather than as gauge bosons.  This is in accord with the
GBET.

\begin{figure}
\begin{center}
\includegraphics[width=0.30\hsize]{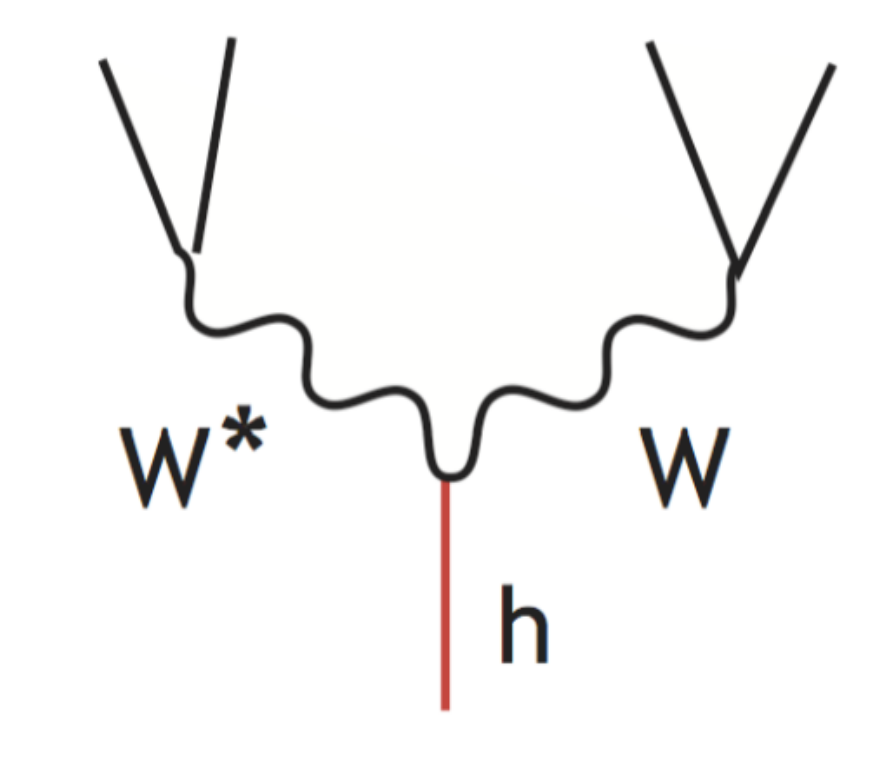}
\end{center}
\caption{Feynman diagram for $h\to WW$ or $h\to ZZ$ decay with the
  vector bosons off-shell.}
\label{fig:WWoff}
\end{figure}

\begin{figure}
\begin{center}
\includegraphics[width=0.70\hsize]{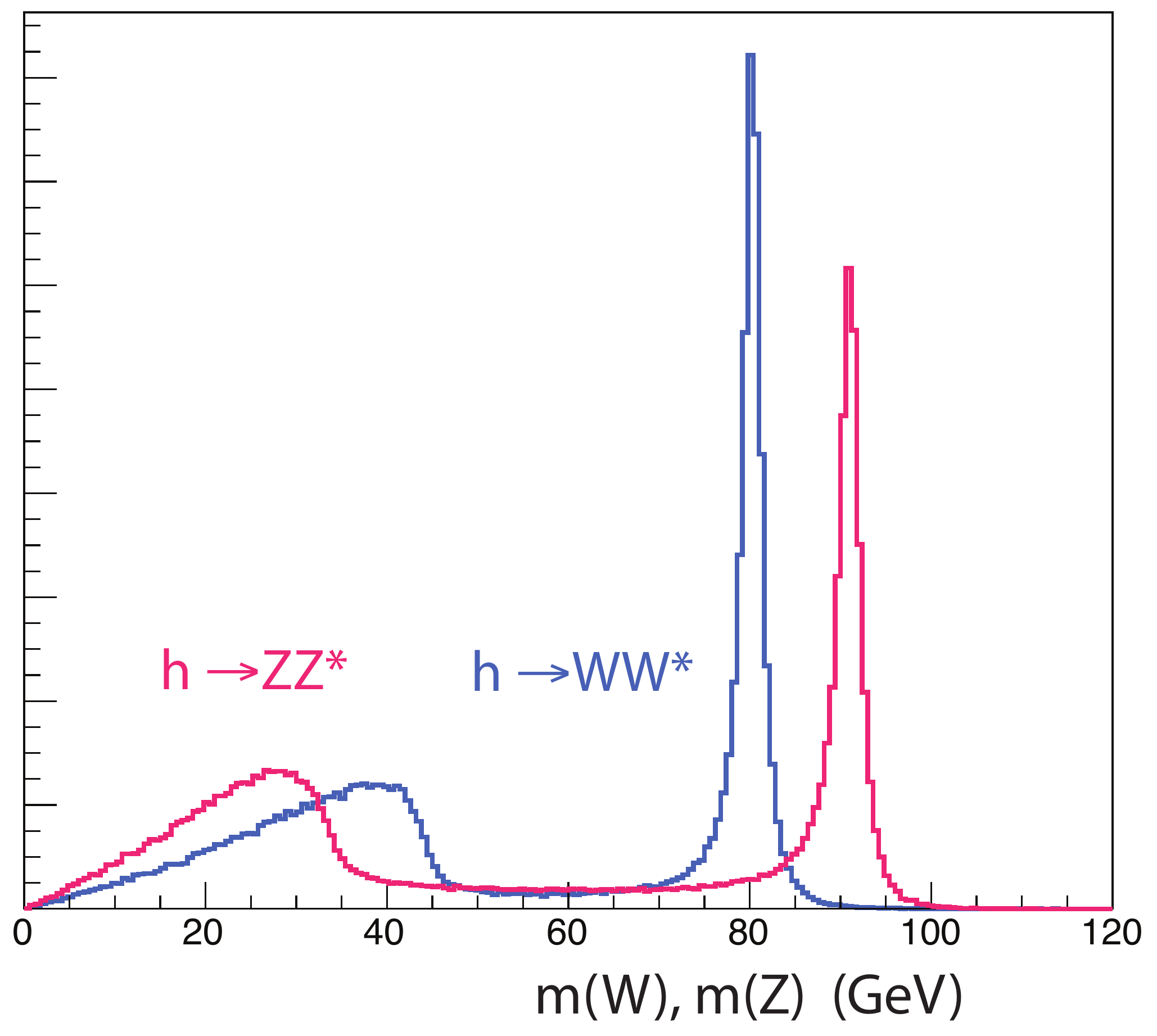}
\end{center}
\caption{Mass distributions of the off-shell $W$ and $Z$ bosons in the
  decay of a 125~GeV Higgs boson.}
\label{fig:WWZZmass}
\end{figure}

For the actual situation of a 125~GeV Higgs boson, one or both of the
$W$ and $Z$ bosons must be off-shell.   Then the decay is best
described as a Higgs decay to 4 fermions, as shown in
Fig.~\ref{fig:WWoff}.
The rate is suppressed by a factor of $\alpha_w$ and by the off-shell
$W$ or $Z$ propagator.   The result is that the rate is competitive
with $b\bar b$ for the $WW$ mode and a factor 10 smaller for $ZZ$.
The SM branching fractions for these off-shell vector boson modes are
\beq
     BR(h\to WW^*) = 22\%  \qquad   BR(h\to ZZ^*) = 2.7\% \  .
\eeqn
The $W$ and $Z$ mass distributions in these decays are shown in
Fig.~\ref{fig:WWZZmass}.

\begin{figure}
\begin{center}
\includegraphics[width=0.90\hsize]{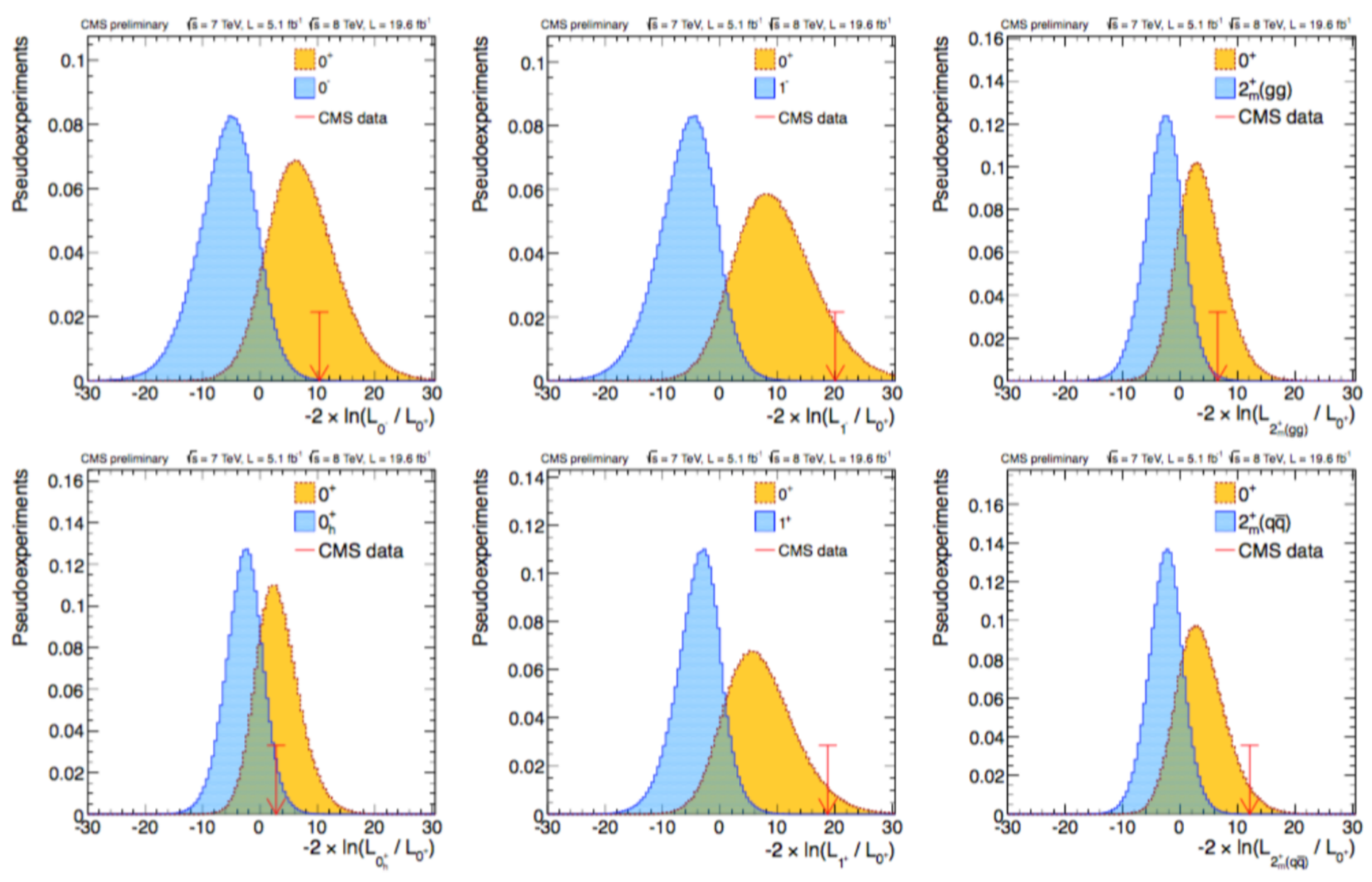}
\end{center}
\caption{Likelihood distributions for tests of the spin and parity of
  the Higgs boson, from \cite{CMSPASZZ}.}
\label{fig:CMSlike}
\end{figure}

The Higgs boson decay to $ZZ^*$ is exceptionally interesting because
it is completely reconstructable in LHC events in which both $Z$s decay to charged
leptons.    The angular distribution of the leptons permits an
analysis of the spin and parity of the Higgs resonance.   In the SM,
where the Higgs boson must have $J^P = 0^+$, the two $Z$ bosons are
predicted to be longitudinally polarized with the two decay planes
parallel. The polarization of the $Z$ can be measured from the decay
angular distribution, as we have discussed for $W$ bosons in
\leqn{Wpolstates}.   This prediction 
contrasts with that for  other possible spin 0 assigments, in
which the Higgs boson couples to $ZZ^*$ through the interactions
\beq
    0^- \ :\  h \eps^{\mu\nu\lambda\sigma} Z_{\mu\nu}
    Z_{\lambda\sigma} \qquad   0^+_h \ : \   h Z_\mu\nu Z^{\mu\nu}   \ .
\eeq{altzero}
For the intereractions in \leqn{altzero}, the $Z$ bosons are
preferentially transversely polarized; also, with   the $0^-$ type
interaction,
 the two decay  planes tend to be orthogonal.    The SM prediction was
 tested even with the relatively small sample of about 15 $Z\to
 4$~lepton events collected by each LHC experiment in run 1 of the LHC.
 Figure~\ref{fig:CMSlike} shows the expectred distributions of the
 likelihood
for tests
 of the predicted SM coupling structure against the  coupling
 structures in \leqn{altzero} and 4 other structures for which the
 resonance has spin 1 or spin 2.  The actual value of the likelihood
 found by CMS experiment  is shown by the arrow.   In all cases, the results
 strongly favor the SM hypothesis~\cite{CMSPASZZ}.

\begin{figure}
\begin{center}
\includegraphics[width=0.70\hsize]{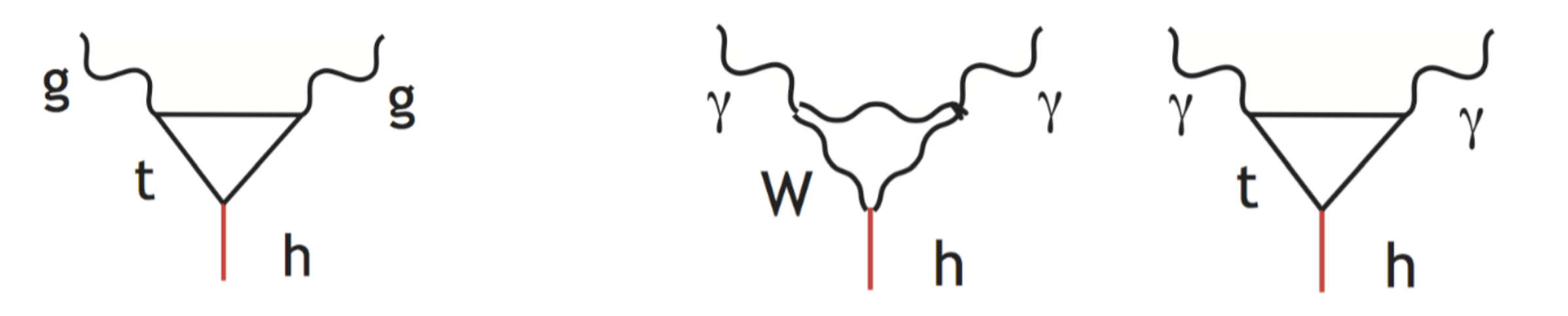}
\end{center}
\caption{Loop diagrams contributing the the $h\to gg$ and $h\to
  \gamma\gamma$ decays.}
\label{fig:toploop}
\end{figure}

Finally, there are loop processes that allow the Higgs boson to decay
to a pair of massless vector bosons, $gg$ or $\gamma\gamma$, or to
$Z\gamma$.    The most straightforward of these to analyze is the
$hgg$ vertex.  This is generated by loop diagrams that involve quarks,
such as the diagram shown on the left in Fig.~\ref{fig:toploop}.

If we compute these loop diagrams, we obtain a local operator that
gives an effective description of the Higgs boson coupling to $gg$.
The lowest-dimension operator that is invariant under the $SU(3)$ gauge
symmetry is 
\beq 
      \Delta \L =    {1\over 4} A h  F_{\mu\nu}^a F^{\mu\nu a} \ , 
\eeq{heffgg}
where $F_{\mu\nu}^a$ is the QCD field strength.   The coefficient $A$
has the dimensions  (GeV)$^{-1}$.  This operator yields the $hgg$
vertex
\beq
      - i A \delta^{ab} (k_1 \cdot k_2 g^{\mu\nu}-
      k_2^\mu k_1^\nu ) \ . 
\eeq{hggeff}

I will compute the coefficient $A$ in a moment, but, first I will
estimate the order of magnitude of the contribution from a quark of
mass $m_q$.   There is a surprise here.   This contribution is
proportional to the Higgs Yukawa coupling, so it must be of the form
\beq
                   \alpha_s {m_f\over v}  {1\over M} \ , 
\eeq{fcontrib}
where $M$ is the momentum that flows in the loop.   For $2m_q \ll m_h$,
$M$ will be of order $m_h$ and so the contribution \leqn{fcontrib}
will be suppressed by a factor $m_f/m_h$.  On the other hand, if 
$2m_q \gg m_h$, $M$ will be of order $m_q$.  In this case, the factors
of $m_q$ cancel and the diagram is at full strength no matter how
large $m_q$ is.   This is bizarre but correct:   The $hgg$ vertex gets
only small contributions from quarks to which the Higgs boson can
decay and obtains full-strength constributions  from quarks to which the Higgs
boson {\it cannot} decay because they are too heavy.

In the SM, the only quark that contributes to the $hgg$ vertex at full
strength is the top quark.   If there were a fourth generation of
quarks that obtained their masses from the SM Higgs boson, each quark would
produce an equal contribution to the $hgg$ coupling, so that the total
decay rate $\Gamma(h\to gg)$ would be $3^2 = 9$ times the SM
prediction~\cite{Vysotsky}.
Such a large shift is already excluded by the LHC Higgs measurements.
This is a much stronger constraint on a fourth generation than the one
that we found from precision electroweak measurements at the end of
Section 3. 

We can compute the contribution to the $hgg$ vertex from a heavy
quark $t$ from the starting point of the QCD vacuum polarization.   The
1-loop quark vacuum polarization diagram has the value
\beqa
    & &   i (k^2 g^{\mu\nu} - k^\mu k^\nu)  \ \tr[t^at^b] {\alpha_s\over
       3\pi}   \log{\Lambda^2\over m_t^2}\CR
  & & \hskip 0.8in    i (k^2 g^{\mu\nu} - k^\mu k^\nu)  \ \tr[t^at^b]
 {\alpha_s\over
       3\pi}   \log{\Lambda^2\over m_t^2}\ .
\eeqa{qvp}
We can produce the top quark loop 
diagram in Fig.~\ref{fig:toploop}, adding a zero-momentum
Higgs boson, by shifting $v\to  v+h$ as in \leqn{hrescale}. 
  The expression \leqn{qvp}
depends on $v$ through $m_t = y_tv/\sqrt{2}$.   This yields  a
contribution to the $hgg$ vertex that is finite and equal to 
\beq
       i (k^2 g^{\mu\nu} - k^\mu k^\nu)  \ \delta^{ab} {\alpha_s\over
       3\pi}   {1\over v} \ . 
\eeqn
Comparing to   \leqn{hggeff}, we find
\beq
     A =   {\alpha\over 3\pi v} =   { g \alpha_s\over 6 \pi m_W} \ . 
\eeqn
From this expression, we can compute the partial width $\Gamma(h\to
gg)$ in the limit $m_h^2 \ll 4 m_t^2$,
\beq
    \Gamma(h\to gg) =  {\alpha_w\alpha_s^2\over 72 \pi^2} {m_h^3\over
      m_W^2} \ . 
\eeqn
The full expression can be shown to be
\beq
    \Gamma(h\to gg) =  {\alpha_w\alpha_s^2\over 72 \pi^2} {m_h^3\over
      m_W^2} \cdot \biggl| {3\over 2} \tau (1 - (\tau-1)( \sin^{-1}
    {1\over \sqrt{\tau}})^2 ) \biggr|^2 \ , 
\eeqn
 where $\tau = 4m_t^2/m_h^2$. 

Another way to interpret this argument is that the shift of $v$ in
\leqn{hrescale} is a change of scale for the SM.   Then the 1-loop
Higgs couplings to a gauge boson should be proportional to the 1-loop
contribution to the renormalization group $\beta$ function. The
calculation just performed satisfies this, since \leqn{qvp} give the
contribution of a quark to the QCD $\beta$ function.  Changing what
needs to be changed, we can obtain the coupling of a Higgs boson to
$\gamma\gamma$.   The contribution from  the top quark and the $W$ boson
to the QED vacuum polarization is
\beq
   i (k^2 g^{\mu\nu} - k^\mu k^\nu)  {\alpha\over
       4\pi} \biggl[ - {22\over 3} + {1\over 3} + {4\over 3}\cdot
     3\cdot \bigl( {2\over 3}\bigr)^2 \biggr]  \log{\Lambda^2\over
       m_{t,W}^2} \ .
\eeq{tWvp}
The first term here is contribution from the $W$, it is just 
the standard vector boson contribution to the
$\beta$ function for an $SU(2)$ gauge theory.   The second term comes
from the Higgs boson that the $W$ boson must eat to become massive.
The third term comes from the top quark; the last two factors are the
top quark color factor and electric charge.  In all, we find, for $m_h
\ll 2m_W, 2m_t $, 
\beq
    \Gamma(h\to \gamma\gamma) =  {\alpha_w\alpha^2\over 144
 \pi^2} {m_h^3\over
      m_W^2} \biggl| {21\over 4} - {4\over 3}\biggr|^2 \ . 
\eeqn
Careful evaluation, including all finite mass effects and the QCD
corrections to the gluon width, gives
\beq
     BR(h\to gg) = 8.6\% \qquad   BR(h\to \gamma\gamma) = 0.23\%   \ .
\eeqn

\begin{figure}
\begin{center}
\includegraphics[width=0.65\hsize]{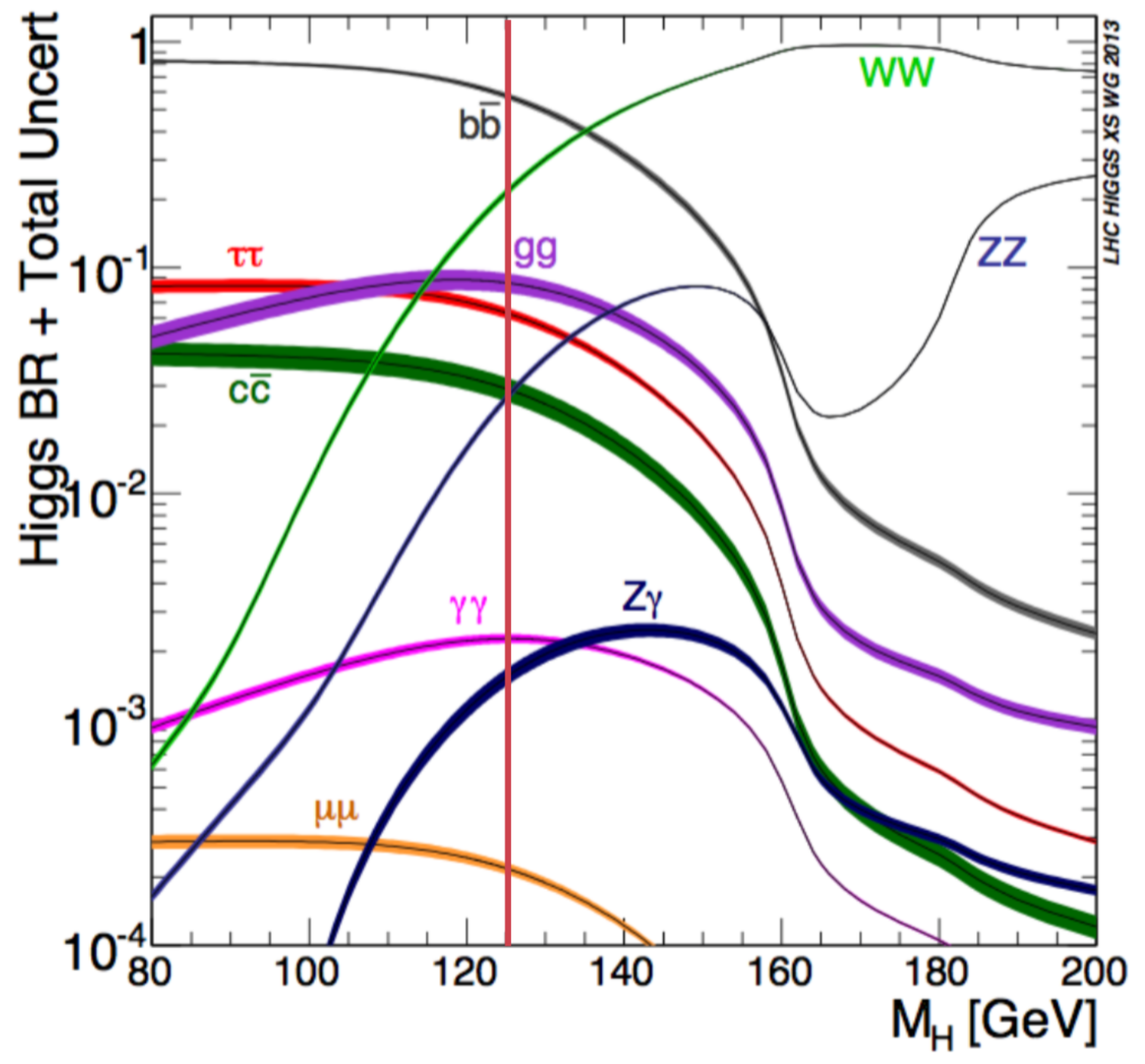}
\end{center}
\caption{Standard Model predictions for the branching ratios of the
  Higgs
boson as a function of the its mass, from \cite{HiggsHandbook}.}
\label{fig:HiggsBRs}
\end{figure}

We are now ready to put all of the pieces together to compile the SM
predictions for the various Higgs boson branching ratios.
Figure~\ref{fig:HiggsBRs} shows the predictions as a function of the
Higgs boson mass.   It is a useful exercise to understand the shape of
the curves based on the physics discussed in this section.   The
position of the observed Higgs resonance is shown by the vertical
line.  At this mass value, there are 10  distinct final states with
branching fractions larger than $10^{-4}$, including the $s\bar s$
channel not shown on this plot.

\subsection{Study of the Higgs boson at the LHC}

\begin{figure}
\begin{center}
\includegraphics[width=0.70\hsize]{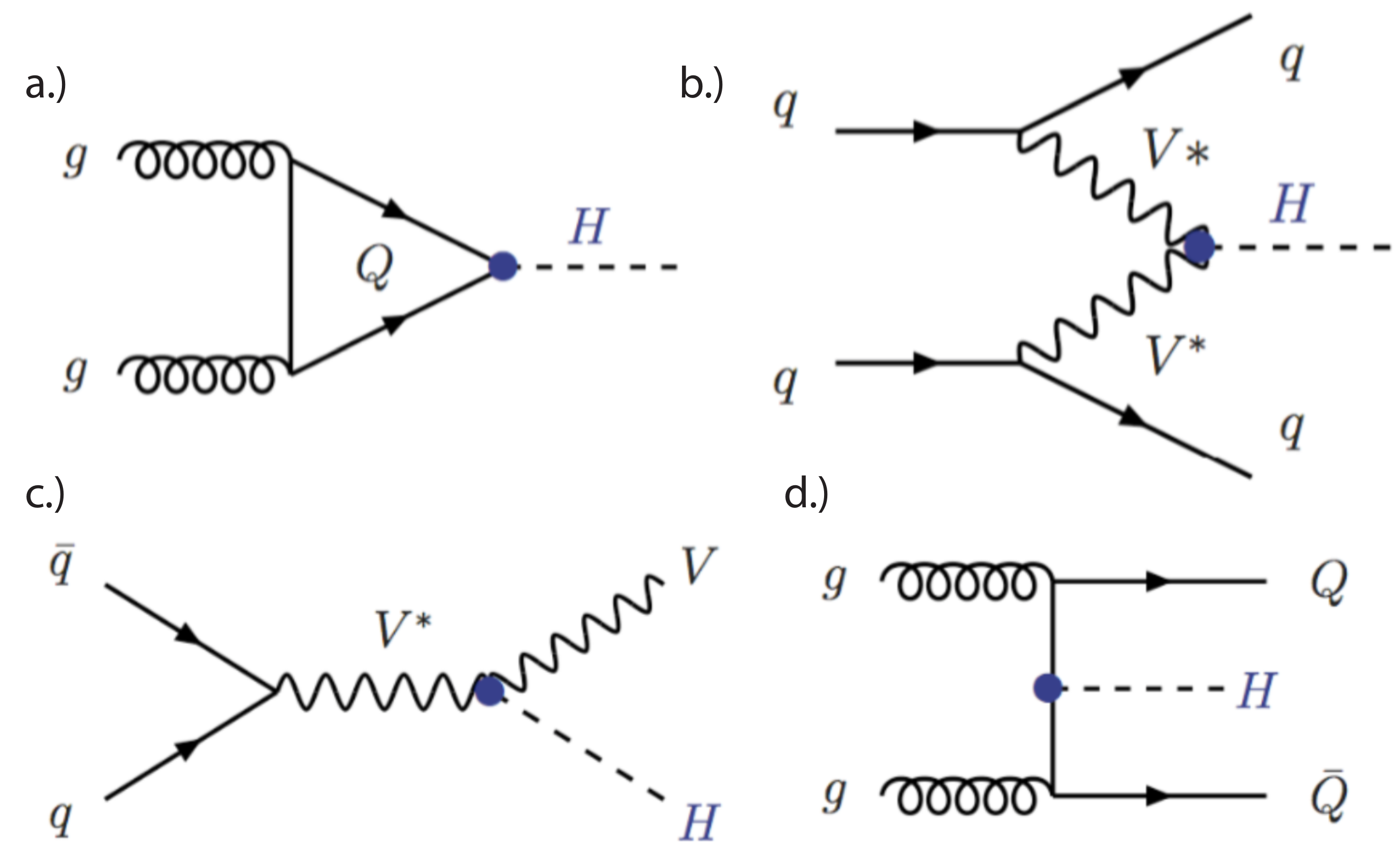}
\end{center}
\caption{Reactions producing the Higgs boson in $pp$ collisions}
\label{fig:Hprocesses}
\end{figure}

\begin{figure}
\begin{center}
\includegraphics[width=0.70\hsize]{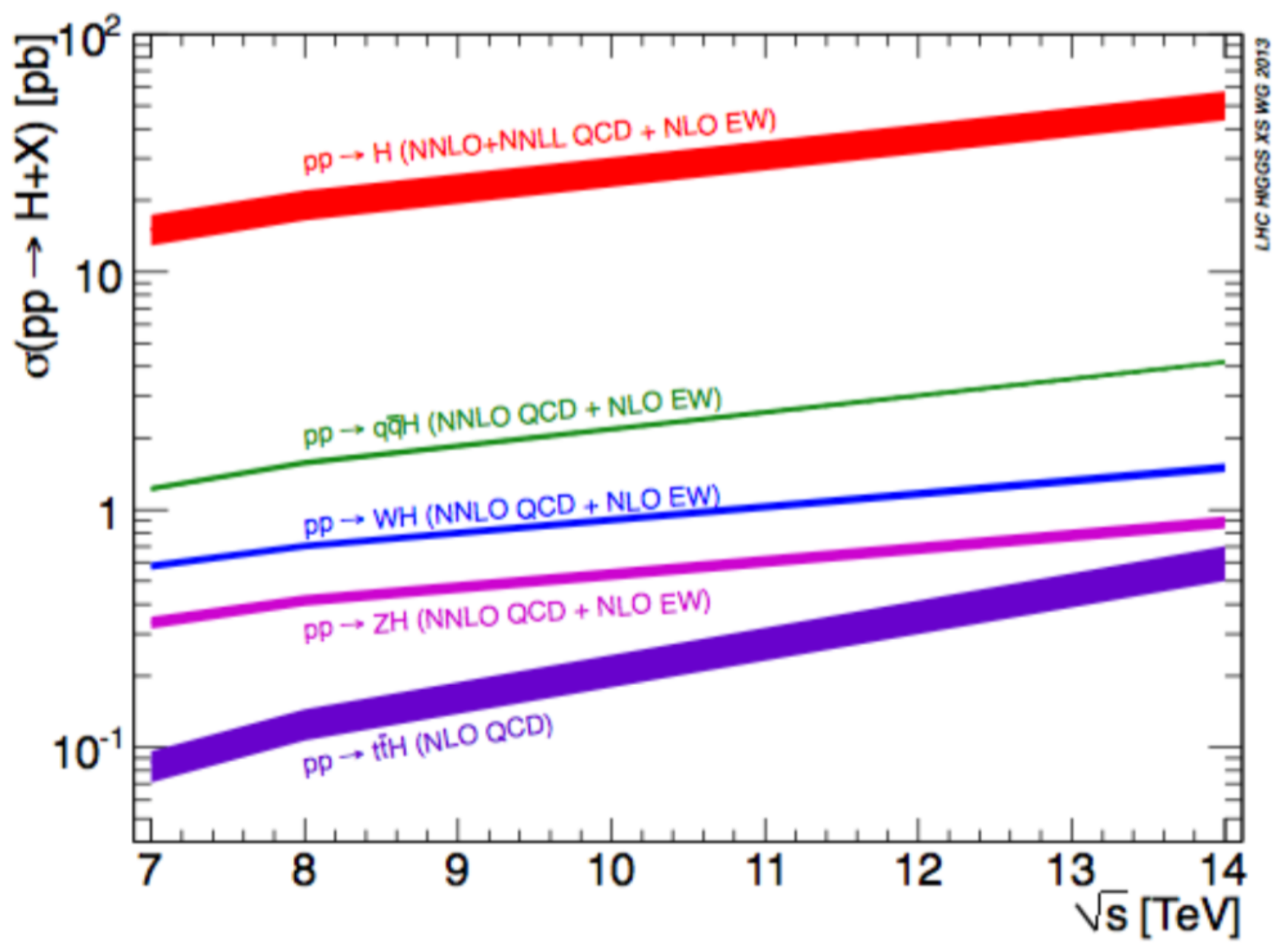}
\end{center}
\caption{Cross sections for Higgs production in $pp$ collisions 
for a 125~GeV Higgs
  boson, from \cite{PDGHiggsfourteen}.}
\label{fig:Higgscs}
\end{figure}

With this understanding of the Higgs boson couplings, I will review
very briefly the results for Higgs boson couplings obtained by the
ATLAS and CMS
experiments.  The most important processes for the production of a
Higgs boson at the LHC are those shown in Fig.~\ref{fig:Hprocesses}: 
gluon-gluon fusion, vector boson fusion, radiation of the Higgs boson
from a $W$ or $Z$ (``Higgsstrahlung''), and associate production of a
Higgs boson with a pair of top quarks.   The cross sections predicted
for these processes for a 125~GeV Higgs boson are shown in
Fig.~\ref{fig:Higgscs}.

The four reactions have different advantages for the study of Higgs
decays.
Gluon-gluon fusion has the highest cross section, so it gives access
to rare Higgs decays.  In vector boson fusion, Higgs events are tagged
by the presence of forward quark jets, reducing the background from
non-Higgs SM processes.  This reaction also has the smallest
theoretical error on the predicted 
cross section.   Higgsstrahlung also gives tagged Higgs decays.  It
also can lead to highly boosted Higgs bosons, which is an advantage
for isolating the $h\to b\bar b$ decay.   Finally, the top  associated
production process gives access to the $ht\bar t$ coupling.

In all cases, what is measured is a combination of the cross section
for Higgs production and the branching fraction for Higgs decay into
the observed final state.    This observable is related to the Higgs
couplings through
\beq
     \sigma(pp\to A\bar A\to h) BR(h\to B\bar B) \sim  {\Gamma(h\to A\bar A)
       \Gamma (h\to B\bar B)\over \Gamma_h }  \ .
\eeq{sigmatimesBR}
In this relation, $A\bar A$ is the parton combination used to produce
the Higgs boson---$gg$, $WW$ or $ZZ$, and $tt$, respectively,
 for the processes in 
Fig.~\ref{fig:Higgscs}.  The measured rates are quoted in terms of the 
{\it signal strength} $\mu$
\beq
    \mu =   \sigma(pp\to h\to B\bar B)/ (\mbox{SM \ prediction})\ .
\eeqn
Note that, if a departure from the SM value $\mu = 1$ is seen, this
might be due to a nonstandard value of the $hA\bar A$ coupling, the
$hB\bar B$ coupling, or the Higgs total width.  Multiple measurements
would be  needed to resolve this ambiguity.

\begin{figure}
\begin{center}
\includegraphics[width=0.95\hsize]{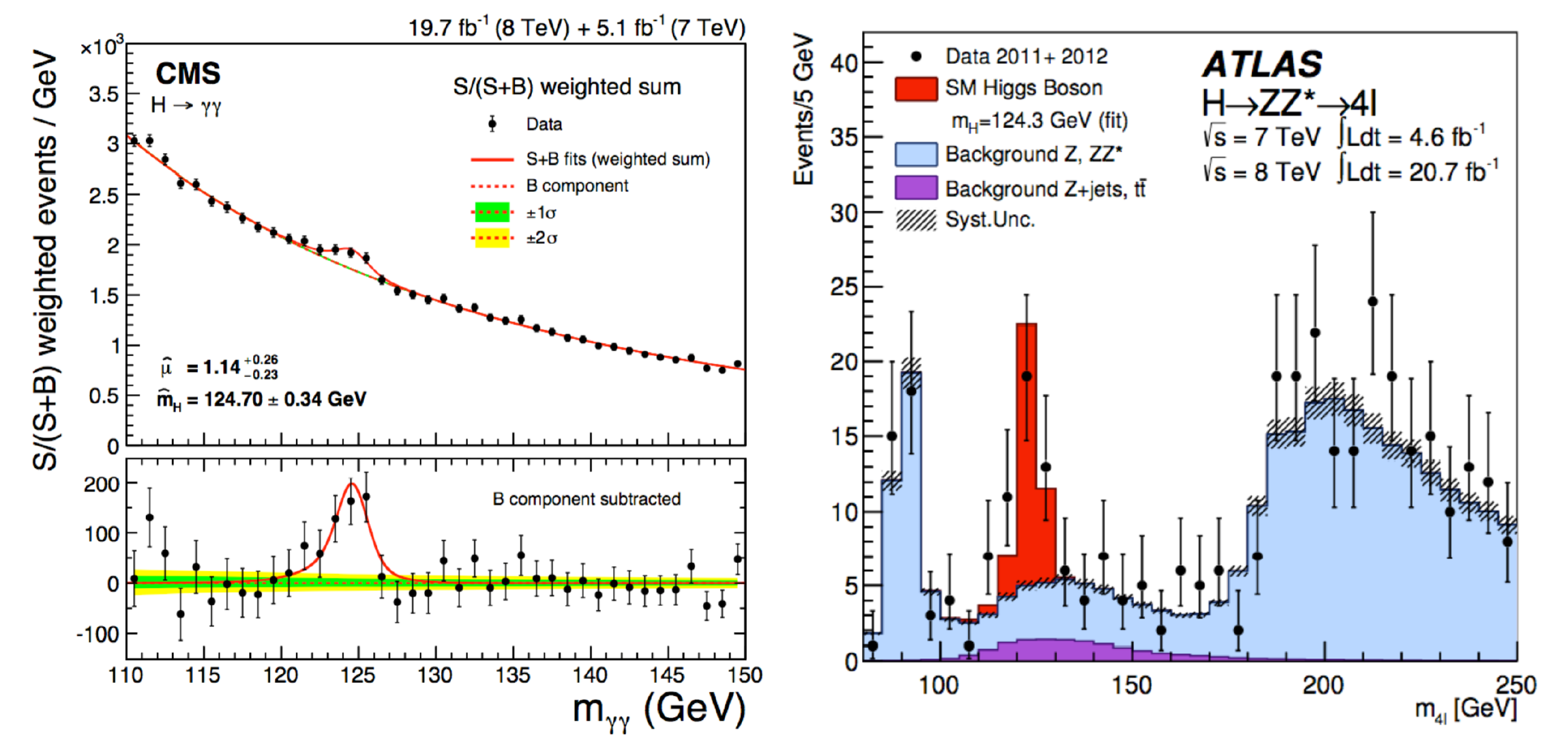}
\end{center}
\caption{Signals of the Higgs boson resonance at the LHC in run 1: left:
  Higgs resonance in the $m(\gamma\gamma)$ distribution, from \cite{CMSgamgam};
   right: Higgs resonance in the $m(4\ell)$ distribution~\cite{ATLASPLZZ}.}
\label{fig:Higgsres}
\end{figure}

The original strategy for observing the Higgs boson at the LHC used
the characteristic decay modes in which this particle could be
reconstructed as a resonance.
\beq
    h \to \gamma\gamma  \ ,   \qquad h \to ZZ^* \to 4~\mbox{leptons}
\eeqn
These modes correspond to branching fractions of 
\beq
         0.23\%   \qquad \mbox{and}  \qquad  0.012\%
\eeqn 
With production cross sections of about 20 pb at 7~TeV,
these processes have rates coresponding to fractions
\beq
        4\times 10^{-13} \qquad \mbox{and}  \qquad  2 \times 10^{-14}
        \ , 
\eeqn
respectively, of the $pp$ total cross section.  The observation of
these very tiny components of the total reaction rate at the LHC is
quite an achievement!  Signals of the
Higgs resonance in LHC run~1 data are shown in
Fig.~\ref{fig:Higgsres}. 

Once we are convinced that the Higgs resonance is actually present at
a mass of 125~GeV, we can look for the signatures of this resonance in
other decay modes.   Higgs decays to these channels give  larger total
rates than the decays to the discovery modes.  But, these channels produce events
that are not obviously distinguishable from other SM reactions.

\begin{figure}
\begin{center}
\includegraphics[width=0.95\hsize]{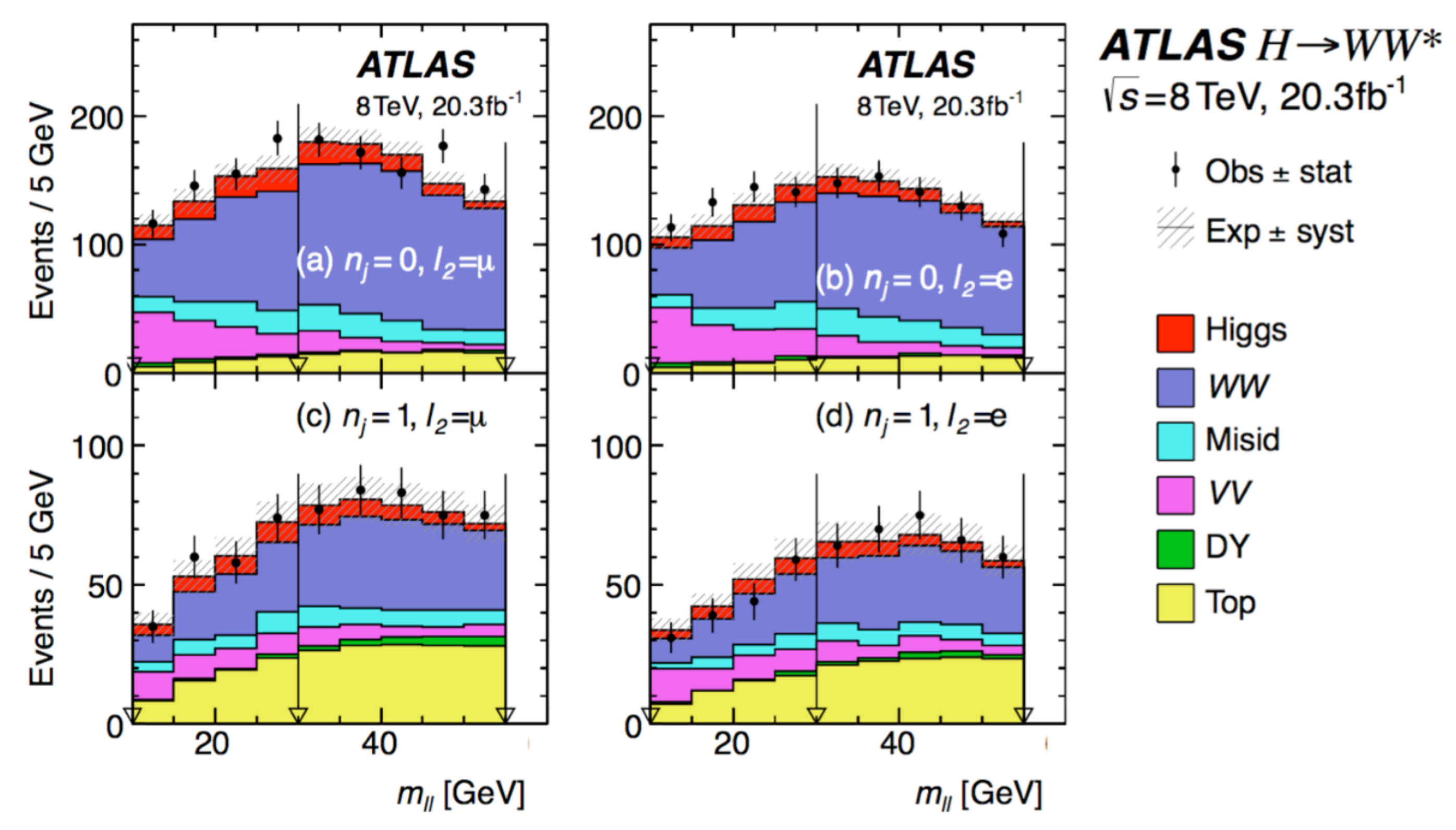}
\end{center}
\caption{Evidence for the Higgs boson in its decay to $WW^*$, from
  \cite{ATLASWW}. }
\label{fig:ATLAShWW}
\end{figure}

An example is 
\beq
   pp \to h \to W^+W^- \to \ell^+\ell^-\nu\bar \nu  \ .
\eeqn
The observable properties of these events overlap strongly with events
from 
\beq
    pp \to W^+W^- \to  \ell^+\ell^-\nu\bar \nu  \ . 
\eeqn
The signal to background ratio can be enhanced by selecting the region
where $m(\ell^+\ell^-)$  and the angle between the two leptons are
both relatively small.   It is also necessary to apply a jet veto
(that is, to select events with at most 1 high-$p_T$ jet) in order to
avoid background from 
\beq
 pp \to t\bar t \to b\bar b \ell^+\ell^- \nu \bar\nu \ .
\eeqn
Figure~\ref{fig:ATLAShWW} shows the distributions in $m(\ell^+\ell^-)$
for four event selections from  the ATLAS analysis at 8~TeV. The
histograms
 show the SM simulation of this event sample, with the various colored
 bands indicating the contributions of expected processes.   The
 largest event rates come from $pp\to WW$ and, for the 1-jet events
 shown in the bottom row,
 $pp\to t\bar t$.   The data points indicate a 10\% excess rate over
 the SM expectation from processes that do not involve a Higgs boson,
 which is well accounted for by the expected rate for Higgs
 production.

\begin{figure}
\begin{center}
\includegraphics[width=0.60\hsize]{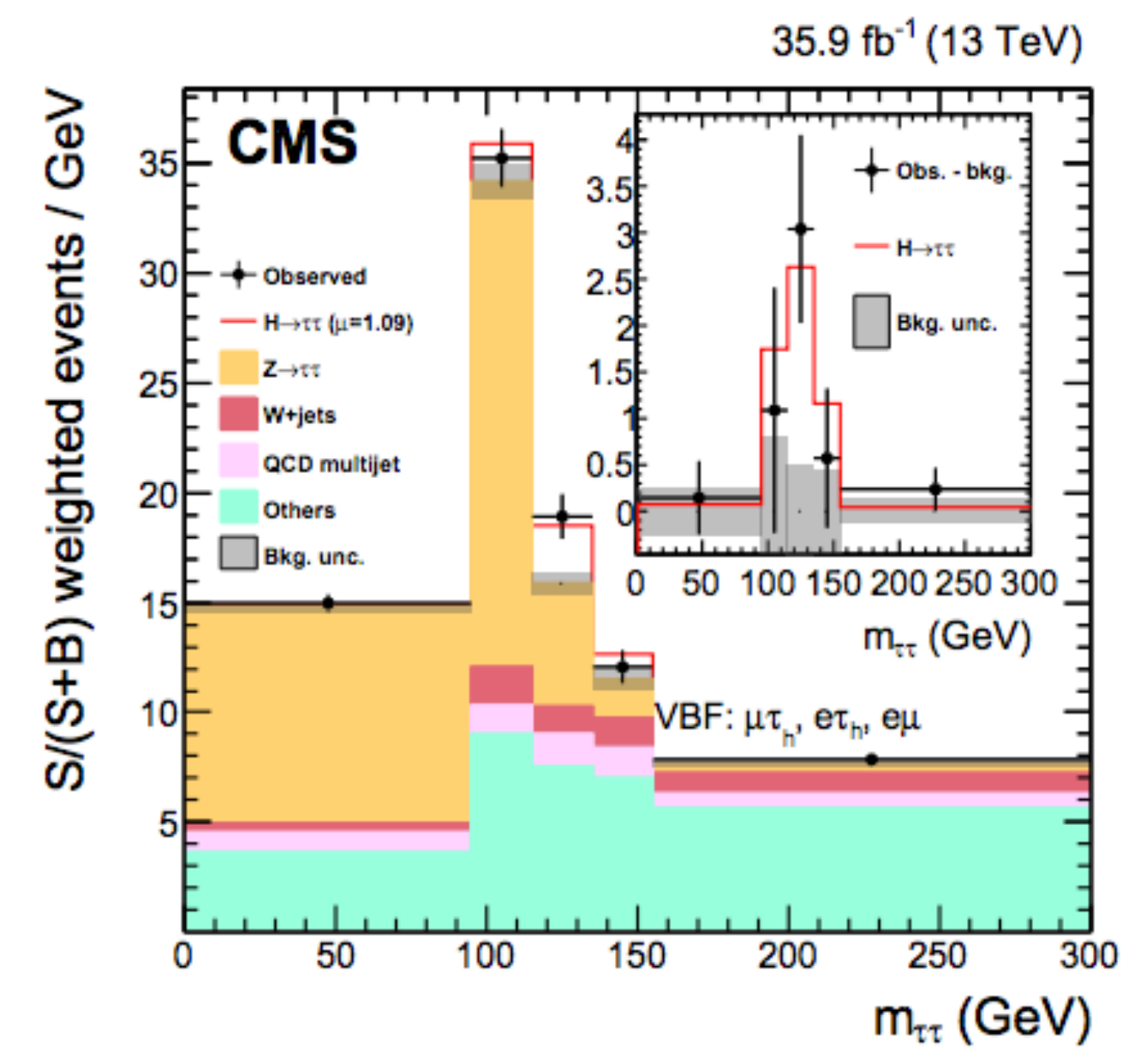}
\end{center}
\caption{Evidence for the Higgs boson decay to $\tau^+\tau^-$, from \cite{CMStau}.}
\label{fig:htautau}
\end{figure}

\begin{figure}
\begin{center}
\includegraphics[width=0.70\hsize]{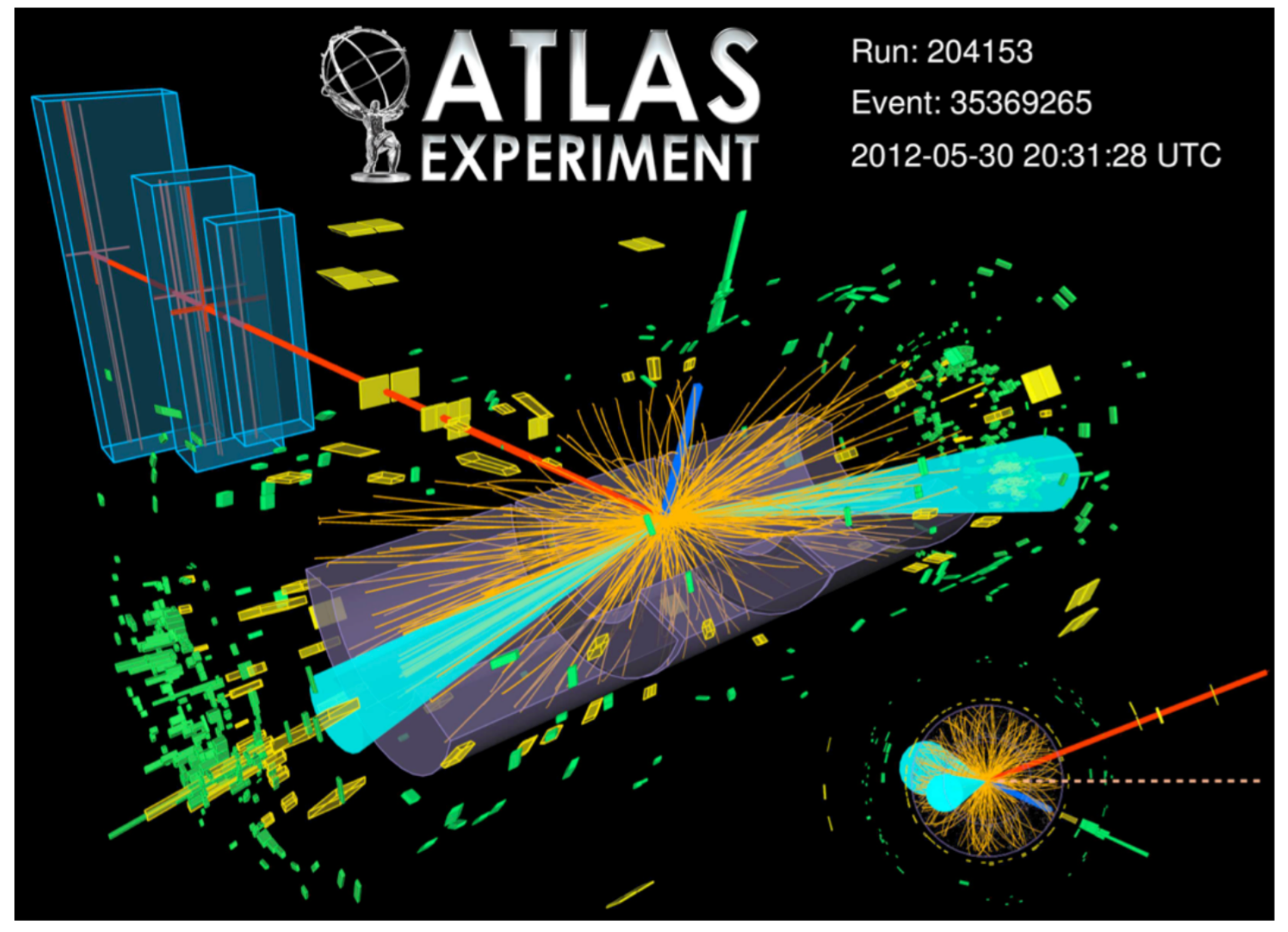}
\end{center}
\caption{A candidate event for vector boson fusion production of a
  Higgs boson decaying to $\tau^+\tau^-$, from \cite{ATLAStau}.}
\label{fig:htautauevent}
\end{figure}

Similar analyses support the presence of Higgs boson production and
decay to $\tau^+\tau^-$.   The most important backgrounds are 
\beq
     pp \to Z \to \tau^+\tau^- \ , \qquad   pp \to W^+W^-\ , 
\eeqn
and QCD reactions where two jets in the final state fake the $\tau$
signatures.    The strongest evidence for the reaction comes from
vector boson fusion, since the tagging by forward jets helps to
minimize the QCD background.   Figure~\ref{fig:htautau}  shows the
very recent 
CMS run 2 analysis with data from 13 TeV.   These events are dominated
by the 
large background from 
$Z\to \tau^+\tau^-$.  However,  this background can be understood  using the
observed distribution of $Z\to \mu^+\mu^-$ events.   The backgrounds
from $WW$ and QCD are more challenging to estimate.
Fig.~\ref{fig:htautauevent} shows a candidate vector boson fusion
$h\to \tau^+\tau^-$ event from ATLAS.   I use the word ``candidate''
advisedly; probably this event is  a $Z\to \tau^+\tau^-$ event produced by vector boson
fusion. 

\begin{figure}
\begin{center}
\includegraphics[width=0.99\hsize]{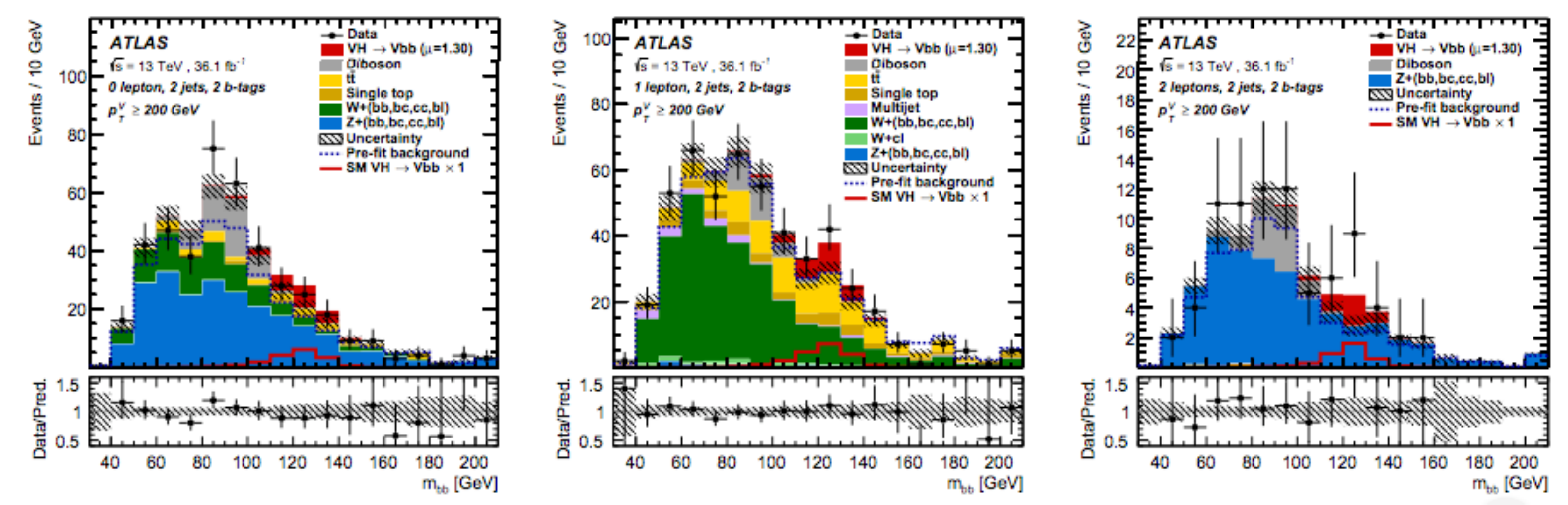}
\end{center}
\caption{Evidence for the Higgs boson decay to $b\bar b$, from
  \cite{ATLAShbb}. The three distributions show 0, 1, and 2-lepton
  events.  The red (dark) boxes near the mass value of 125 GeV show  the expectation from $pp\to Vh$, $h\to
b\bar b$. }
\label{fig:hbb}
\end{figure}

The most challenging of the major modes of Higgs decay is the one with
the highest branching ratio, $h\to b\bar b$.   It is probably hopeless to observe this mode in
gluon fusion at low Higgs $p_T$, since $gg\to b\bar b$ with $m(b\bar b) \sim 125$~GeV has
a cross section about a million times larger that that of the Higgs
process.
Current analyses use the Higgsstrahlung process with a tagging $W$ or
$Z$
\beq
   pp \to V h \ , \quad h \to b\bar b
\eeq{Vh}
where $V$ is $W$ or $Z$. 
  However, there are other SM processes with similar signatures
that do not involve a Higgs boson,
\beqa
     pp \to V  Z \ , &\quad &  Z\to b\bar b \CR
     pp \to V g\ , &\quad & g\to b\bar b    \ . 
\eeqa{Vother}
The second reaction involves an off-shell gluon with a mass near
125~GeV that converts to $b\bar b$.  Convincing evidence for this
decay has been obtained only very recently, in the 13 TeV data~\cite{ATLAShbb}. .   The
current evidence from the ATLAS run 2 data is shown in
Fig.~\ref{fig:hbb}.   It is expected that discrimination of  the three processes \leqn{Vh},
\leqn{Vother} can be improved  in an event sample in which the state recoiling against
the vector boson is highly boosted, using techniques that measure the
dijet mass and color flow.    A recent analysis by CMS shows a small
signal for $h\to b\bar b$ in a sample of high $p_T$ jets recoiling
against a gluon jet~\cite{CMShbb}.

\begin{figure}
\begin{center}
\includegraphics[width=0.70\hsize]{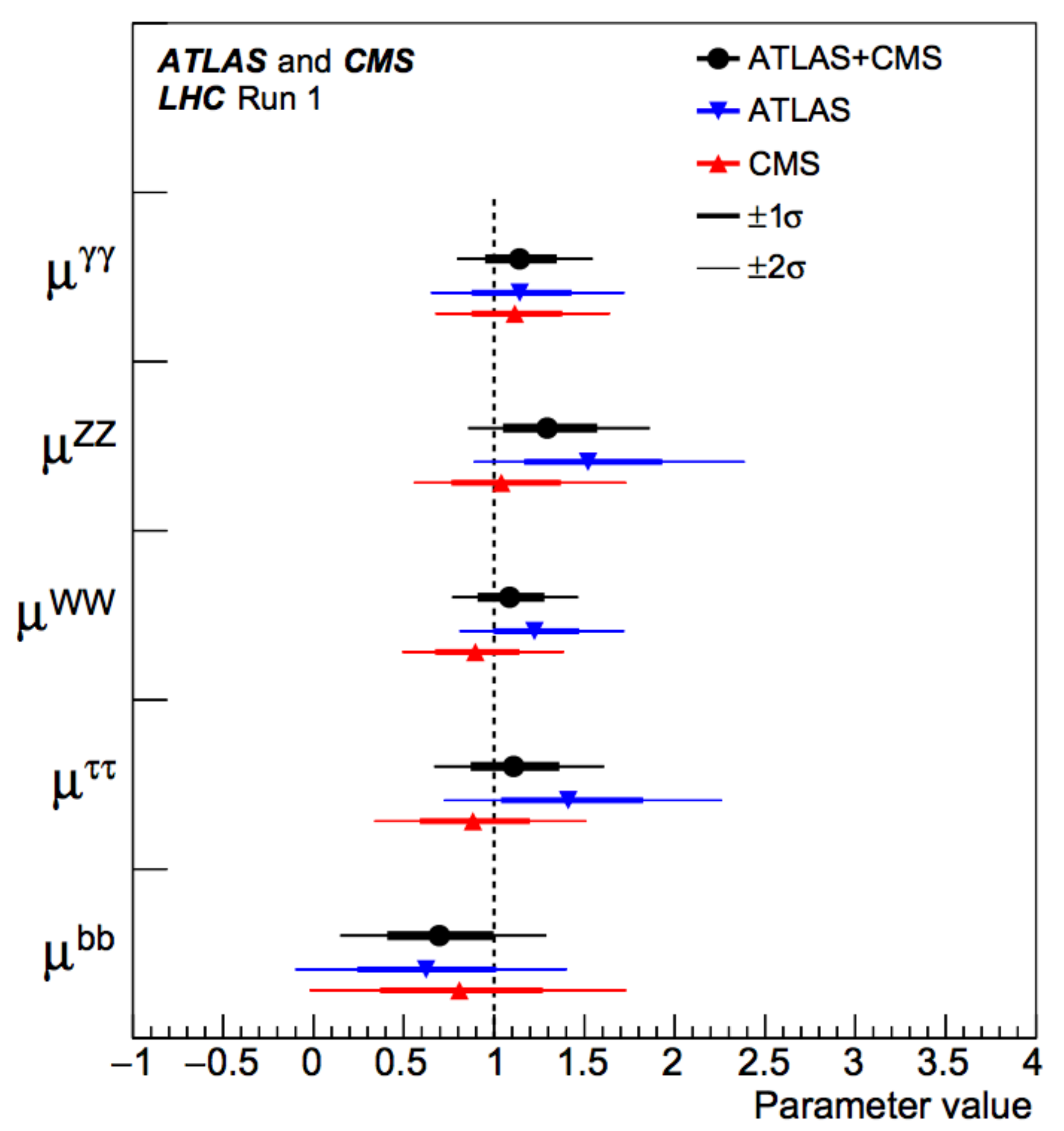}
\end{center}
\caption{Summary of Higgs $\mu$ measurements, from \cite{PDGHiggs}.}
\label{fig:hsumm}
\end{figure}

Figure~\ref{fig:hsumm} shows a summary of the measurements of the
Higgs boson signal strengths made by ATLAS and CMS in run 1 of the
LHC~\cite{PDGHiggs}.   A signal strength of 0 indicates no presence of
the Higgs boson.  This hypothesis is excluded by run~1 data for all of the modes
considered except $h\to b\bar b$. I have discussed above the more
significant evidence for $h\to \tau^+\tau^-$ and $h\to b\bar b$ found
already in run~2.   A signal strength of 1 is the
prediction of the SM.   The measured rates agree with this
prediction within about 30\% accuracy.   So the quantitative study of
the Higgs boson has begun and will be improved as the LHC accumulates data.

\section{Precision measurements of the Higgs boson properties}

In the last segment of these lectures, I take a step outside the
Standard Model.  In this section, I will discuss the expectations for
the couplings of the Higgs boson in theories beyond the Standard
Model.   This is an interesting story that motivates a dedicated
experimental campaign to measure the couplings of the Higgs boson with 
high precision.   First, though, I will explain why I believe there
must be new interactions of physics waiting to be discovered.

\subsection{The mystery of electroweak symmetry breaking}

I have shown in the previous lectures that the SM of weak
interactions is an extremely successful theory in its own domain.   It
is not a complete theory of nature, but we can supplement it by adding
gravity, quantum chromodynamics (QCD) as the theory of the strong
interactions,
and some model of dark matter and dark energy.   It is also not
difficult to add neutrino masses to the model, either by introducing
three generations of right-handed neutrinos or by adding
lepton-number-violating Majorana mass terms.  Each of these additions
accounts for some set of observed phenomena that is outside the range
of topics considered in these lectures.

But this is not enough.   A key part of the explanation for the
structure of the weak interactions and the generation of masses for
quarks, leptons, and gauge bosons is the spontaneous symmetry breaking
of $SU(2)\times U(1)$ and the generation of the Yukawa couplings that
link the symmetry-breaking Higgs field to the quarks and leptons.
The structure that I have described leads immedately to questions
about all of these ingredients:
\begin{itemize}
\item Why just quarks and leptons?  What is the origin of the quantum
  number assignments $(I,Y)$ for the matter particles seen in nature?
\item What explains the spectrum of quark and lepton masses?   The
  SM gives the relation
\beq
    m_f =   { y_f v \over \sqrt{2}}   \ ,
\eeqn
where $v$ is the Higgs field vacuum expectation value.   But the $y_f$
are renormalized parameters that cannot be predicted with the Standard
Model.  The presence of nonzero CKM angles---and, with neutrinos, PMNS
angles---adds further difficulty  to this problem.
\item What is the origin of the Higgs field?  Is there only one such
  field, or are there multiplets of scalar fields with different
  quantum numbers?   The SM makes the minimal choice of
  one Higgs multiplet.   Is this necessary?
\item  Why is $SU(2)\times U(1)$ spontaneously broken?   The shape of
  the Higgs potential energy function is an input for which the
  SM gives no explanation.
\end{itemize}

\begin{figure}
\begin{center}
\includegraphics[width=0.30\hsize]{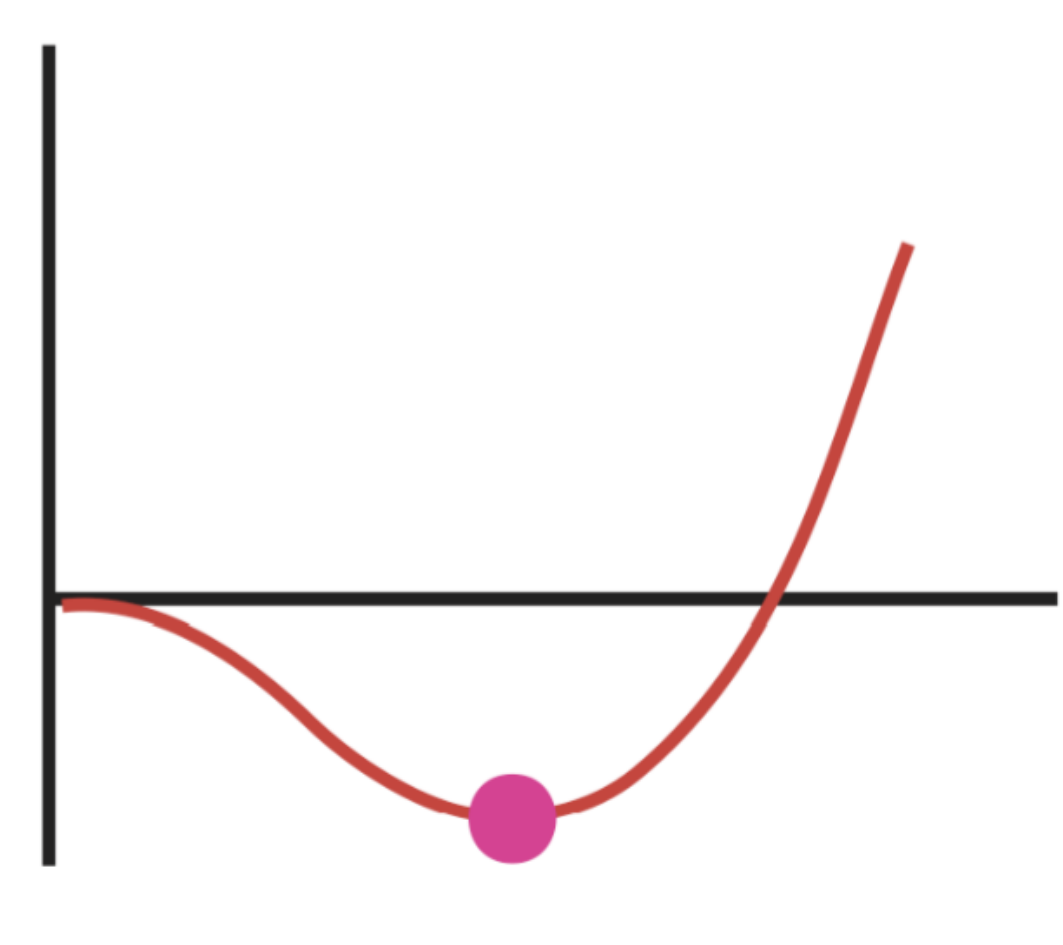}
\end{center}
\caption{The Higgs potential $V(|\varphi|)$.}
\label{fig:Higgspot}
\end{figure}

This last question merits more discussion.   Here is the explanation
for electroweak symmetry breaking given in the SM:   The
model instructs us to write the most general renormalizable potential 
for the Higgs field $\varphi$,
\beq
    V(\varphi) =  \mu^2 |\varphi|^2 + \lambda |\varphi|^4 \ . 
\eeq{simpleHiggspot}
We assume that $\mu^2 < 0$.  Then the potential has the correct shape,
shown in Fig.~\ref{fig:Higgspot}, to
drive
spontaneous symmetry breaking.

Why must $\mu^2$ be negative?  That question cannot be addressed
within the model.   It is just a choice, perhaps a random one.

\begin{figure}
\begin{center}
\includegraphics[width=0.5\hsize]{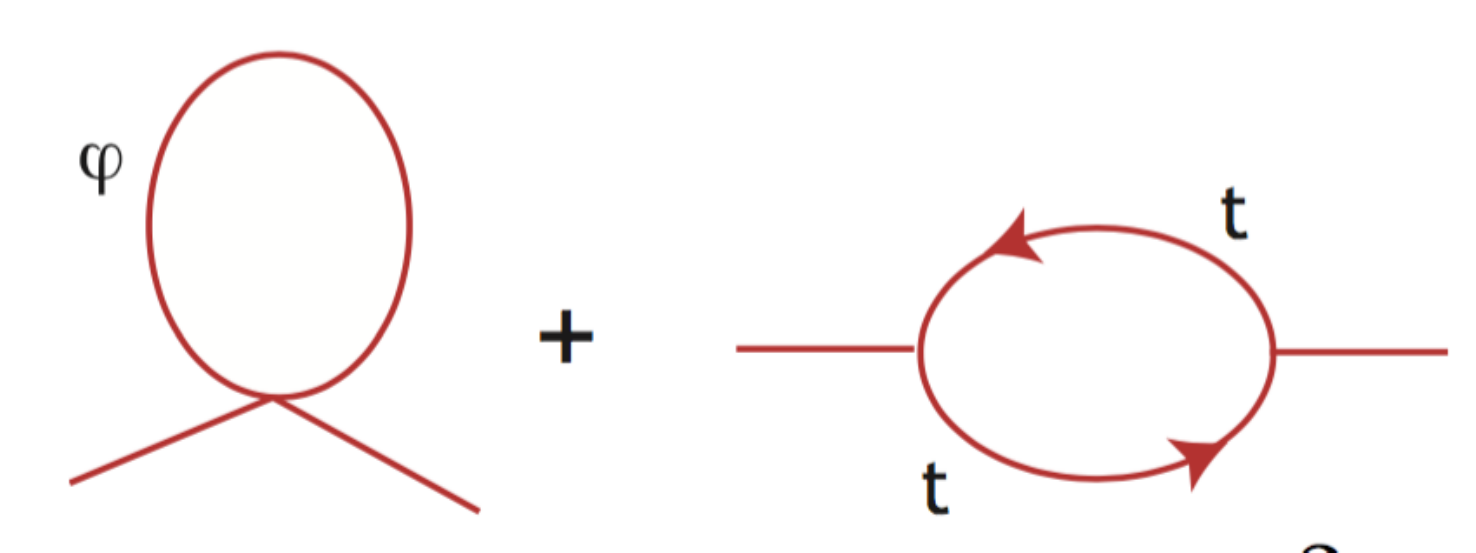}
\end{center}
\caption{One-loop corrections to the $\mu^2$ parameter from the Higgs
  field coupling to the top quark and from the Higgs field self-coupling.}
\label{fig:muoneloop}
\end{figure}

We get into deeper trouble if we try to take this explanation  to a higher
level of precision by computing the radiative corrections to the
parameter $\mu^2$.   The leading one-loop corrections, from loops
containing the Higgs and top quark fields, are shown in
Fig~\ref{fig:muoneloop}.  They give
\beq
 \mu^2 = \mu^2_{\rm bare} + {\lambda\over 8\pi^2} \Lambda^2 -
{ 3y_t^2\over 8\pi^2} \Lambda^2 + \cdots \ . 
\eeq{mucorrect}
The diagrams are ultraviolet divergent.   I have regularized them by
cutting off their momentum integrals at a mass scale $\Lambda $,
arbitrarily chosen to be the same for Higgs and top.    The final
value of $\mu^2$ needed to produce the observed Higgs boson mass is
$\mu^2 \approx - (100~\mbox{GeV})^2$.    So if $\Lambda$ is much
larger than 1~TeV, this formula requires large cancellations among the
ingredients with no obvious explanation.   If we assert that the
SM is correct up to the Planck scale,   the first 33
significant figures must cancel.   It is also apparent that the
right-hand side contains both positive and negative contributions, so
it is not obvious without invoking a much deeper explanation why the
final answer after the cancellation should turn out to be negative.

The simplest resolution of this set of problems would be that there
are new particles, not yet known to us, that generate additional
diagrams contributing to the calculation of $\mu^2$.   If these
particles have masses of TeV size, they might cancel the
divergences seen in \leqn{mucorrect} and---in the best case---leave
over a calculable answer for $\mu^2$.   However, we have not yet been
able to discover these particles in high-energy experiments.

The general problem of the uncalculability of the parameter $\mu^2$ is
not new to high-energy physics.   It is encountered in all systems in
which a symmetry is spontaneously broken.   Condensed matter physics
gives many examples.

The most direct analogy to the Higgs theory comes in the phenomenon of
superconductivity seen in most metals at cryogenic temperatures.   The
original papers on the Higgs mechanism by Englert and Brout, Higgs, 
and Guralnik, Hagen, and Kibble~\cite{BE,Higgs,GHK}  all used the analogy to
superconductivity to motivate their arguments.  However, they used
only a piece of the complete theory.   Supercondutivity was discovered
in 1911 by Kamerlingh Onnes and was quickly seen to be associated with
a sharp phase transition~\cite{Onnes}.  However, the explanation for this phase
transition was not understood for another 45 years.

In 1950, Landau and Ginzburg proposed a phenomenological theory of
superconductivity based on a scalar field with the potential
\leqn{simpleHiggspot}~\cite{Landau}.    They assumed that the parameter $\mu^2$
would be a function of temperature, taking negative values below the
phase transition temperature  $T_C$.   Coupling this theory to
electromagnetism, they found that the photon acquires a mass by the
Higgs mechanism and that the scalar fields in the vacuum can transmit
electric current frictionlessly. 
  This theory turned out to be extremely successful in
explaing many aspect of superconductivity, including the Meissner
effect in which superconductors repel magnetic flux, the existence of
Type I and Type II superconductors, and the systematics of the
destruction of superconductivity by high currents or high magnetic
fields.

However, this theory could not address the most important problem of
why superconductivity occured in the first place.  The answer to that
question waited until 1957, when Bardeen, Cooper, and Schrieffer
discovered the mechanism that causes electrons in a metal to pair up
into bound states and form a boson condensate with the properties of the Landau-Ginzburg
scalar field~\cite{BCS}.

In our understanding of the phase transition to symmetry breaking of
$SU(2)\times U(1)$, we are now at the Landau-Ginzburg stage.

In the case of superconductivity, physicists knew that there must be a
deeper explanation that had to be given in terms of the interactions
of elecrons and atoms.   For the symmetry-breaking of the weak
interactions, any analogous explanation must involve new elementary
particles outside the SM.  We do not know what these
particles are.  We only know that we have not discovered them yet.

\subsection{Expectations for the Higgs boson in theories 
beyond the
  Standard
Model}

Even if we cannot discover new heavy particles responsible for the
Higgs potential energy, we can hope to find clues to the nature of
these new particles and interactions by looking more deeply into the
properties of the Higgs boson itself.   In the previous lecture, I
emphasized that the SM makes precise predictions for the couplings
of the Higgs boson to all particles of the SM in terms of
the measured masses of those particles.  Any deviation from these
predictions must indicate the presence of new interactions beyond the
SM.  In this and the next two sections, I will trace out the
expectations for corrections to the Higgs properties in different
classes of models of new physics.

To begin, I will present two sets of expectations for the properties
of new physics models.   The first is guidance from the concept that
these models should solve the problem of the calculability of the
Higgs potential.   The second comes from a constraint that is
well-satisfied in the precision electroweak measurements.

I have already explained that the parameter $\mu^2$ in the Higgs
potential cannot be computed within the SM.  To construct
a model in which $\mu^2$ can be computed, that model must satisfy some
special properties.   In particular, some structure in the theory msut
require the cancellation  of quadratically divergent Feynman diagrams
which would otherwise add large, arbitrary terms to the final result
for $\mu^2$.

There are two strategies to achieve this.   The first is to include in
the model a symmetry that forbids the appearance of the 
\beq
                     \mu^2 |\varphi|^2 
\eeq{baremuphi}
term in the Lagrangian.   It is not so obvious how to construct such a
symmetry, since the operator \leqn{baremuphi} seems to be conpletely
neutral.   It would be forbidden in a scale-invariant theory, but in
quantum field theory scale invariance is usually explicitly broken by
the running of coupling constants.   Two schemes that do forbid such
as term are supersymmetry, the spacetime symmetry that links fermions
and bosons, and the identification of $\varphi$ with a Goldstone boson
of some spontaneous symmetry breaking at a very high mass scale.   The
computation of the Higgs potential in models of supersymmetry is
reviewed in \cite{Martin,mySUSY}.    The computation of the Higgs
potential in models in which the Higgs boson is a Goldstone boson is
reviewed in \cite{Contino,Gherghetta}.   There are also other 
 proposed generalizations of the SM Higgs sector in which  the Higgs potential is not calculable.

One of the properties of mass generaion in the SM is the
relation $m_W = m_Z c_w$, as we saw in \leqn{WZrel}.   This property
can be derived from a symmetry of the Higgs potential assumed in the
SM.  Since the relation works so well, it is suggested
that generalizations of the SM Higgs sector should also
have this property. 

The origin of the relation \leqn{WZrel} can be seen as follows:
Look at the form of the vector boson mass matrix acting on the
original $SU(2)\times U(1)$ fields, 
\beq
 m^2 =   \pmatrix{ g^2 & & & \cr & g^2 & & \cr & & g^2 & - gg'\cr & &
   -gg' & g^{\prime 2} } \qquad \mbox{on} \qquad
\pmatrix{A^1\cr A^2 \cr A^3 \cr B\cr} \  . 
\eeqn
The form of the matrix is dictated by the requirement that the matrix
have a zero eigenvalue, associated with the massless photon, and that
the part of the matrix acting on the $SU(2)$ fields  $(A^1, A^2, A^3)$
should be symmetric among these fields.  The
requirement for 
the latter statement is
that the theory contains an $SO(3)$ transformation that rotates the $SU(2)$
gauge fields into one another and is unbroken even when the $SU(2)$ gauge symmetry is
spontaeously broken.   This extra transformation is called {\it
  custodial symmetry}~\cite{SSVZ}. 

Custodial symmetry is an accidental property of the SM
Higgs potential.  If we write
\beq
      \varphi = {1\over \sqrt{2}}  \pmatrix{ \varphi^1 + i
        \varphi^2\cr \varphi^0 + i \varphi^3\cr}
\eeqn
the Higgs potential depends only on the combination
\beq
   | \varphi|^2 =  (\varphi^0)^2 +  (\varphi^1)^2 +  (\varphi^2)^2 +
   (\varphi^3)^2 \ .
\eeqn
A vacuum expectation value for $\varphi^0$ preserves the $SO(3)$
symmetry that acts on  $(\varphi^1, \varphi^2, \varphi^3)$.   From
this observation, we understand why the SM satisfies
\leqn{WZrel}.

There are many generalizatios of the SM Higgs theory that
also satisfy this condition.  For example, we could introduce two or
more scalar field multiplets with $(I,Y) = (\half, \half)$.  In the
most general case, a different Higgs boson can be used to give mass to
the charged leptons, $d$ quarks, and $u$ quarks, by writing the Higgs 
Yukawa interactions as 
\beq
 \L = - y_e L^\dagger \cdot \varphi_1 e_R - y_d Q^\dagger \cdot\varphi_2
 d_R - y_u Q^\dagger_a \eps_{ab} \varphi^\dagger_{3b} u_R + h.c. . 
\eeqn
In this equation, $L$ is the left-handed lepton doublet, $Q$ is the
doublet of left-handed quarks, and all three Higgs multiplets have 
$I = \half$, $Y = \half$.  The three Higgs fields should have a
potential that aligns their vacuum expectation values so that the
$U(1)$ symmetry giving electromagnetism remains unbroken.
  This structure can be  extended to three
generations by replacing the three Yukawa couplings by three
$3\times 3$ matrices.  The resulting theory shares with the Standard 
Model the property that,
after a change of variables, the Higgs couplings are all CP even and
flavor diagonal.

It can be shown that the Yukawa coupling with a complex
conjugated
field $\phi_3^\dagger$ is inconsistent with supersymmetry.   Then, in
models of supersymmetry, we must introduce at least two Higgs double
fields, one with $I= \half, Y = +\half$, to give mass to the $d$
quarks and leptons, and a different field with $I = \half, Y =
-\half$, to give mass to the $u$ quarks. 

More complex Higgs field multiplets are also possible.   Georgi and
Machacek found a way to preserve custodial symmetry with Higgs bosons
in higher  representations, corresponding to spin $I$ under
 the weak interaction $SU(2)$
symmetry~\cite{GandM,Logan}.
For example, for $I=1$, we could introduce a $3\times 3$ matrix of
fields
\beq
  X =   \pmatrix{ \chi^{0*} & \xi^+ & \chi^{++} \cr
             - \chi^{+*}   & \xi^0  &  \chi^+ \cr
             \chi^{++*} & - \xi^{+*} & \chi^0 \cr } \ ,
\eeqn
in which the rows are $SU(2)$ triplets and the columns have $Y = -1,
0, 1$, respectively.  The potential for this field can be arranged to
have 
$SU(2)\times SU(2)$ symmetry and a minimum at
\beq
      \VEV{X} =  V \cdot {\bf 1}_3
\eeqn
that preserves the diagonal $SU(2)$ as a global symmetry.   We need at
least one $I = \half$ Higgs  multiplet to give mass to the quarks and
leptons, but we can supplement this with additional Higgs fields with
any value of $I$. 

The criterion of custodial symmetry also provides guidance in
constructing models of composite Higgs bosons that satisfy current
phenomenological constraints.  To provide examples of such models, let
me begin by describing the Technicolor model introduced in 1978 by
Weinberg
and Susskind~\cite{WeinbergT,Susskind}. These authors introduced a copy
of QCD with two massless techni-quark flavors $(U,D)$, and with a
strong
interaction mass scale corresponding to a techni-$\rho$ meson mass at
2~TeV.    This model has $SU(2)\times SU(2)$ chiral symmetry,
analogous to that in the known strong interactions.   Just as happens
there, the theory should have a spontaneous breaking of this symmetry
to a diagonal $SU(2)$ symmetry, dynamically generating masses for the
techni-quarks and creating three techni-pions as Goldstone bosons. The
diagonal $SU(2)$ symmetry remains unbroken, and this plays the role of
the custodial symmetry.   If
this model is coupled to the $SU(2)\times U(1)$ gauge symmetry of the
SM, the $W$ and $Z$ bosons eat the Goldstone techni-pions
and acquire mass through the Higgs mechanism.    The $W$ and $Z$
masses obey  \leqn{WZrel}, with 
\beq
       m_W = {g F_\pi\over 2} \ , 
\eeqn
where $F_\pi$ is the analogue of the pion decay constant in the
technicolor interactions.   We obtain the observed $W$ and $Z$ masses 
for $F_\pi = 246$~GeV, the Higgs field expectation value in the
SM.   In this model, the Higgs boson would be a  spin zero, isoscalar
bound state of the $U$ and $D$ quarks and their antiquarks.

The Weinberg-Susskind technicolor model is now excluded.   The model
predicts a Higgs boson mass at about 1~TeV, and also too large an $S$
parameter to be consistent with precision electroweak measurements.
However, it points the way to more sophisticated models that also
build the Higgs boson as a composite state.

An example is given by the following scenario, which uses the strong
interaction chiral symmetry breaking in a different way:  Introduce
new QCD-like strong interactions at a mass scale of 10~TeV, with
4 associated quarks in real, rather than complex, representations of
the gauge group.   This theory has a chiral symmetry
$SU(4)$, which is spontaneously broken to $SO(4)$ when the quarks
dynamically acquire mass. $SU(4)$ has 15 generators, and
$SO(4)$ has 6, so the symmetry-breaking creates $15-6 = 9$
Goldstone bosons. We might  take two of the four quarks to transform as a
doublet under the weak interaction $SU(2)$ and the other two to be
weak interaction singlets  that form a doublet under another $SU(2)$.
Then the Goldstone boson multiplet will contain 4 bosons that
transform as  $(\half, \half)$ under this $SU(2)\times SU(2)$.   We
can identify this multiplet with the Higgs boson doublet.  This
scenario
realizes the idea of the Higgs doublet as a set of Goldstone bosons
that, by Goldstone's theorem, stay massless while the strong
interaction chiral symmetry is broken.   In a set of models called
{\it Little Higgs}, it is possible to perturb the strong interaction
theory to produce a nonzero, calculable Higgs 
potential~\cite{LittleH,LittleHreview}.

\subsection{The Decoupling Theorem}

Through the strategies described in the previous section, it is
possible to build many models of the Higgs field that are more complex
than the SM and yet compatible with all current
experimental constraints.  One's first instinct is that these models
will lead to wildly different predictions for the properties of the
Higgs boson that are easily distinguished experimentally.   However,
this is not correct.   To distinguish models of the Higgs sector, it
is necessary to make detailed measurements reaching a relatively high
degree of precision.   This is a consequence of the Decoupline
Theorem, enunciated by Howard Haber in \cite{HaberDec}.

The Decoupling Theorem states: If the spectrum of the Higgs sector
contains one Higgs boson of mass $m_h$, with all other Higgs particles
having masses at least $M$, then the influence of these particles on
the properties of the light Higgs boson is proportional to 
\beq
                 m_h^2/M^2  \ .
\eeqn
If the Higgs sector contains additional particles, but these particles
have masses of 1~TeV, they shift the properties of the known Higgs
boson by corrections to the Higgs couplings at the percent level.

The proof of this theorem is quite straightforward.   It uses the
viewpoint of effective Lagrangians described in Section 5.4.    As I
have explained above, once we have measured the mass of the Higgs
boson, the parameters of  the SM relevant to the Higgs
field are fixed, and the SM makes precise predictions for
the Higgs couplings.   On the other hand, I have also explained that
the SM Lagrangian is the most general renormalizable
Lagrangian
with the known quark and lepton fields and the gauge symmetry 
$SU(3)\times SU(2)\times U(1)$.    So, in an effective Lagrangian
description, any perturbation of the Higgs couplings away from the 
SM predictions must be associated with operators of
dimension 6.  These operators have dimensionalful coefficients.  If
they are generated by particles of mass $M$, their coeffcients will be
of order $1/M^2$. 

This situation is challenging but not hopeless.   It implies that the 
current level of agreement of the Higgs boson properties with the predictions
of the SM---to 20-30\%, as described in the previous
section---is absolutely to be expected no matter how complex the Higgs
sector might be.  But, it offers the opportunity that, with
measurements of higher precision, an picture of the
Higgs boson entirely different from that of the SM might be revealed.

\subsection{Effects on the Higgs boson couplings 
from models of new physics}

To amplify this discussion of the effects of new physics on  the SM
Higgs couplings,  I will now review some specific
examples of those effects.

To begin, consider models with two Higgs scalar doublets.   I remind
you that supersymmetric models necessarily contain these effects,
since supersymmetry requires two  different Higgs doublets 
$\varphi_u$, $\varphi_d$ to give mass
to the $u$ and $d$ quarks.

In a model with two Higgs doublets, there are a total of 8 Higgs
degrees of freedom.    When the Higgs fields acquire vacuum
expectation values, 3 of these bosons are eaten by $W$ and $Z$ when
these particles obtain mass through the Higgs mechanism.  The
remaining physical Higgs particles include two CP-even neutral Higgs
bosons $h^0$ and $H^0$, a neutral pseudoscalar bosons $A^0$, and a
pair of charged Higgs bosons $H^\pm$.   Most of the parameter space
for such particles to have masses below 200~GeV has been excluded by
searches at the LHC~\cite{LHCHiggssearchesC,LHCHiggssearchesA}.  

  In general, these
particles correspond to  mixtures of the fields  in the original two
Higgs doublets.    The mixing angle that
defines the CP-even mass eigenstates is called $\alpha$.    For the 
CP-odd states, one mixture gives the eaten Goldstone bosons and
orthogonal combination gives
 the physical boson mass eigenstates.   The mixing angle that
defines these linear combinations is called $\beta$, with 
\beq
     \tan \beta =    \VEV{\varphi_u}/\VEV{\varphi_d} \ . 
\eeqn
The properties of the observed Higgs boson are then predicted to be
modified as a result of these mixings.  At the lowest order,
\beq
      g(hdd) = - {\sin \alpha\over \cos\beta} {m_d\over v} \qquad
       g(huu) =  {\cos \alpha\over \sin\beta} {m_d\over v}  \ .
\eeqn
The first of these modifications applies to the $b$ quark-Higgs coupling,
the second to the $c$ and $t$ couplings.

The Decoupling Theorem requires that the angles $\alpha$, $\beta$
cannot take arbitrary values but rather must be correlated.   For
example, in the minimal supersymmetric model,
\beq
   - {\sin \alpha\over \cos\beta}  = 1 + {\cal O}({m_Z^2\over
          m_A^2}) \ ,
\eeqn
consistent with the expected decoupling.

In supersymmetric models, the Higgs couplings also receive corrections
from loop diagrams involving the partners of the quarks and leptons.  Typically,
the largest effects come from diagrams with the $b$ squarks and the
gluino.  These diagrams obey decoupling, but they are enhanced  when
$\tan \beta$ is large.

\begin{figure}
\begin{center}
\includegraphics[width=0.70\hsize]{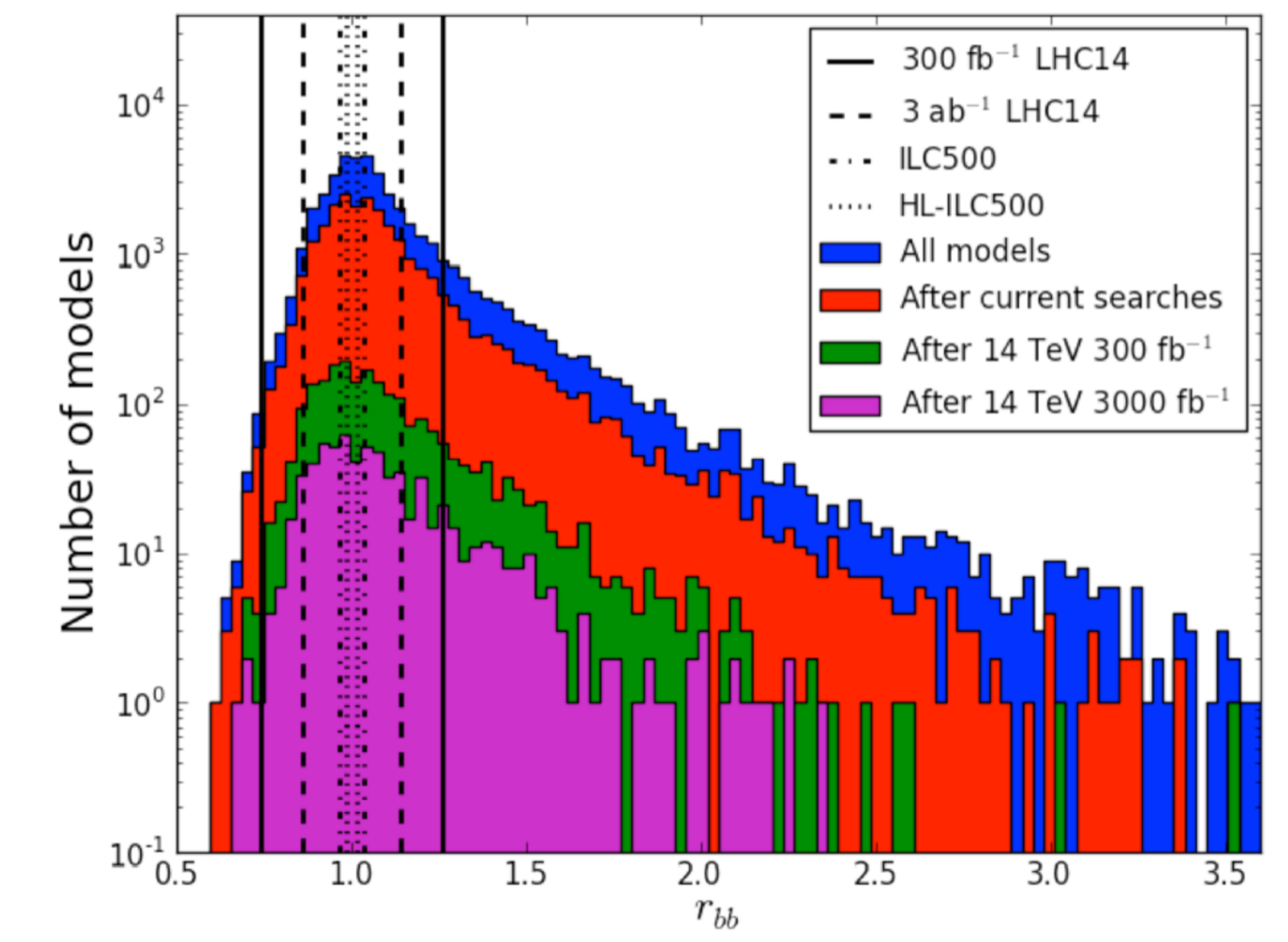}
\end{center}
\caption{Values of $r_{bb} = \Gamma(h\to b\bar b)/SM$ in a
  collection of about 250,000 allowed parameter points of the Minimal
  Supersymmetric  Standard Model, from \cite{PMSSM}.    The colored
  bands show models that can be discovered in new particle searches in
  the various stages of the LHC and the HL-LHC.}
\label{fig:PMSSMb}
\end{figure}

\begin{figure}
\begin{center}
\includegraphics[width=0.70\hsize]{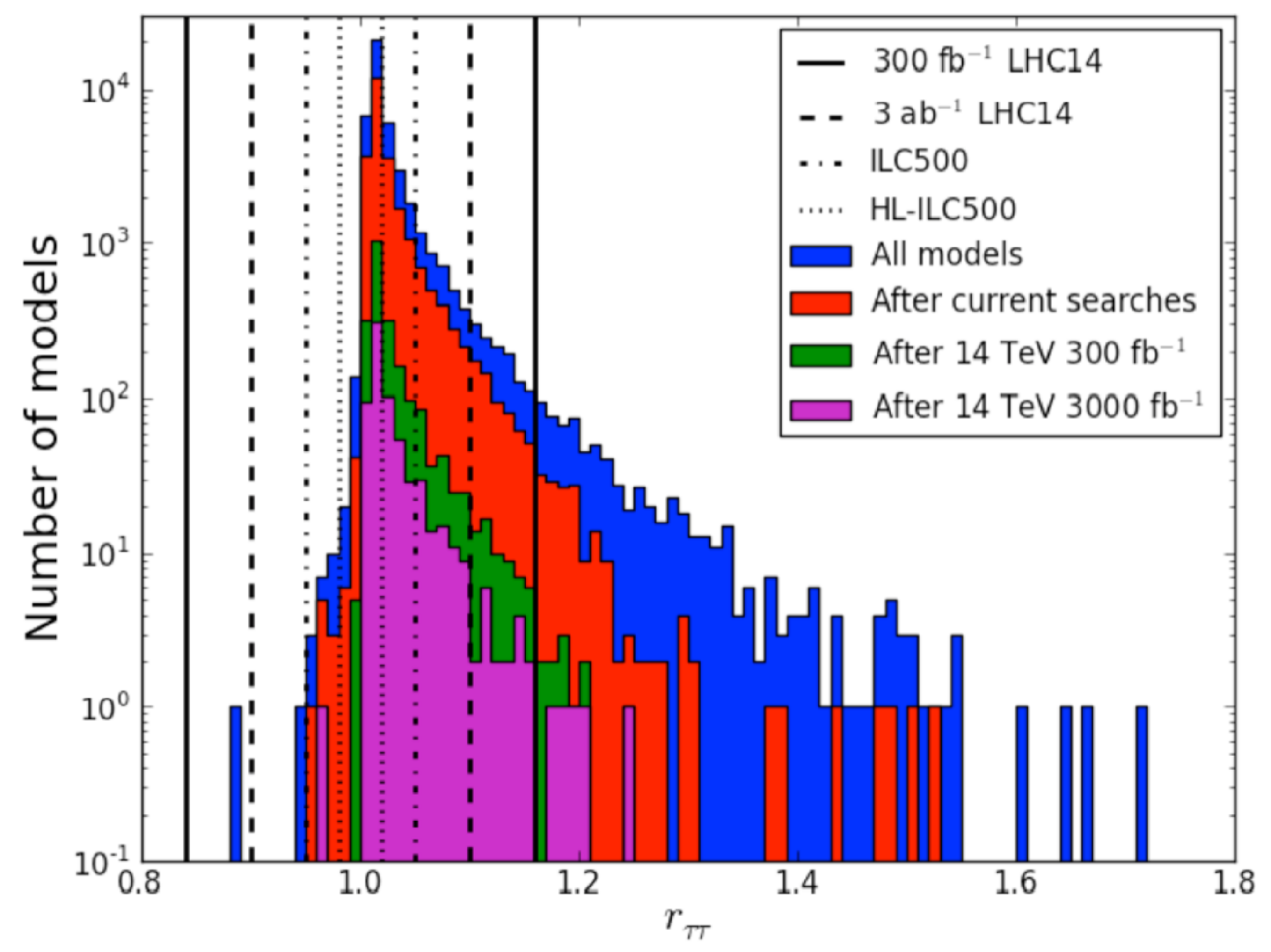}
\end{center}
\caption{Values of $r_{\tau\tau} = \Gamma(h\to \tau^+\tau^-)/SM$ in a
  collection of about 250,000 allowed parameter points of the Minimal
  Supersymmetric  Standard Model, from \cite{PMSSM}.    The colored
  bands show models that can be discovered in new particle searches in
  the various stages of the LHC and the HL-LHC.}
\label{fig:PMSSMtau}
\end{figure}

Figure~\ref{fig:PMSSMb}  shows the distribution of effects on the
Higgs couplng $g(hbb)$ seen in a large collection of supersymmetric
models constructed by Cahill-Rowley, Hewett, Ismail, and Rizzo~\cite{PMSSM}.   The
colored panels in the figure show the sensitivity of the models to 
searches for supersymmetric particles at the LHC.  It is interesting
that the constraint from a precision measurement of the Higgs coupling
to $b\bar b$ is essentially orthogonal to the current and expected
constraints from LHC searches.   Thus, the precision study of Higgs
couplings gives us a new and different way to probe for new physics.
Figure~\ref{fig:PMSSMtau} shows the comparable distribution for
perturbations of the coupling $g(h\tau\tau)$. 

It is important to note that, while the presence of multiple Higgs
doublets can have significant effects on the Higgs couplings to
fermions, it typically has a smaller effect on the Higgs couplings to
the $W$ and $Z$ bosons.  In the minimal supersymmetric model,
\beq
         g(hVV) =  { 2m_V^2\over v} \cdot \bigl(1 + {\cal O}({m_Z^4\over
         m_A^4})   \bigr)
\eeqn
for $V = W,Z$. 

However, there are many other scenarios in which the Higgs couplings
to $W$ and $Z$ are shifted as much as possible consistent with the
Decoupling
Theorem.    If the Higgs boson mixes with a Higgs singlet field of
mass $m_s$ by an 
angle $\gamma$, the
whole set of Higgs couplings is shifted by 
\beq
           g(hVV)  =   { 2m_V^2\over v} \cdot \cos \gamma 
\eeqn
where, typically, $\gamma \sim  m_h/m_s$.   A similar effect is produced
by loop corrections from any new particles that modify the Higgs boson 
self-energy diagrams~\cite{HanCH,CraigMcCullough}.

If the Higgs boson is a composite Goldstone boson, the Higgs couplings
are corrected in a similar way by the nonlinear Lagrangian generated
by spontaneous symmetry breaking.  This gives
\beq
           g(hVV)  =   { 2m_V^2\over v} \cdot (1 - v^2/F^2)^{1/2}
           \approx   { 2m_V^2\over v} \cdot (1 -\half  v^2/F^2) \ , 
\eeqn
an effect of 1--3\%. 

We have seen in the previous section that the decays
\beq
   h\to gg\ , \quad  h\to \gamma\gamma\ , \quad h \to \gamma Z 
\eeqn
proceed through loop diagrams in which the dominant contributions come
from particles for which $2M > m_h$.   Tnis means that new heavy
particles have the potential to make large corrections to the rates of
these
decays.  But this would only be true for particles that obtain their
full
mass from electroweak symmetry breaking. 

  As
we have discussed already,  the LHC measurement of $pp\to h\to
\gamma\gamma$ 
already excludes
a conventional fourth generation of quarks and lepton, up the mass at
which the Yukawa coupling exceeds the unitarity bound.   Any fermions
that we have not yet discovered must then be vectorlike fermions,  with
equal
electroweak quantum numbers for the left- and right-handed fields.
Such fermions
can obtain an $SU(2)\times U(1)$-invariant mass term that does not
require the Higgs field vacuum expectation value.
For example, in models with extra space dimensions, excitations in the
extra dimensions lead to separate Dirac fermion partners for the left- and
right-handed states, which obtain masses $M \sim \pi/R$, where $R$ is
the size of the 
extra dimensions.   The Higgs field can mix these states, leading to a
small correction $\delta M$ to the  mass matrix that depends 
on the Higgs vacuum
expectation value.  The relative shift in the masses due to the Higgs vacuum
expectation value is of the order of  $(\delta M)^2/M^2$, and so the
contribution of these particles to loop decays of the Higgs is
suppressed by this factor---just as we would expect from the
decoupling theorem.

\begin{figure}
\begin{center}
\includegraphics[width=0.70\hsize]{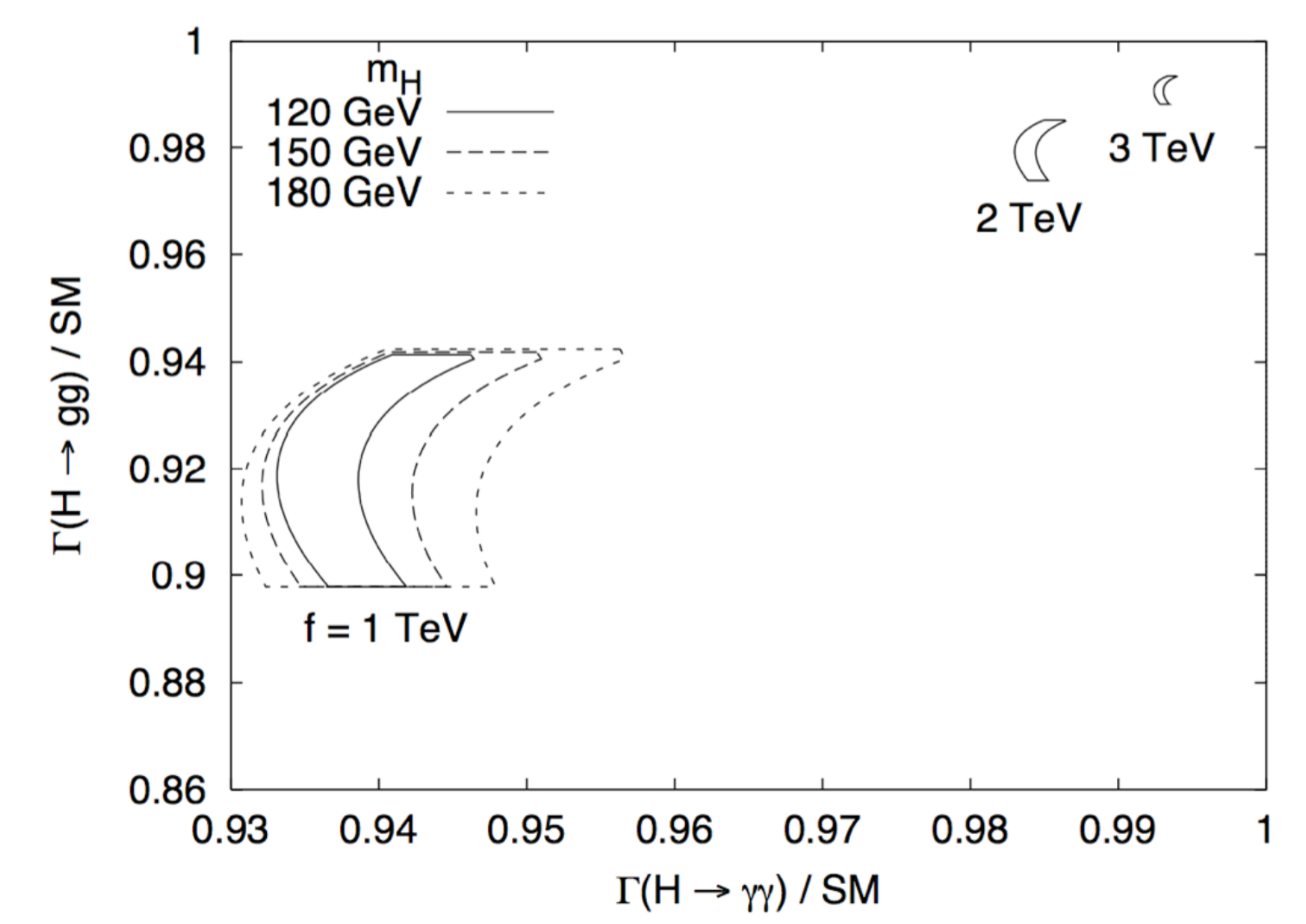}
\end{center}
\caption{Corrections to $\Gamma(h\to \gamma\gamma)$ and $\Gamma(h\to
  gg)$ in the Littlest Higgs model.}
\label{fig:Littlest}
\end{figure}

A similar effect is seen in Little Higgs models.   These models
typically contain several new heavy quarks, which also mix with the
top quark.  An estimate of the corrections to the loop decays in the
``Littlest
Higgs'' model is shown in
Fig.~\ref{fig:Littlest}~\cite{HanLittleHiggs}.   Mixing with heavy
states can also modify the top quark Yukawa coupling.   To fully
understand the origin of the effects, it is important to measure separately
the Higgs-gluon coupling and the Higgs-top coupling.   The LHC might
provide some complementary information by measuring Higgs boson
production  from
gluon fusion  at large $p_T$~\cite{GrojeanHiggsatlargepT}. 

The Higgs boson also has a self-coupling that determines the shape of
the Higgs potential.   This is something of a special case in the general
story of the Higgs couplings. 
 On one hand, the Higgs self-coupling is more difficult to
measure.  While there are realistic proposals to measure the other
Higgs couplings to the percent level, it will already be difficult to
measure the self-coupling to the level of 10--20\%  accuracy.  On the
other hand, there are models that require very large deviations of the
Higgs self-coupling form its SM value.   Theories of baryogenesis, the
origin of the matter-antimatter asymmetry of the universe, require a
period when the early universe was out of thermal equilibrium.  We are 
confident that the nonzero Higgs field expectation value was
established at a phase transition from a hot symmetric phase just
after the Big Bang.   In the SM, this phase transition is predicted to
be second-order and thus too smooth for substantial out-of-equilibrium
effects.    If the Higgs phase transition were strongly first-order,
then it is possible the the universe might have developed a
baryon-antibaryon asymmetry through CP- and baryon number violating
interactions available at that time~\cite{EWbaryoreview}.  This
requires values of the Higgs self-coupling substantially different
from that in the SM, a 50\% increase or more~\cite{NoblePerelstein}.

\begin{figure}
\begin{center}
\includegraphics[width=0.90\hsize]{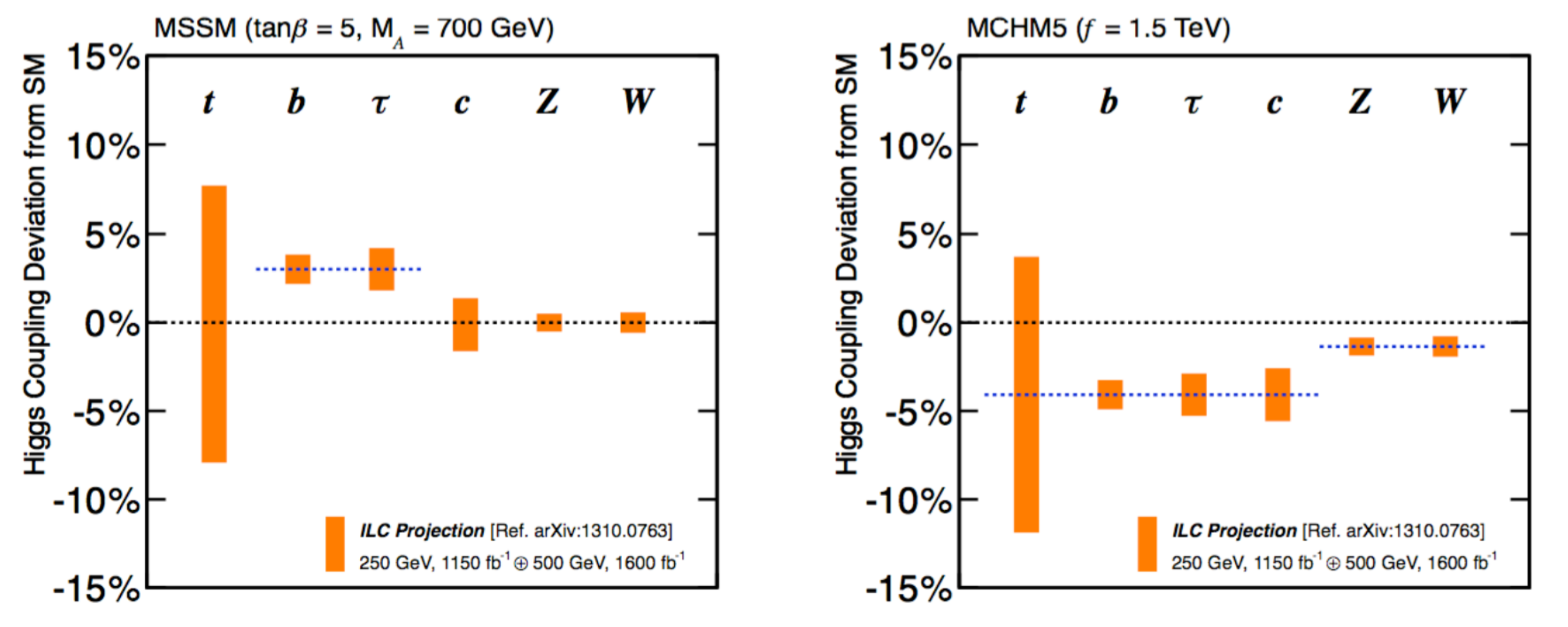}
\end{center}
\caption{Patterns of deviations in Higgs couplings, from
  \cite{ILCcase}.   These examples of nonstandard Higgs effects are
  taken from a broader survey in \cite{Kanemura}.}
\label{fig:Kanemura}
\end{figure}

The result of this survey of new physics effects is that each
individual Higgs coupling has its own personality and is guided by
different types of models.   In very broad terms:
\begin{itemize}
\item The Higgs couplings to fermions are sensitive to the presence of
  multiple Higgs doublets.
\item The Higgs couplings to $W$ and $Z$ are sensitive to the presence
  of Higgs singlets and to compositeness of the Higgs boson.
\item The Higgs couplings to $gg$ and $\gamma\gamma$ are senstive to
  the presence of new vectorlike fermions.
\item The Higgs coupling to $t\bar t$ is sensitive to new heavy
  fermions that mix with the top quark and to composite structure of
  the top quark.
\item The Higgs self-coupling  has large deviations from its SM
  value in models of baryogenesis at the electroweak scale.
\end{itemize}
Each model of new physics predicts is own pattern of deivations of the
Higgs couplings from the predictions of the SM.  Two examples of these
patterns, for specific supersymmetric and composite Higgs models, is 
shown in Fig.~\ref{fig:Kanemura}~\cite{Kanemura}. The challenge for us
to is measure the full suite of couplings with sufficient accuracy
that we can read this pattern and use it to gain information about
physics beyond the SM.

\subsection{Measurement of the Higgs boson properties at $\ee$
  colliders}

Given the interest in obtaining precise knowledge of the couplings of
the Higgs boson and the difficulty of reaching a sufficient level of
accuracy at the LHC, it is not surprising that there are a number of
proposals
for new $\ee$ colliders that would specifically address the
measurement of the Higgs couplings.   It would be very valuable to
study the Higgs boson with precision, in the same way that, in the
1990's, experiments at $\ee$ colliders carried out the precision study of
the $Z$ boson that I reviewed in Section 4 of these lectures.

\begin{figure}
\begin{center}
\includegraphics[width=0.70\hsize]{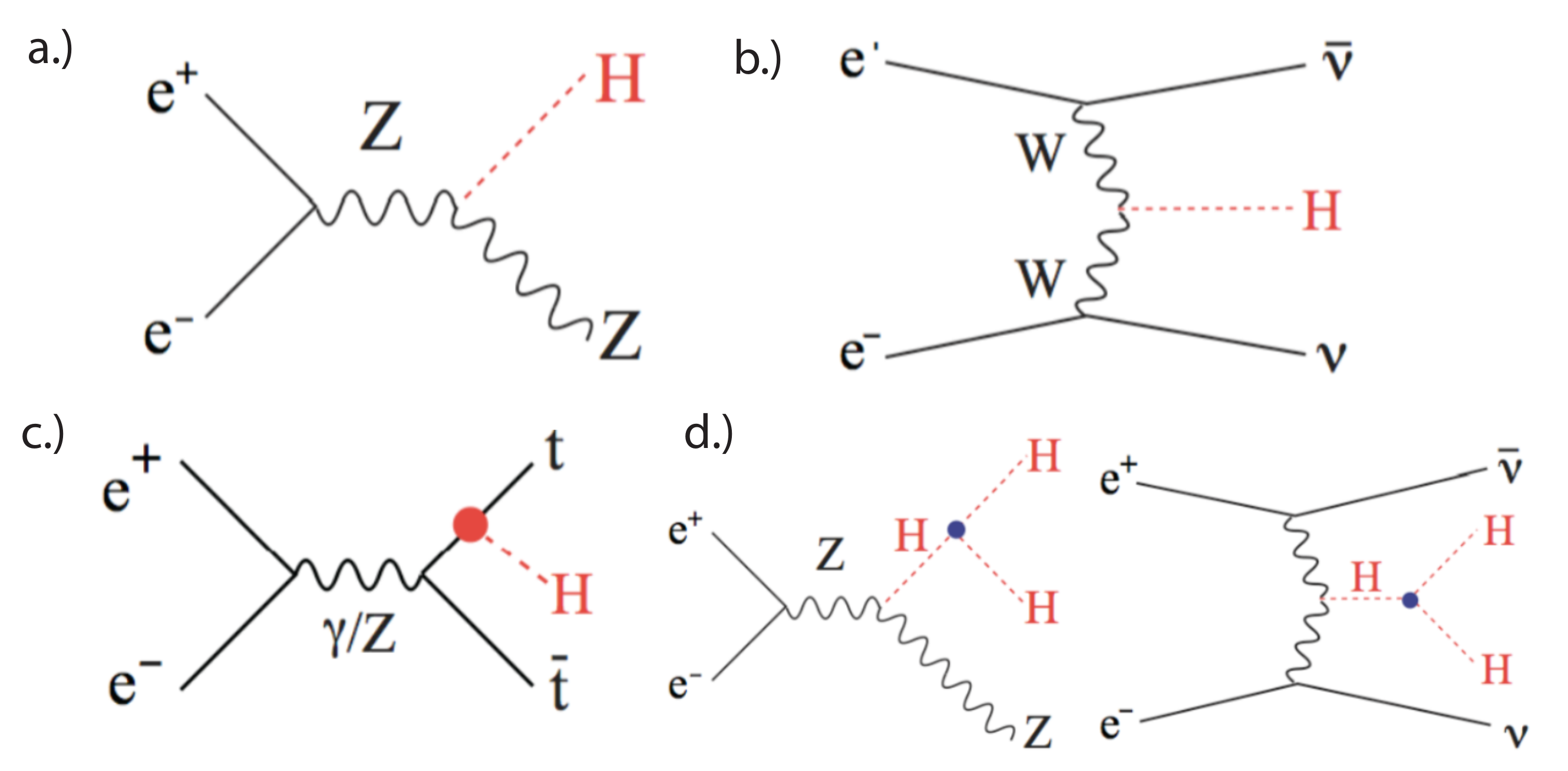}
\end{center}
\caption{Reactions producing the Higgs boson in $\ee$ collisions}
\label{fig:Hprocessesee}
\end{figure}

\begin{figure}
\begin{center}
\includegraphics[width=0.55\hsize]{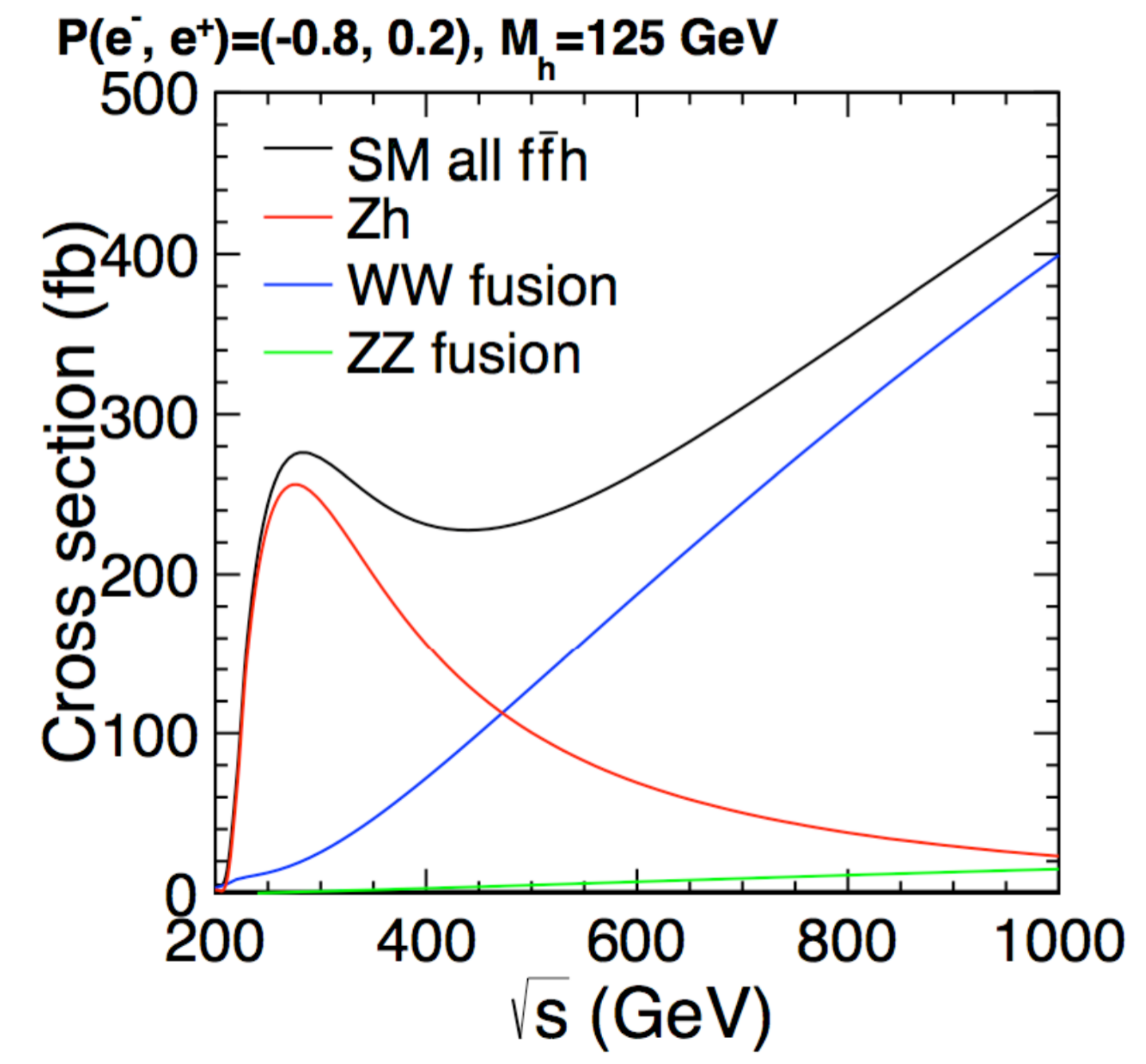}
\end{center}
\caption{Cross sections for Higgs production in $\ee$ collisions 
for a 125~GeV Higgs
  boson.}
\label{fig:Higgscsee}
\end{figure}

The most important processes for the production of a
Higgs boson at $\ee$ colliders  are those shown in
Fig.~\ref{fig:Hprocessesee}.
These are analogous to the corresponding processes in hadron-hadron
collisions shown in Fig.~\ref{fig:Hprocesses}.   The most important
reaction
near the Higgs threshold is radiation of the Higgs boson
from a $W$ or $Z$ (``Higgsstrahlung'').   At higher energies, Higgs
bosons are also produced by vector boson fusion, associated production
of a Higgs with a pair of top quarks, and the double Higgs production
reactions shown in the last line of the figure.  The cross sections predicted
for the Higgsstrahlung and fusion reactions  for a 125 GeV Higgs boson 
are shown in
Fig.~\ref{fig:Higgscsee}. 

Just as at hadron-hadron colliders, the different reactions available
at $\ee$ colliders have different advantages for the study of Higgs
boson decays.   Higgsstrahlung is available at the lowest center of
mass energy.  In this reaction, the Higgs boson is produced in association with
a $Z$ boson at a fixed  energy.    At 250~GeV in the center of
mass, the $Z$ boson has a lab frame energy of 110~GeV.  To a first
approximation, any $Z$ boson observed at this energy arises from the 
reaction $\ee\to Zh$, and whatever particles are  on the other side of the event
are the decay products of the Higgs boson.    This is an ideal setup
for measuring the branching ratios of the Higgs boson and for
discovering and identifying Higgs decays into exotic modes not
expected in the SM.  Also, since $\ee\to
Zh$ events can be recognized without reconstruction of the Higgs boson, this 
reaction allows a measurement of the absolute cross section rather
than a $\sigma\cdot BR$ as in \leqn{sigmatimesBR}.   Then this
reeaction
 can be used to determine the absolute magnitude of the
$Z$-Higgs
coupling.

The remaining reactions have complementary advantages.  Using the
Higgs branching ratio to $b\bar b$ measured with Higgsstrahlung, the
$WW$ fusion reaction can complement and firm up the measurement of the
absolute normalization of Higgs couplings.   As we see from
Fig.~\ref{fig:Higgscsee}, this reaction also gives higher statistics
for Higgs decays
at energies well above the threshold.    The remaining processes allow
the measurement of the Higgs coupling to top quarks and the Higgs
self-coupling.

\begin{figure}
\begin{center}
\includegraphics[width=0.65\hsize]{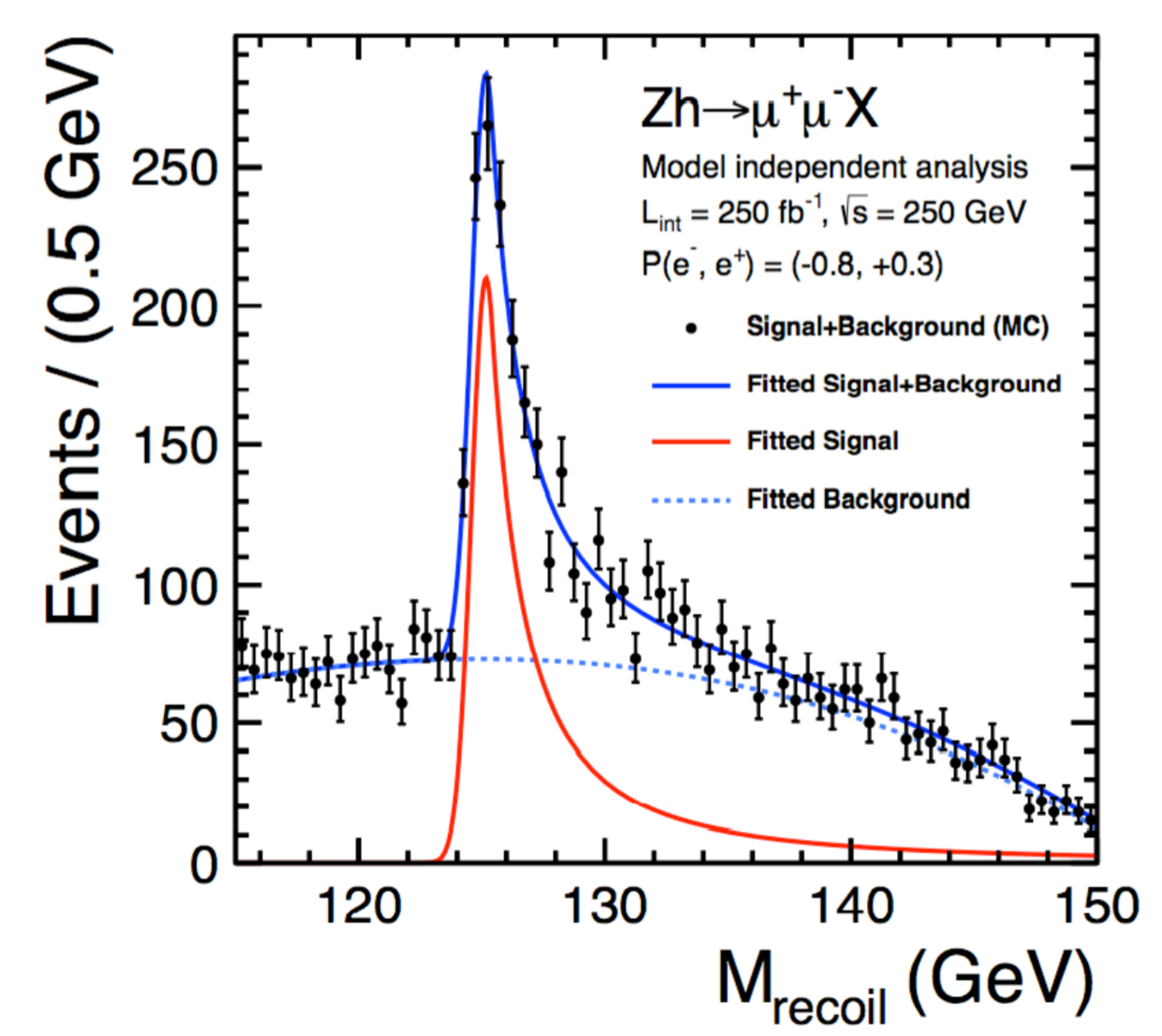}
\end{center}
\caption{Recoil mass distribution in $\ee\to Zh$, $Z\to \mu^+\mu^-$,
  from \cite{Recoil1}}
\label{fig:mumumass}
\end{figure}

\begin{figure}
\begin{center}
\includegraphics[width=0.6\hsize]{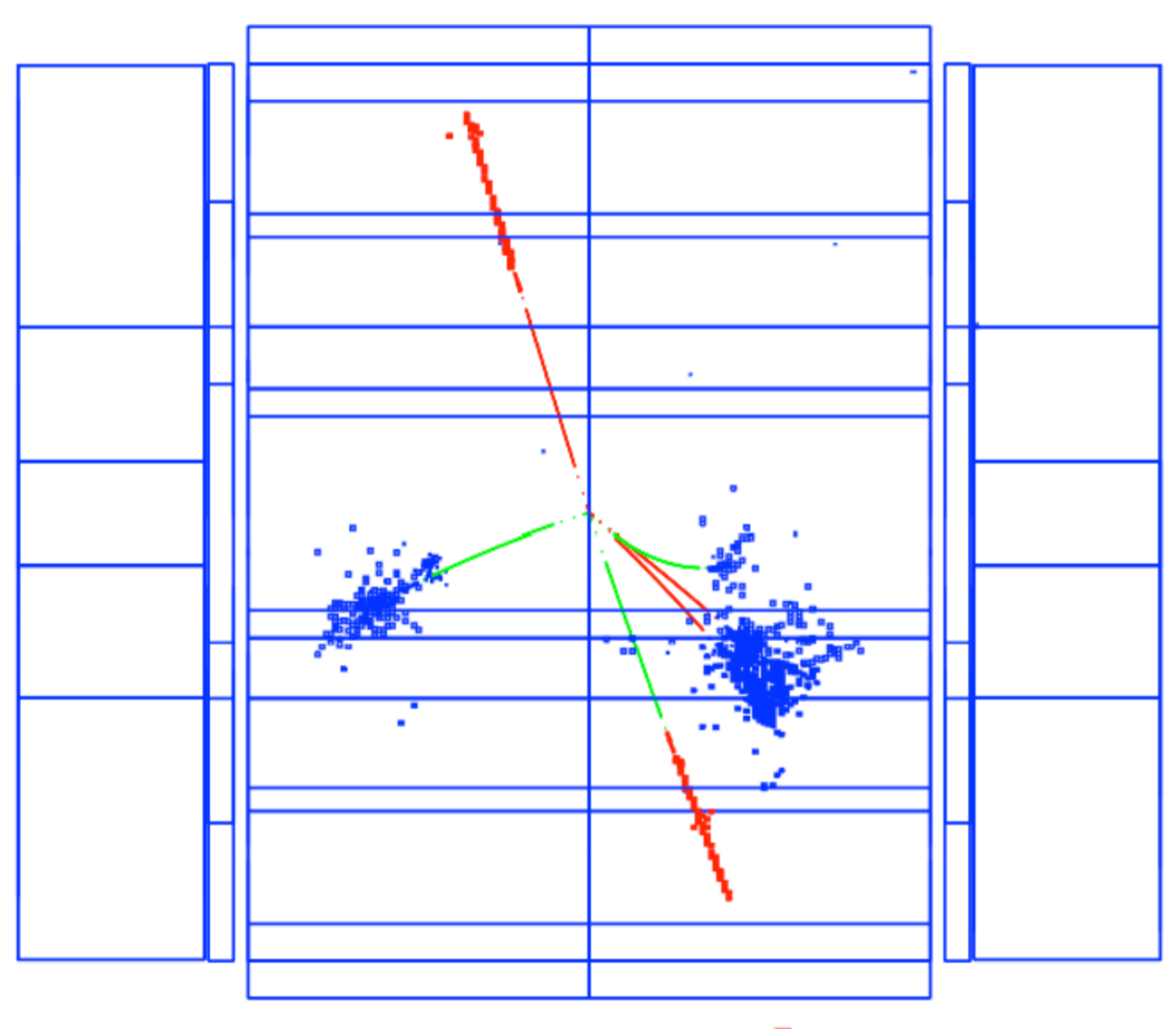}
\end{center}
\caption{Event display of an $\ee\to Zh$, $h \to
  \tau^+\tau^-$ event simulated in the ILD detector~\cite{Ruan}.}
\label{fig:tautau}
\end{figure}

\begin{figure}
\begin{center}
\includegraphics[width=0.99\hsize]{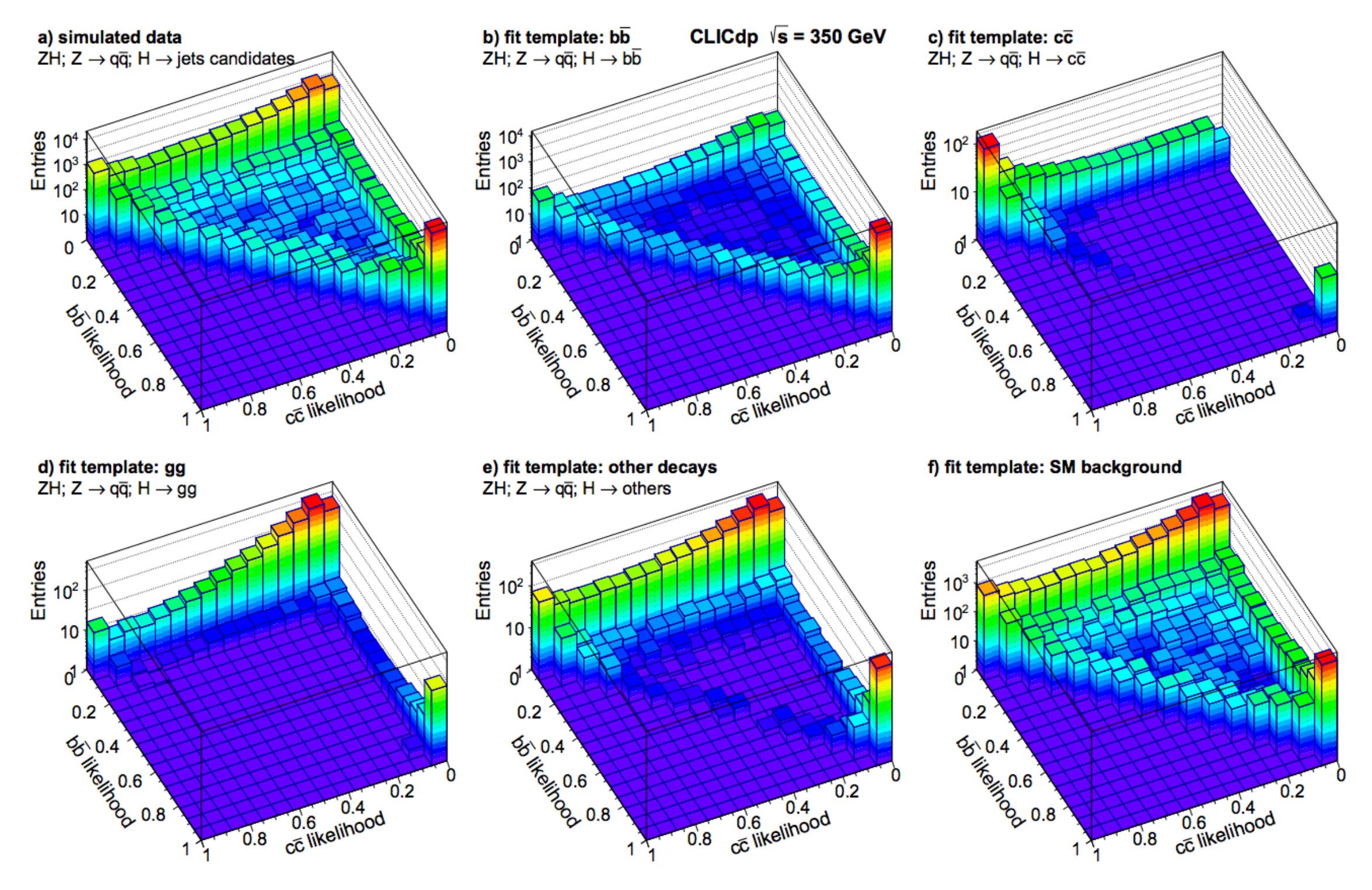}
\end{center}
\caption{Identification of Higgs boson decays to hadronic final states
  by template fitting, from \cite{CLIC}. Note in particular the sharp
  discrimination of the modes $h\to b\bar b$, $h\to c\bar c$, $h\to gg$.}
\label{fig:separateHiggs}
\end{figure}

\begin{figure}
\begin{center}
\includegraphics[width=0.55\hsize]{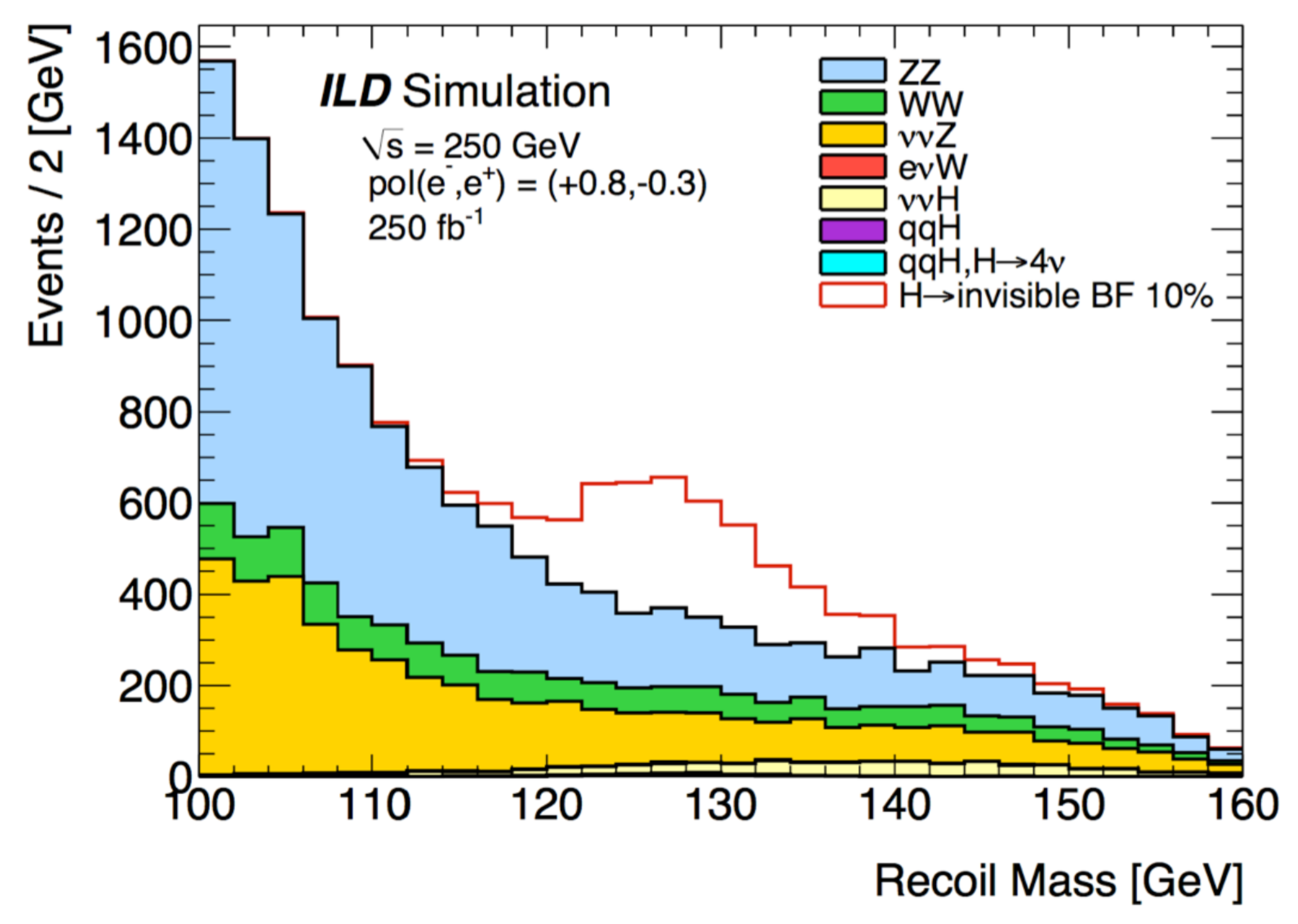}
\end{center}
\caption{Recoil mass distribution for $\ee\to Z +$~missing events,
  assuming a 10\% branching ratio of the Higgs boson into invisible
  modes, from \cite{invisible}.}
\label{fig:hinvisible}
\end{figure}

A compete description of the program of Higgs studies at $\ee$
colliders can be found in \cite{ILCTDR}.   Here I will just provide
some snapshots of this program.   The recoil mass spectrum in the
reaction $\ee\to Zh$, $Z\to \mu^+\mu^-$ is shown in
Fig.~\ref{fig:mumumass}.    The main background is $\ee\to ZZ$ plus
initial state radiation, a reaction that is understood to very high
accuracy.
We estimate that this measurement gives the Higgs boson mass with an
accuracy of 15~MeV~\cite{Recoil1}.   The precision Higgs coupling program actually
needs a Higgs boson mass with this high accuracy. The partial
widths for $h\to WW$ and $h\to ZZ$ depend strongly on the Higgs mass,
so that this accuracy already corresponds to a 0.1\% systematic error
on the SM predictions.   Figure~\ref{fig:tautau} shows a
Higgsstrahlung event with Higgs decay to $\tau^+\tau^-$.   In general,
these events are very characteristic of the various $Zh$ event
topologies.
Figure~\ref{fig:separateHiggs}, from the physics study for the CLIC
accelerator,
 shows the separation of Higgs
events$\ee$ annhiliation events at 250 GeV into 4 Higgs categories and
one background category by template fitting~\cite{CLIC}.   The figure shows that
the modes $h\to gg$ and even $h \to c\bar c$, which has a 3\%
branching ratio in the SM, can be cleanly extracted.
Figure~\ref{fig:hinvisible} shows the recoil mass distribution for
events with a $Z$ boson plus missing momentum.   The simulation
assumes a high value (10\%) for the Higgs branching ratio to invisible 
decay products, but the figure makes  clear that this process is visible at much
smaller values of the branching ratio, well below 1\%~\cite{invisible}.

\begin{figure}
\begin{center}
\includegraphics[width=0.85\hsize]{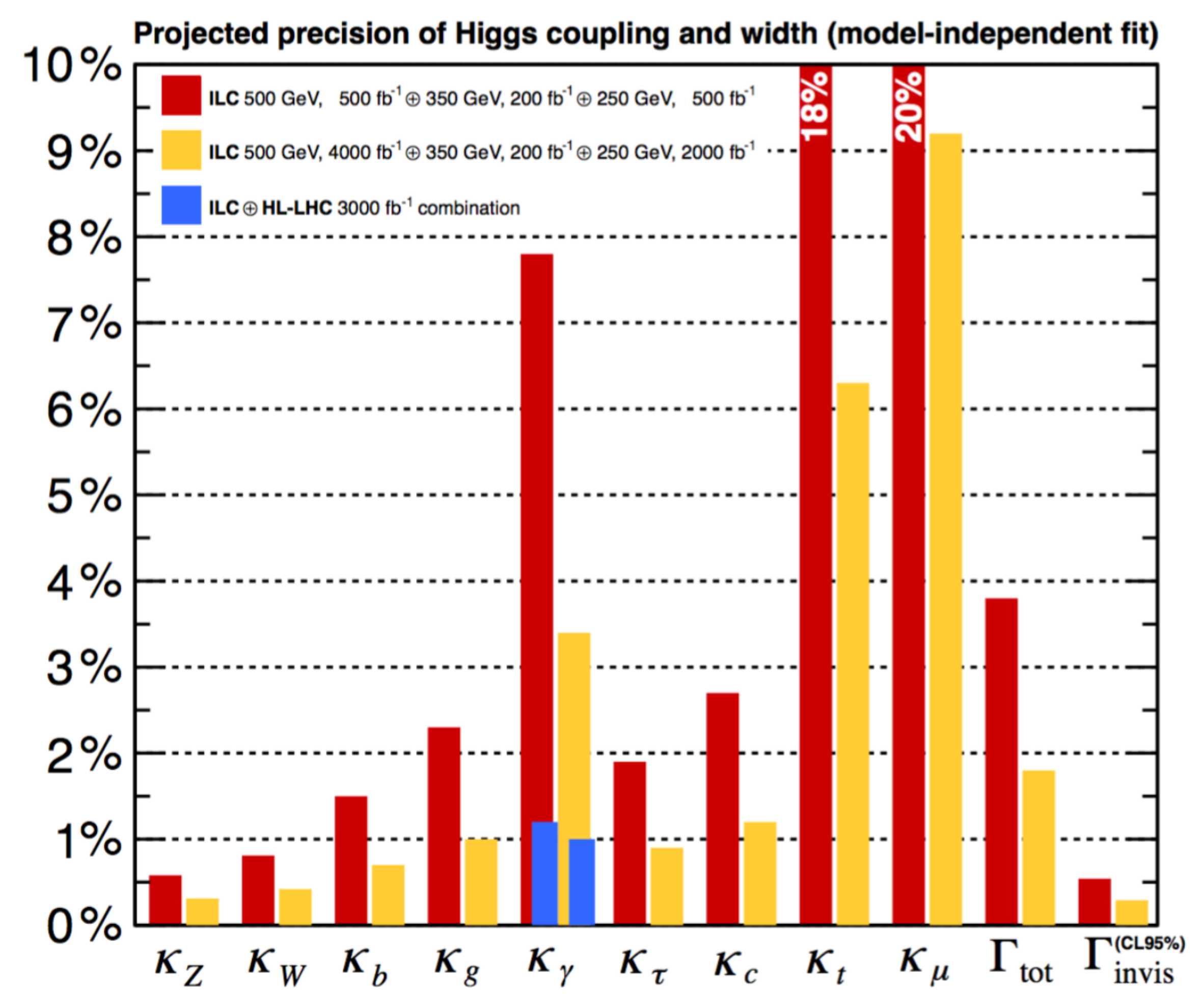}
\end{center}
\caption{Higgs coupling uncertainties projected for the ILC, from \cite{ILCcase}.}
\label{fig:Higgsacc}
\end{figure}

Finally,
Figure~\ref{fig:Higgsacc}, from \cite{ILCcase}, shows the 
accuracies for the determination of Higgs couplings to the full range
of SM particles projected for the complete program of the
International Linear Collider  (ILC).   For the Higgs decay to
$\gamma\gamma$, the blue histograms show the result of combining the
ILC data with the LHC measurement of $BR(h\to \gamma\gamma)/BR(h\to
ZZ^*)$.   The accuracy of the measurement of the Higgs coupling to the
top quark is limited by the fact that this figure considers only ILC
running at 500~GeV and below.   Even an energy increase to 550~GeV
would improve the accuracy of this measurement to 3\%.

The precision study of Higgs boson couplings at an $\ee$ collider will
then yield a wealth of information about the properties of this
particle.  Through the logic of the previous section, that information
will give us insight not only into the existence of new physics beyond
the SM but also into its qualitative nature.
I look forward to this program as the next great
 project in the future of particle physics.

\section{Conclusions}

In these lectures, I have developed the theory of the weak interaction
from its experimental foundations in the \VmA\ effective theory,
through the precision study of $SU(2)\times U(1)$ couplings at the $Z$
resonance, to the present and future study of the couplings of the
Higgs boson.   We have learned much about this fundamental interaction
of nature, but there is much more that we need to learn, and that we can
learn from future experiments.   The study of the weak interaction is
not a closed subject but one that still contains tantalizing questions
and promises to open new chapters in our exploration of particle
physics.

\end{document}